\documentclass[fleqn,usenatbib]{mnras}

\usepackage{newtxtext,newtxmath}

\usepackage[T1]{fontenc}
\usepackage{ae,aecompl}

\usepackage{graphicx}	
\usepackage{amsmath}	
\usepackage{amssymb}	
\usepackage{xcolor}

\title[Variability from molecular clouds and gas inflow]{Stochastic modelling of star-formation histories II: star-formation variability from molecular clouds and gas inflow}

\author[Tacchella, Forbes \& Caplar]{
Sandro Tacchella,$^{1}$\thanks{E-mail: sandro.tacchella@cfa.harvard.edu}
John C. Forbes,$^{2}$\thanks{E-mail: jforbes@flatironinstitute.org}
Neven Caplar$^{3}$\thanks{E-mail: ncaplar@princeton.edu}
\\
$^{1}$Center for Astrophysics $|$ Harvard \& Smithsonian, 60 Garden St, Cambridge, MA 02138, USA\\
$^{2}$Center for Computational Astrophysics, Flatiron Institute, 162 Fifth Avenue, New York, NY 10010, USA \\
$^{3}$Department of Astrophysical Sciences, Princeton University, 4 Ivy Ln.,  Princeton, NJ 08544, USA\\
}

\date{Draft version of \today}

\pubyear{2020}

\begin{document}
\label{firstpage}
\pagerange{\pageref{firstpage}--\pageref{lastpage}}
\maketitle

\begin{abstract}
A key uncertainty in galaxy evolution is the physics regulating star formation, ranging from small-scale processes related to the life-cycle of molecular clouds within galaxies to large-scale processes such as gas accretion onto galaxies. We study the imprint of such processes on the time-variability of star formation with an analytical approach tracking the gas mass of galaxies (``regulator model''). Specifically, we quantify the strength of the fluctuation in the star-formation rate (SFR) on different timescales, i.e. the power spectral density (PSD) of the star-formation history, and connect it to gas inflow and the life-cycle of molecular clouds. We show that in the general case the PSD of the SFR has three breaks, corresponding to the correlation time of the inflow rate, the equilibrium timescale of the gas reservoir of the galaxy, and the average lifetime of individual molecular clouds. On long and intermediate timescales (relative to the dynamical timescale of the galaxy), the PSD is typically set by the variability of the inflow rate and the interplay between outflows and gas depletion. On short timescales, the PSD shows an additional component related to the life-cycle of molecular clouds, which can be described by a damped random walk with a power-law slope of $\beta\approx2$ at high frequencies with a break near the average cloud lifetime. We discuss star-formation ``burstiness'' in a wide range of galaxy regimes, study the evolution of galaxies about the main sequence ridgeline, and explore the applicability of our method for understanding the star-formation process on cloud-scale from galaxy-integrated measurements.
\\
\end{abstract}




\section{Introduction}
The past two decades have been momentous in understanding the buildup of galaxies through most of cosmic time, especially in terms of their global properties and behaviour. Today, the unknown physics of star formation and feedback represent the main uncertainty in our understanding of galaxy formation \citep[e.g.][]{naab17}. It is challenging to gain further insight on star formation and feedback because these processes act on multiple spatial and temporal scales. In this paper, we address the question of how different processes -- external and internal to galaxies -- shape the variability of star formation on a wide range of timescales.

On galactic and cosmic scales, star formation as manifested in the cosmic star-formation rate (SFR) density and the star-forming main sequence can be understood as an interplay between the buildup of dark matter haloes and self-regulation, as demonstrated by empirical and semi-analytical models as well as more detailed hydrodynamical models that link recipes of star formation to $\Lambda$CDM structure formation \citep[e.g.][]{somerville15, wechsler18}. The cosmic SFR density increases from the early Universe to the peak at Cosmic Noon (redshift $z\sim1-3$) because of an increase in both the gas accretion rate of individual galaxies and the number density of galaxies that are able to form stars efficiently \citep[e.g.][]{hernquist03, springel03_csfrd, tacchella13, behroozi13b, birrer14}. Toward lower redshifts, the cosmic SFR density declines because the accretion rates decline and some galaxies fully cease their star formation (``quenching'', e.g., \citealt{faber07, peng10_Cont, schaye10, renzini16}). 

Closely related, a wide range of models reproduce the observed evolution of the star-forming main sequence (correlation between SFR and stellar mass $M_{\star}$; \citealt{brinchmann04, noeske07, daddi07, whitaker14, boogaard18}), where the decline of the specific SFR ($\mathrm{sSFR}=\mathrm{SFR}/M_{\star}$) with time is consistent with the decline of the gas accretion rate onto galaxies, which itself is closely related to the evolution of the cosmological specific accretion rate into dark matter haloes \citep{neistein06, noeske07b, bouche10, dutton10, lilly13_bathtub}. The observed, rather small scatter of the star-forming main sequence suggests that galaxies self-regulate their growth and propagate along this SFR$-M_{\star}$ ridgeline. Although galaxies' dark matter haloes build hierarchically, most stars from in ``normal'', main-sequence galaxies, which sustain their SFRs for extended periods of time in a quasi-steady state of gas inflow, gas outflow, and gas consumption \citep{bouche10, daddi10, genzel10, tacconi10, dave12, dekel13, lilly13_bathtub, forbes14a, forbes14b, hopkins14, mitra15, tacchella16_MS}. 

The key question -- which timescale is encoded in the main sequence scatter? -- remains observationally unanswered. As discussed in \citet[][see also \citealt{munoz15}]{abramson15}, if the scatter arises due to short-term fluctuations in the star-formation history, similar-mass galaxies mostly grow-up together \citep[e.g.][]{peng10, behroozi13b}. On the other hand, if the main sequence scatter arises due to long-term fluctuations, similar-mass galaxies do not grow-up together and the star-forming sequence is not an attractor solution, but is just an observed coincidence \citep{gladders13}. In the latter case, we expect that the star-formation histories of galaxies are correlated throughout the age of the Universe \citep{kelson14}.

Theoretically, based on the cosmological zoom-in simulations VELA, \citet{tacchella16_MS} proposed that the self-regulated evolution of galaxies through phases of gas compaction, depletion, possible replenishment, and eventual quenching, leads to an attractor state of the main sequence. While galaxies evolve about the main sequence ridgeline, they build up their bulge component \citep{tacchella16_profile}. \citet{rodriguez-puebla16} found that the scatter of the halo mass accretion rate in the Bolshoi-Planck simulation is comparable to the scatter of the main sequence, suggesting that halo accretion histories play an important role in determining the scatter. Based on the EAGLE simulations, \citet[][see also \citealt{katsianis19}]{matthee19} show that the main sequence scatter originates from a combination of fluctuations on short timescales (ranging from $0.2-2$ Gyr) that are associated with self-regulation and on longer timescales that are related to differences in the halo accretion. Models such as these forecast how galaxies oscillate about the main sequence ridgeline, making predictions concerning oscillation timescales and star-formation distribution on spatially resolved scales. However, an important caveat in numerical simulation studies is the effect of stochasticity of the SFR due to small-scale dynamical chaos, seeded by numerical round-off and through explicit randomness in many subgrid models \citep{genel19, keller19}.

In more general terms, the scatter of the main sequence is directly related to the variability of star-formation histories of individual galaxies, which encodes a wealth of information about the star-formation process, baryon cycle, galaxy-galaxy mergers, strength of stellar and black hole feedback and dark matter accretion histories. On short timescales ($<100~\mathrm{Myr}$), the strength of the fluctuations may encode the formation and destruction of individual giant molecular clouds (GMCs) where feedback is locally too weak to prevent gravitational collapse \citep{scalo84, scalo86, quillen08, kruijssen14_center, krumholz15_center, torrey17, faucher-giguere18, orr19}, or even individual massive stars when the SFR is sufficiently low that the IMF is not fully sampled \citep[e.g.][]{fumagalli11,da-silva12, da-silva14}. On intermediate timescales ($\sim0.1-1~\mathrm{Gyr}$), galaxy mergers, disk instabilities, galactic winds, bar-induced inflows and environmental effects are thought to drive the variability of star formation \citep{gunn72, hernquist89, mihos96, robertson06, oppenheimer08, mcquinn10, dekel14_nugget, zolotov15, tacchella16_MS, sparre17_bursty, torrey18, wang19}. On the longest timescales ($>1~\mathrm{Gyr}$), dark matter haloes of galaxies, as well as galaxy quenching, play crucial roles \citep{behroozi13b, birrer14, behroozi19, moster13, moster18, rodriguez-puebla16, tacchella18}. 

It is challenging to invalidate these predictions observationally based solely on estimates of the SFR from individual bands or lines of an individual galaxy because these quantities are only measured at one point in a galaxy's lifetime. The maturing approach of spectral energy distribution modelling allows a more robust measurement (in particular of the associated uncertainties) of the star-formation history of individual galaxies on long timescales \citep{pacifici16, iyer17, iyer19, leja17, leja19_nonparm, carnall19_sfh}. However, star-formation histories of individual galaxies are still poorly determined from observations on short timescales ($<100~\mathrm{Myr}$). The short term variability of star formation, i.e. ``burstiness'', is usually inferred by comparing a shorter to a longer SFR tracer, such as nebular emission line versus continuum flux \citep{weisz12, kruijssen14, kruijssen18, guo16, emami19, broussard19, faisst19, caplar19}. The main difficulty is to break the degeneracy between these measures of burstiness, and variation in the initial mass function, ionization parameter, dust attenuation, stellar population model, or metallicity \citep[e.g.][]{johnson13, shivaei18, haydon18,  broussard19}.

To quantify the variability of star formation on a \textit{range} of different timescales, \citet{caplar19} introduced the framework of modelling star-formation histories as a stochastic process, which can be described through a power spectral density (PSD). The PSD framework, widely used for stochastic modelling of the short-term variability ($\sim1~\mathrm{yr}$) of active galactic nuclei (AGN; \citealt{kelly09, macleod10, macleod12, dexter11, caplar17_ptf}), is a powerful tool to measure the amount of power contained in SFR fluctuations on a given timescale and therefore encodes the variability or ``burstiness'' on that timescale. Iyer et al. (in prep.) show that the PSD contains a wealth of information about different physical processes and that different galaxy formation models (ranging from empirical models to hydrodynamical simulations) predict widely different PSDs for similar-mass galaxies, i.e. measuring and constraining the PSD parameters can be used to differentiate and constrain theoretical models. Based on $z\sim0$ observations of the main sequence scatter \citep{davies19}, star-formation histories of galaxies with $M_{\star}\approx10^{10}~M_{\odot}$ show a break in their PSD at $\sim200~\mathrm{Myr}$, indicating that galaxies lose ``memory'' of their previous activity on this timescale. A similar conclusion was reached by \citet{hahn19}, who studied the scatter of the stellar-to-halo mass relation. \citet{wang20} used MaNGA data to constrain the PSD of star formation in local galaxies, finding slopes between 1.0 and 2.0, which indicates that the power of the star-formation variability is mostly contributed by longer timescale variations. 

In this paper, we link this PSD framework to physical processes that take place in and around galaxies. Specifically, processes acting on different spatial scales, including formation and distribution of GMCs, spiral arms, galaxy-galaxy mergers, and galaxy- and halo-scale cosmological inflows and outflows, drive variations in the SFR on different timescales, which are reflected and measurable through the PSD. \textit{What does the variability of star formation tell us about the physics of galaxies?} Is the variability mainly related to gas physics acting on large scales (e.g. accretion and inflow) or to small-scale cloud physics where the star-formation takes place? How does this depend on galaxy stellar mass and cosmic time? Bottom line: we aim at building a physical intuition for the PSD in this work.

Although hydrodynamical simulations have made tremendous progress in simulating large populations of galaxies at high resolution \citep[e.g.][]{hopkins14, schaye15, ceverino17, tremmel17, hopkins18_FIRE2, pillepich19, nelson19}, these models are still limited by exploring a small range of parameter space (it is difficult to re-run simulation with different input physics) and galaxy regimes (high versus low redshifts, low- versus high-mass galaxies). In order to bridge the gap between small and large scale physics and to explore a wide range of galaxy regimes, our approach is based on the idea of regulator models, where we model the gas cycle through galaxies with analytical equations that we numerically solve. The model is described in detail in Section~\ref{sec:model}. Our simplistic approach allows us to study how different physical processes drive the SFR variability, i.e. the shape of the PSD (Section~\ref{sec:PSD}). We discuss in Section~\ref{sec:discussion} how in the future, properties such as GMC lifetimes can be constrained from integrated observations and how key aspects of our model can be tested by both more detailed hydrodynamical models and observations. Furthermore, we show how gradients across the main sequence depend on the timescale of the SFR indicator. Finally, we conclude in Section~\ref{sec:conclusions}. Throughout this paper, we denote with $\langle \cdot \rangle$ the per-GMC average (approximately the same as the number-density-weighted average $\langle \cdot \rangle_{\rm n}$), while $\langle \cdot \rangle_{_{\rm M}}$ denotes the GMC mass-weighted average and $\langle \cdot \rangle_{\rm t}$ is the average over a long time interval relative to the dynamical timescale of the galaxy.

\section{From gas accretion to the process of star formation}
\label{sec:model}

In this section, we describe our model to understand the variability of star formation on different timescales, arising from both galaxy internal and external processes. We build on the considerations of the regulator model (Sections~\ref{subsec:regulator} and \ref{subsec:basic_regulator}), which uses mass conservation to derive the evolution of the gas mass and the SFR with cosmic time \citep[e.g.][]{bouche10, dave11, dave12, lilly13_bathtub, birrer14, forbes14b, dekel14_bathtub}. We call this model the basic regulator model throughout this work. In this basic regulator model, the variability of the SFR is driven by a stochastic inflow rate processed through a single gas reservoir. In Section~\ref{subsec:GMC}, we introduce the extended regulator model, which addresses a puzzle associated with the basic regulator model, namely exactly what gas is included in the gas reservoir. Specifically, we assume that the SFR is sustained by a population of individual GMCs, which form from the diffuse reservoir of gas (which we may now roughly identify as HI), and which have a certain star-formation efficiency and lifetime distribution. In this extended regulator model, the source of variability in the SFR is driven by both the variability of the inflow rate and the stochasticity of GMC formation. An overview of the key functions and parameters of our model are given in this section and in Table~\ref{tab:parameters}.

\begin{figure*}
    \centering
    \includegraphics[width=\textwidth]{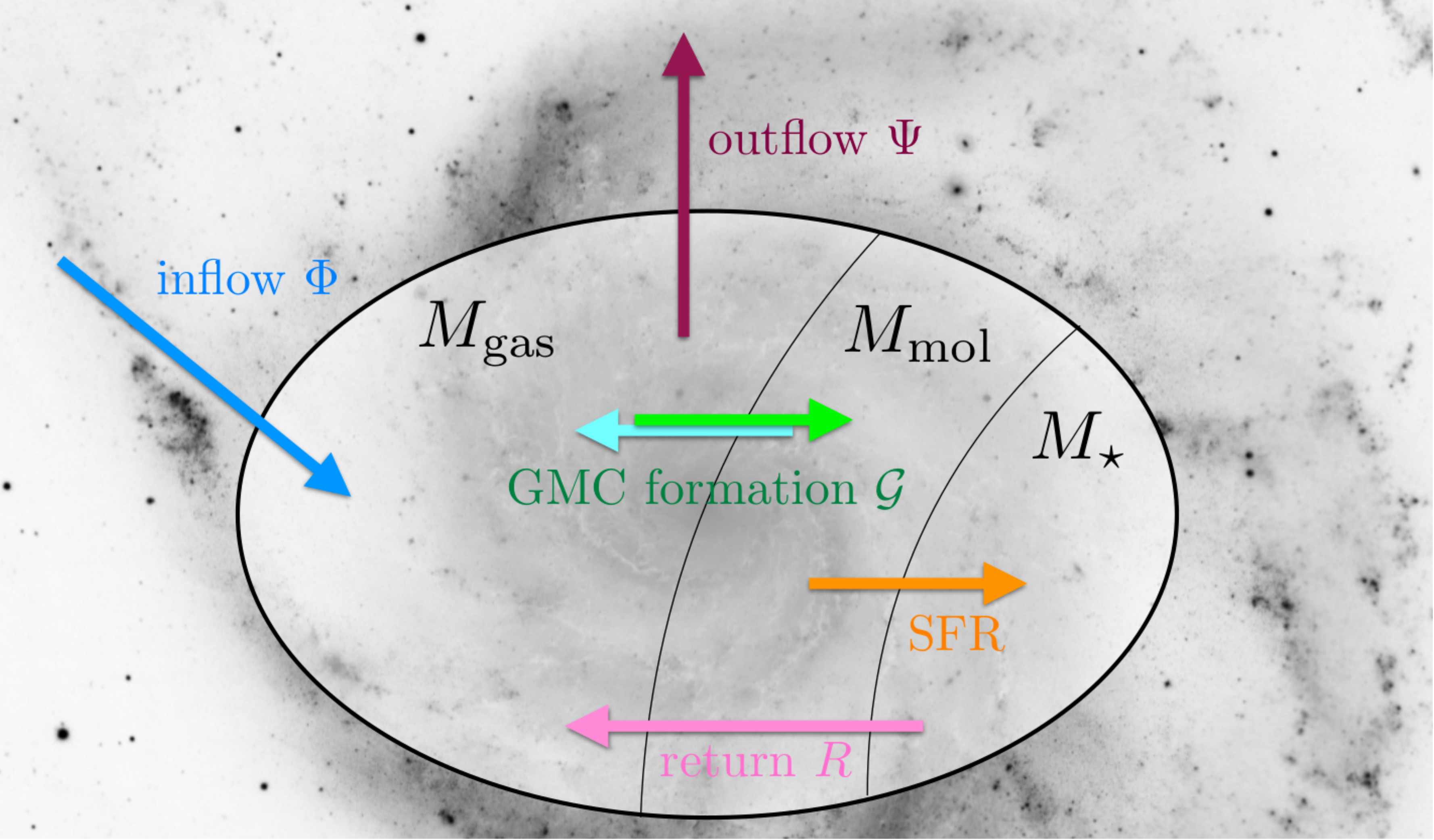}
    \caption{Illustration of how gas cycles through galaxies in our extended regulator model. Gas flows into the gas reservoir $M_{\rm gas}$ with an inflow rate $\Phi$ (blue arrow). Giant molecular clouds (GMCs) form out of this reservoir (green arrow) and constitute the molecular gas mass $M_{\rm mol}$. These GMCs sustain a SFR that builds up the stellar mass $M_{\star}$ (orange arrow), while some of this GMC mass is returned to the gas reservoir due to inefficient star formation (cyan arrow). Some of the stellar mass, which is not locked up in long-lived stars, is returned to the gas reservoir at a rate $\mathcal{R}$ (pink arrow). Finally, some gas is expelled from the galaxy by an outflow rate $\Psi$ that is assumed to be proportional to the SFR (red arrow). The mass of gas in the reservoir is free to vary and this regulates the formation of GMCs, and hence star formation. The main goal of this model is to understand the source of variability of the SFR due to gas inflow and stochastic GMC formation.}
    \label{fig:cartoon}
\end{figure*}

\begin{table*}
	\centering
	\caption{Functions and parameters our model, roughly split into regulator and GMC specific quantities. The columns indicate the variable names, give a short description, and provide additional information.}
	\label{tab:parameters}
	\begin{tabular}{lll}
		\hline \hline
 Function/parameter & Description & Additional comments \\ \hline \hline
 \textbf{Regulator specifics:} & & \vspace{0.1cm} \\ 
 
 $M_{\rm gas}(t)$ & gas mass of the reservoir & given by Eq.~\ref{eq:gas_regulator_general} \vspace{0.1cm}\\
 $M_{\rm mol}(t)$ & molecular gas mass  & given by the sum of all active GMCs \vspace{0.1cm}\\
 $M_{\star}(t)$ & stellar mass content of the galaxy & --- \vspace{0.1cm}\\
 $\mathrm{SFR}(t)$ & star-formation rate of the galaxy & given by Eq.~\ref{eq:depletion_time_basic} (basic) and Eq.~\ref{eq:SFR_full} (extended) \vspace{0.1cm} \vspace{0.1cm}\\
 
 $\Phi(t)$ & gas inflow rate into the reservoir & stochastic process, given by Eq.~\ref{eq:inflow_regulator} \\
 depends on: $x(t)$ & random variable (distributed standard normal) & described by PSD$_{\rm x}$ (Eq.~\ref{eq:PSD_x} with $\tau_{\rm x}$, $\beta_{\rm l}$, and $\beta_{\rm h}$) \\
 \hspace{1.53cm} $\mu$ & normalization of the gas inflow & typically within [0.1, 10.0] \\
 \hspace{1.53cm} $\sigma$ & normalization of variability of the gas inflow rate & typically within [0.5, 2.0] \vspace{0.1cm} \\

 $\Psi(t)$ & mass outflow rate & given by Eq.~\ref{eq:outflow} \\
 depends on: $\lambda$ & mass-loading factor & typically within $[0.1, 10.0]$ \vspace{0.1cm}\\

 $\mathcal{G}(t)$ & star-formation process & given by Eq.~\ref{eq:depletion_time_basic} (basic) and Eq.~\ref{eq:SF_process_full} (extended) \vspace{0.1cm}\\
 $\mathcal{R}(t)$ & mass return rate from stars to reservoir & specified in Appendix~\ref{app:mass_return} \vspace{0.1cm}\\
 $\tau_{\rm dep}$ & depletion time & given by Eq.~\ref{eq:depletion_time_basic} \vspace{0.1cm}\\
 $\tau_{\rm eq}$ & equilibrium timescale & given by Eq.~\ref{eq:eq_timescale} \vspace{0.1cm}\\ \hline

 \textbf{GMC specifics:} & & \vspace{0.1cm}\\ 
 
 $m_{\rm b}$ & mass of an individual GMC at formation (birth) & --- \vspace{0.1cm} \\
 $m$ & mass of an individual GMC & given by Eq.~\ref{eq:mass_GMC} \vspace{0.1cm} \\
 $\mathrm{SFR}_{\rm GMC}(m_{\rm b})$ & SFR of a GMC with birth mass $m_{\rm b}$ & given by Eq.~\ref{eq:GMC_SFR} \vspace{0.1cm} \\
 $N_{\rm birth}(t)$ & number of GMCs drawn in a given time step & given by Eq.~\ref{eq:N_birth} \vspace{0.1cm} \\
 $N_{\rm GMC}(t)$ & number of active / star-forming GMCs & given by Eq.~\ref{eq:relation_Nbirth_Ngmc} \vspace{0.1cm} \\
 $\tau_{\rm mol}$ & molecular gas formation timescale & calibrated via Eq.~\ref{eq:total_sfr_ext} \vspace{0.1cm}\\

 $n_{\rm b}(m_{\rm b})$ & GMC birth mass function & given by Eq.~\ref{eq:GMC_MF_birth} \\
 depends on: $\alpha_{\rm b}$ & power-law slope & typically within $\alpha_{\rm b}=[-1.5, 2.5]$ \\
 \hspace{1.53cm} $m_{\rm min}, m_{\rm max}$ & minimal/maximal mass of GMCs & $m_{\rm min}=10^{4}-10^{5}~\mathrm{M}_{\odot}$ and $m_{\rm max}=10^{7}-10^{9}~\mathrm{M}_{\odot}$ \vspace{0.1cm} \\
 
 $n_{\rm obs}(m)$ & observed GMC mass function & given by Eq.~\ref{eq:MF_conversion} \\
 depends on: $\alpha_{\rm obs}$ & low-mass power-law slope of observed mass function & $\alpha_{\rm obs}=\alpha_{\rm b}+\alpha_{\rm l}$ \vspace{0.1cm} \\

 $\tau_{\rm L}(m_{\rm b})$ & GMC lifetime & given by Eq.~\ref{eq:GMC_LT} \\
 depends on: $\tau_0$ & normalization & typically within [3.0, 50.0] Myr \\
 \hspace{1.53cm} $\alpha_{\rm l}$ & power-law slope & typically within [0, 0.5] \vspace{0.1cm} \vspace{0.1cm} \\
 
 $\varepsilon(m_{\rm b})$ & GMC star-formation efficiency & given by Eq.~\ref{eq:GMC_SFE}, upper limit of 1 \\
 depends on: $\varepsilon_0$ & normalization & typically within [0.01, 1.0] \\
 \hspace{1.53cm} $\alpha_{\rm e}$ & power-law slope & typically within [0, 0.5] \vspace{0.1cm} \\
 
	\hline \hline
	\end{tabular}
\end{table*}

\subsection{Regulator model -- general considerations}
\label{subsec:regulator}

The time-evolution of the gas and stellar mass content of a galaxy can be described with the regulator model. We build on the model described in \citet{lilly13_bathtub}, in which the SFR is regulated by the gas mass ($M_{\rm gas}$) in a reservoir within the galaxy. In the basic regulator model, stars form directly out of this reservoir, while in the extended regulator model, the gas mass of this reservoir only includes the neutral gas from which the molecular gas ($M_{\rm mol}$) forms, i.e. it does not include $M_{\rm mol}$, which we track separately via a population of GMCs. The key ingredients of this model are as follows.

The fundamental equation originates from mass conservation of the gas reservoir within the galaxy where we have separated the source and sink terms as in Fig.~\ref{fig:cartoon}:

\begin{equation}
    \frac{dM_{\rm gas}}{dt} = \Phi(t) + \mathcal{R}(t) - \Psi(t) - \mathcal{G}(t),
    \label{eq:gas_regulator_general}
\end{equation}

\noindent
where $\Phi(t)$ is the gas inflow rate to the gas reservoir, $\mathcal{R}(t)$ is the rate of mass returned from stars to the gas reservoir, $\Psi(t)$ is the outflow rate, and $\mathcal{G}(t)$ describes the rate at which mass from the reservoir is converted to molecular gas and stars, i.e. star-formation processes. 

We describe $\Phi(t)$ in detail in Section~\ref{subsec:phi}. It is important to note that the inflow rate $\Phi(t)$ not only includes pristine gas inflow into $M_{\rm gas}$ from the \textit{outside} of the dark matter halo, but describes all gas inflow into the gas reservoir. Specifically, $\Phi(t)$ describes the whole interface between the gas reservoir and the circumgalactic medium, which is not well-understood and rather complex \citep{tumlinson17}. This includes processes such as pristine gas inflow into the halo, gas cooling within the halo, cold streams feeding the galaxy, galaxy-galaxy mergers,  recycling of outflows, and flows of gas within the galaxy from regions of inefficient star formation (e.g. interarm regions or extended HI disks). Therefore, we parameterize $\Phi(t)$ as a general stochastic process, allowing it to describe a wide range of physical mechanisms. As we show later in the paper, $\Phi(t)$ is one of two key drivers of the time-variability of the SFR in our model. 

The rate at which mass is returned from stellar populations to the gas reservoir, $\mathcal{R}(t)$, is obtained from stellar population models \citep[e.g.][]{bruzual03}. We highlight the details in Appendix~\ref{app:mass_return}. Briefly, for a simple stellar population (SSP), $\mathcal{R}(t)$ exponentially declines with time, leading to a linear increase of the returned mass fraction ($f_{R}$) with log-time. When considering long-term evolution of the gas reservoir, instantaneous return is often assumed so that a fraction $(1-f_{R})$ steadily builds up a population of long-lived stars and stellar remnants \citep{tinsley80}. However, this assumption will break down in galaxies with young ages, such as high-redshift galaxies \citep[$z>4$; see][]{tacchella18}. Furthermore, \citet{leitner11} show that gas from stellar mass loss can provide most or all of the fuel required to sustain today's level of star formation in late-type galaxies. Therefore in our numerical implementation we track $\mathcal{R}(t)$ and compute it based on the star-formation history at each timestep self-consistently throughout this work.

Furthermore, star formation may drive a wind out of the galaxy, which we characterize as outflow rate throughout this paper as

\begin{equation}
    \Psi(t) = \lambda \cdot \mathrm{SFR}(t),
    \label{eq:outflow}
\end{equation}

\noindent
where $\lambda$ is the mass-loading factor. The mass-loading factor $\lambda$ is still uncertain, both observationally and theoretically, depending critically where it is measured (around star-forming regions or at a certain fraction of the virial radius) and how it is defined. In theoretical models, there is a significant difference between a measured mass-loading factor (ratio of the ouflow rate to the SFR) and the injected mass-loading factor in the sub-grid model \citep[e.g.][]{torrey19}. Recent observations \citep[e.g.][]{bouche12, newman12, bolatto13, kacprzak14, schroetter15, schroetter19,  davies19_outflow, forster-schreiber19, kruijssen19, chevance20} measure mass-loading factors of between 0.1 and 30.0 for star-forming galaxies at $z=0-2$, only depending weakly on stellar mass and redshift, which is in rough agreement with theoretical models that measure the mass-loading factor \citep[e.g.][]{muratov15, barai15, torrey19}. Connecting the regulator model with metallicity yields, \citet{zahid12} and \citet{lilly13_bathtub} constrain the mass-loading factor to be between 0.1 and 0.8 for star-forming galaxies in the local universe\footnote{The exact value depends on the adopted yield and $f_R$.}. Because of the large uncertainty in the mean value of $\lambda$, we assume it to be fixed in time in our model, but we note that this is an assumption that could be relaxed. In this work, we assume a fiducial value for the mass-loading factor of $\lambda=1.0$, and investigate variation between 0.1 and 30.0.

Finally, $\mathcal{G}(t)$ describes the star-formation process. As described in the next section (Section~\ref{subsec:basic_regulator}), in the basic regulator model, the SFR is directly related to gas mass in the reservoir via the depletion time. In Section~\ref{subsec:GMC}, we expand on this by assuming that star formation is sustained by GMCs, which constitute the molecular gas phase $M_{\rm mol}$. In this extended regulator model, we assume that the formation of GMCs is stochastic and, hence, $\mathcal{G}(t)$ is the second key driver of the variability of the SFR.

\subsection{Basic regulator model}
\label{subsec:basic_regulator}

In the basic regulator model, we assume that the SFR in the galaxy is determined by the mass of gas in the internal reservoir, which is motivated by a Kennicutt-Schmidt-type relation \citep{schmidt59, kennicutt89}:

\begin{equation}
    \mathcal{G}(t) = \mathrm{SFR}(t) = \frac{M_{\rm gas}(t)}{\tau_{\rm dep}},
    \label{eq:depletion_time_basic}
\end{equation}

\noindent
where $\tau_{\rm dep}$ is the depletion time. Equation \ref{eq:depletion_time_basic} is in some sense the definition of the depletion time, where again for simplicity we have assumed that the depletion time is constant. With these specifications (Eqs.~\ref{eq:outflow} and \ref{eq:depletion_time_basic}), Eq.~\ref{eq:gas_regulator_general} can be solved numerically.

In order to gain further intuition and link to earlier works, we can momentarily adopt a single constant value of $f_R$ and $\tilde{\Phi}(t) \equiv \Phi(t)/\tau_{\rm dep}$, which together with $\lambda$ and $\tau_\mathrm{dep}$ define the equilibrium timescale

\begin{equation}
    \tau_{\rm eq} = \frac{\tau_{\rm dep}}{1-f_R+\lambda}.
    \label{eq:eq_timescale}
\end{equation}

\noindent
We can then re-write Eq.~\ref{eq:gas_regulator_general} as:

\begin{equation}
    \frac{d\mathrm{SFR}(t)}{dt} + \frac{1}{\tau_{\rm eq}} \cdot \mathrm{SFR}(t) = \tilde{\Phi}(t).
    \label{eq:sfr_regulator}
\end{equation}

The equilibrium timescale $\tau_{\rm eq}$ describes the timescale by which the SFR converges towards a stable-state, equilibrium solution. This can directly be seen by solving Eq.~\ref{eq:sfr_regulator} analytically (assuming that $\Phi$, $f_R$, $\lambda$, and $\tau_{\rm dep}$ are all constant or only change slowly with time, see also \citealt{peng14_gasreg}):

\begin{equation}
\begin{split}
    \mathrm{SFR}(t) = & \left( \mathrm{SFR}(t_0) - \frac{\Phi(t)}{1-f_R+\lambda} \right) \cdot e^{-\tau_{\rm eq}^{-1}(t-t_0)} \\ & + \frac{\Phi(t)}{1-f_R+\lambda}.
\end{split}
\end{equation}

\noindent
The equilibrium SFR is given by the last term of this equation and it can be reached for $t>>\tau_{\rm eq}$. These considerations motivate calling this framework the ``regulator model'': the SFR of a galaxy regulates itself as described by Eq.~\ref{eq:sfr_regulator}. If the gas inflow rate goes up, the gas mass of the reservoir increases, and hence the SFR becomes larger. However, as the SFR increases, so does the consumption rate from the gas reservoir both via star formation and outflows, which slows the growth of the gas mass, and hence the SFR.

In order to solve Eq.~\ref{eq:sfr_regulator} generally and compute the time-evolution of the SFR, we need to specify the inflow rate $\Phi$, the depletion time $\tau_{\rm dep}$, and the mass-loading factor $\lambda$. In the general case, we expect $\Phi$, $\lambda$, and $\tau_{\rm dep}$ to vary with cosmic time and galaxy properties, such as stellar mass, gas density, and metallicity. For example, \citet{tacchella16_MS}, based on cosmological zoom-in simulations, showed that inflow rate, gas mass, gas distribution, and depletion time play a role in determining the SFR at a given stellar mass: galaxies above the main sequence ridgeline have higher cold gas masses in a more compact configuration and shorter depletion times than galaxies below it. This is consistent with recent observations that show such gradients about the main sequence \citep{genzel15, tacconi18, freundlich19}, though uncertainties remain in the determination of SFRs as well as gas masses that could lead to spurious correlations. In case of variable inflow, the observed SFR will be both function of this variable inflow rate and the parameters of the regulator. We refer reader to further discussion of this interplay in \citet{wang19}. \citet{wang19} also argued that variation of the depletion time plays only a secondary role in determining the variation of SFR with respect to the variation of inflow rate based on the regulator model.

From these considerations, we assume for the basic regulator model that the variations in $\Phi(t)$ are dominant, and variations in $\tau_{\rm dep}$ and $\lambda$ with time (and galaxy property) are negligible for the evolution of individual galaxies. Specifically, for a certain realization of the model (i.e. computation of $\mathrm{SFR}(t)$), we vary $\Phi(t)$, while holding the other parameters of the basic regulator model ($\tau_{\rm dep}$ and $\lambda$) fixed. As shown by Eq.~\ref{eq:sfr_regulator}, the key quantity that regulates the SFR of the galaxy in the basic regulator model is the equilibrium timescale (Eq.~\ref{eq:eq_timescale}), which itself depends to first order on $\tau_{\rm dep}$ and $\lambda$ when $\lambda \ga 1$, and weakly on $\mathcal{R}(t)$. In the extended regulator model (next section), in which star formation is sustained by GMCs, the depletion time varies self-consistently with time as described in Section~\ref{subsec:connection_basic_full}.

\subsection{Extended regulator model: star formation sustained by GMCs}
\label{subsec:GMC}

We now expand on the above formalism of the basic regulator model by adding more physical realism to the star-formation process $\mathcal{G}(t)$. This is necessary in order to describe the variability of the SFR by a combined effect of the inflow rate $\Phi$ and the galaxy-internal star formation process itself. In particular, at each timestep $\mathrm{d}t$, we assume that a population of GMCs is formed. The sum of all living (i.e. star-forming) GMCs make up the molecular gas mass $M_{\rm mol}$ and they sustain a certain SFR to build up the stellar mass of the galaxy $M_{\star}$, as shown in Fig.~\ref{fig:cartoon}.

\subsubsection{GMC parameterization}

\begin{figure*}
    \centering
    \includegraphics[width=\textwidth]{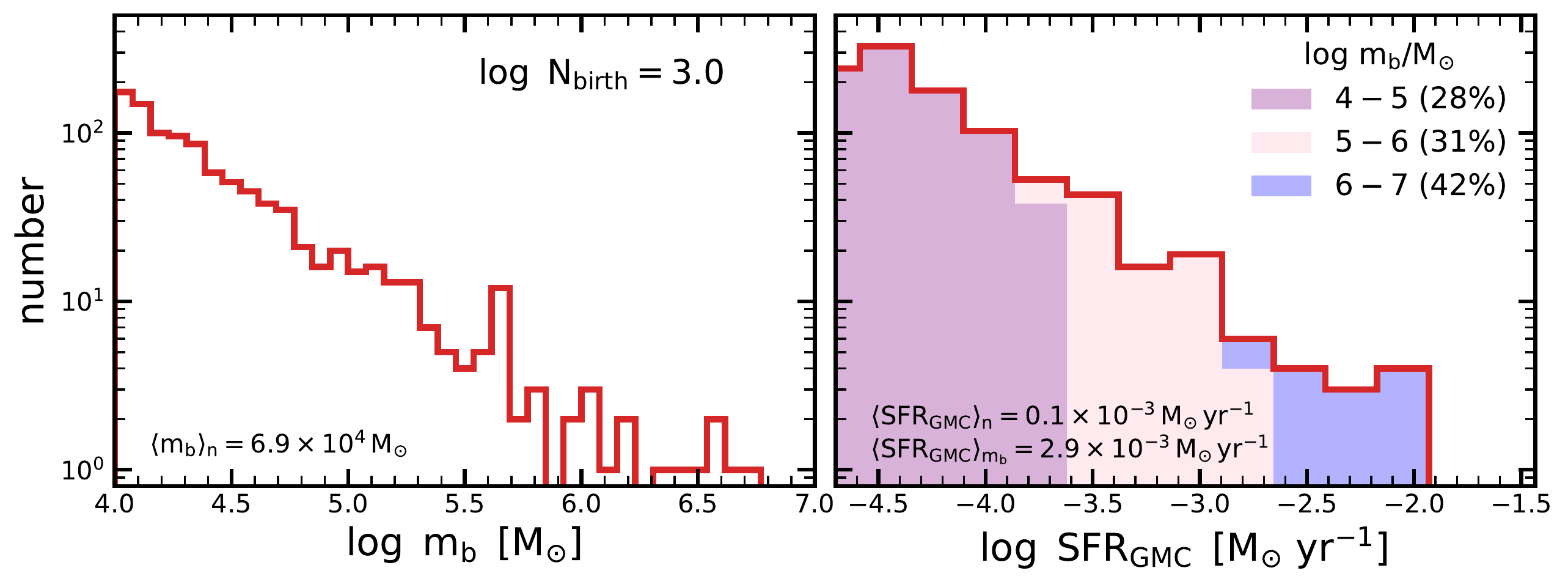}
    \caption{Physical properties of an example draw of 1000 GMCs at the time of formation. We show the distribution of GMC mass $m_{\rm b}$ (left) and SFR (right). We draw the mass of each GMC from the birth mass function, given by Eq.~\ref{eq:GMC_MF_birth} with $\alpha_{\rm b}=-2.0$. We use Eqs.~\ref{eq:GMC_LT} and \ref{eq:GMC_SFE} to obtain lifetimes and star-formation efficiencies, assuming $\tau_0=10.0~\mathrm{Myr}$, $\alpha_{\rm l}=0.25$, $\varepsilon_0=0.02$, and $\alpha_{\rm e}=0.25$. With this parametrization, the mass-weighted average lifetime, star-formation efficiency and SFR is $\langle \tau_{\rm L} \rangle_{\rm m_{\rm b}} = 26.8~\mathrm{Myr}$, $\langle \varepsilon \rangle_{\rm m_{\rm b}} = 0.05$, and $\langle\mathrm{SFR}_{\rm GMC}\rangle_{\rm m_{\rm b}}=2.9\cdot10^{-3}~\mathrm{M}_{\odot}~\mathrm{yr}^{-1}$. Important to note is that this is the GMC population at the time of formation. At later times, after several such draws, the mass function will be shallower because massive GMCs live longer (see Appendix~\ref{app:birth_vs_observed_MF} for details). This leads to an increase in the importance of massive clouds, which are significantly contributing (in some cases dominating) the SFR. Since these massive clouds are rare overall, their contribution to the total SFR fluctuates significantly with time.}
    \label{fig:GMC_example}
\end{figure*}

Our model of star formation sustained by GMCs is primarily based on recent observational results of the GMC population, i.e. we take a rather empirical approach. This is different from the approaches by \citet{faucher-giguere18} based on recent simulations and analytic models of star formation in galaxies, or by \citet[][see also \citealt{scalo86, struck-marcell84, struck-marcell87}]{scalo84} which models the details of the position-velocity-mass phase space of GMCs and the resulting variability of star formation. Specifically, we assume that the GMC population of a galaxy can be characterised by functions describing distributions in mass, lifetime, and star-formation efficiency. We parameterize all these functions as power-laws, which allow for enough flexibility to describe the star-formation processes in local and high-redshift galaxies, and are consistent with current observational constraints. As we highlight below, it is important to differentiate between GMC properties at time of formation (``birth'') and at time of observation, i.e. we model GMCs as time-dependent entities in galaxies. 

But what do we mean by GMCs exactly? The physical properties of GMCs have been studied in the Milky Way and in a growing number of nearby galaxies, including the LMC, M51, M83, and M33 \citep[e.g.][]{rosolowsky07, colombo14, garcia14, rice16, faesi16, freeman17_gmc, sun18, faesi18}. It is well known that stars form in GMCs \citep[e.g.][]{myers86, mooney88, scoville89, williams97} and that the interstellar medium is turbulent and highly structured. As a result, clouds that undergo gravitational collapse generally have several different centres of collapse, which are evident as filamentary structures in both observations and simulations \citep[e.g.][]{goodman90, elmegreen04, padoan12, lee15_sfr, raskutti16, rice16, li18, grudic18}. 

\paragraph{GMC mass function}

In our extended regulator model, we assume that the total SFR of a galaxy is sustained by a population of independent GMCs (see \citealt{kruijssen14, kruijssen18} for approaches employing similar assumptions). We assume that these GMCs are born with a certain molecular gas mass $m_{\rm b}$ and that the fraction of GMCs with birth masses between $m_{\rm b}$ and $m_{\rm b} + dm_{\rm b}$, i.e. the GMC birth mass function $n_{\rm b}(m_{\rm b})$, can described by a simple power-law:

\begin{equation}
    n_{\rm b}(m_{\rm b}) \mathrm{d}m_{\rm b} = A_{\rm b} m_{\rm b}^{\alpha_{\rm b}} \mathrm{d}m_{\rm b},
    \label{eq:GMC_MF_birth}
\end{equation}

\noindent
where $A_{\rm b}$ is the normalization and $\alpha_{\rm b}$ is the power-law slope. In addition, we assume that a negligible fraction of GMCs form below $m_{\rm min}$ ($=10^{4}~M_{\odot}$) or above $m_{\rm max}$ ($=10^{7-9}~M_{\odot}$). We choose the normalization $A_{\rm b}$ so that $n_{\rm b}$ is a probability density function (i.e. $\int_{m_{\rm min}}^{m_{\rm max}} n_{\rm b}(m_{\rm b}) \mathrm{d}m_{\rm b} = 1$). Observationally the upper truncation in the mass function tends to be around $m_{\rm max} \approx 10^6-10^7~\mathrm{M_{\odot}}$ in local galaxies \citep{rice16}, and is believed to be greater in galaxies that have greater gas surface density and gas fraction (i.e. galaxies at higher redshifts) on theoretical grounds \citep[e.g.][]{escala08, reina-campos17}. A population of GMCs that fully populate the birth mass function following Eq.~\ref{eq:GMC_MF_birth} will have an average birth mass of:

\begin{equation}
    \langle m \rangle_{\rm n_b} = \int_{m_{\rm min}}^{m_{\rm max}} n_{\rm b}(m_{\rm b}) m_{\rm b} \mathrm{d}m_{\rm b} \propto m_{\rm max}^{2+\alpha_{\rm b}}-m_{\rm min}^{2+\alpha_{\rm b}}\ \ \ (\alpha_{\rm b}\ne-2).
    \label{eq:GMC_avgM}
\end{equation}

\paragraph{GMC lifetimes}

For a given GMC birth mass $m_{\rm b}$, we assign a certain lifetime $\tau_{\rm L}(m_{\rm b})$ and star-formation efficiency $\varepsilon(m_{\rm b})$. We define the lifetime of GMCs to be the time when the GMC is actively forming stars, i.e. the time between the formation of the first star until the time when the GMC is disrupted and further star formation is prevented. We assume that the lifetime of GMCs depends on the GMC birth mass $m_{\rm b}$ and can be described by the following power-law relation:

\begin{equation} 
    \tau_{\rm L} (m_{\rm b}) = \tau_0 \cdot \left(\frac{m_{\rm b}}{10^4 M_{\odot}}\right)^{\alpha_{\rm l}}.
    \label{eq:GMC_LT}
\end{equation}

\noindent
Assuming that the lifetime is a multiplicative factor of the free-fall time and that GMCs have a characteristic surface density \citep{larson82}, then $\alpha_{\rm l}\approx 0.25$, i.e. more massive GMCs tend to live longer \citep{fall10}. In reality GMCs have substantial scatter in their surface densities, likely resulting from variations in ISM pressure \citep{sun18}, which would presumably introduce a corresponding scatter in their lifetimes which we do not include here for simplicity.

\paragraph{GMC SFRs}

We define the star-formation efficiency as the fraction of the GMC birth mass $m_{\rm b}$ that gets turned into stellar mass over the lifetime of the GMC. This is sometimes called the integrated star-formation efficiency. We assume that this efficiency can be parameterized by the following power-law relation:

\begin{equation}
    \varepsilon(m_{\rm b}) = \varepsilon_0 \cdot \left(\frac{m_{\rm b}}{10^4 M_{\odot}}\right)^{\alpha_{\rm e}}.
    \label{eq:GMC_SFE}
\end{equation}
\noindent
Again, this relation likely has substantial scatter in nearby galaxies. \citet{krumholz12} argues that there is little scatter in the efficiency per freefall time, but this issue remains controversial in the literature \citep[e.g.][]{lee16}. Our model avoids any explicit reference to the freefall time of individual GMCs, relying instead on generic powerlaws.

We assume that the SFR of a GMC is constant throughout its lifetime. Following the above definitions, we can express the SFR of an individual GMC with birth mass $m_{\rm b}$ by

\begin{equation}
    \mathrm{SFR}_{\rm GMC}(m_{\rm b}) = \varepsilon(m_{\rm b}) \cdot \frac{m_{\rm b}}{\tau_{\rm L}(m_{\rm b})} \propto \frac{\varepsilon_0}{\tau_0} \cdot m_{\rm b}^{1+\alpha_{\rm e}-\alpha_{\rm l}}.
    \label{eq:GMC_SFR}
\end{equation}

\noindent
Since GMCs are generally not 100\% efficient at transforming their gas into stars, the question of how GMC mass is returned to the gas reservoir arises. Physically the process of GMC disruption is poorly understood, but may be mass-dependent, closely tied to different forms of feedback, or the large-scale environment of the GMCs \citep[e.g.][]{lopez11, goldbaum11, matzner15, li19_feedback}. For simplicity we assume that these processes act to linearly reduce the cloud mass to zero at the end of its lifetime, so that the current mass $m$ of an individual GMC can be expressed as:

\begin{equation}
    m(t^\prime) = m_{\rm b}\left[1-\frac{t^\prime}{\tau_{\rm L}(m_{\rm b})}\right],
    \label{eq:mass_GMC}
\end{equation}

\noindent
where $t^{\prime}$ is the time since formation. A fraction of $\varepsilon(m_{\rm b})$ of this mass decrease is continuously converted into stars, while the remaining fraction ($1-\varepsilon(m_{\rm b})$) is returned back to the gas reservoir. This treatment neglects important processes, especially accretion \citep{goldbaum11} and potentially slow star formation near the beginning of a GMC's lifetime \citep[e.g][]{lee16, chevance20}. Altering this assumption would change the fine-scale details of the PSD, e.g. the cosine factor derived in Appendix~\ref{app:full_derivation}, but we neglect this factor anyway (see Section \ref{subsec:only_gmc}).

In the case where the GMC lifetimes are mass dependent ($\alpha_{\rm l}\ne0$), the actual GMC mass function at any given moment, $n_{\rm obs}(m)$, is different from the birth mass function $n_{\rm b}(m_{\rm b})$ \citep[see also][for more complex examples of evolving mass functions]{huang13, kobayashi17}. As we show in Appendix~\ref{app:birth_vs_observed_MF}, $n_{\rm obs}(m)$ can also be described by a power-law with a high mass turnover, similar to a \citet{schechter76} function (Eq.~\ref{eq:MF_conversion}). Over most of the mass range, the observed mass function is well described by a power-law ($n_{\rm obs} \propto m^{\alpha_{\rm obs}}$ with a power-law slope of $\alpha_{\rm obs}=\alpha_{\rm b}+\alpha_{\rm l}$. For a positive $\alpha_{\rm l}$, this means that the observed mass function is shallower than the birth mass function because more massive clouds tend to live longer. This parameterization allows us to capture the diversity of observed mass functions between different galaxies as well as within a given galaxy. Observers typically measure a power-law slope $\alpha_{\rm obs}$ ranging from $-2.5$ to $-1.5$ depending on environment -- typically it is more top-heavy (i.e. $\alpha_{\rm obs}$ less negative) in the inner parts of galaxies and more bottom-heavy (i.e. $\alpha_{\rm obs}$ more negative) in the outer parts. For $m$ below $m_{\rm min}$, the mass function turns over, and the overall fraction of molecular mass and star formation in that regime is negligible, but accounted for in our model.

\paragraph{Example}

In order to build an intuition for this parameterization, Fig.~\ref{fig:GMC_example} shows the resulting properties of a random draw of $N_{\rm birth}=10^3$ GMCs. Specifically, these GMCs have been drawn from the birth mass function assuming $\alpha_{\rm b}=-2.0$, $m_{\rm min}=10^4~\mathrm{M_{\odot}}$ and $m_{\rm max}=10^7~\mathrm{M_{\odot}}$. Then the lifetime ($\tau_0=10.0~\mathrm{Myr}$, $\alpha_{\rm l}=0.25$) and star-formation efficiency ($\varepsilon_0=0.02$, $\alpha_{\rm e}=0.25$) for each GMC has been obtained using Eqs.~\ref{eq:GMC_LT} and \ref{eq:GMC_SFE}. We derive the SFR for each GMC (Eq.~\ref{eq:GMC_SFR}) and SFR function is plotted in the right panel. With this parameterization, the mass-weighted lifetime, star-formation efficiency and SFR is $\langle \tau_{\rm L} \rangle_{\rm m_{\rm b}} = 26.8~\mathrm{Myr}$, $\langle \varepsilon \rangle_{\rm m_{\rm b}} = 0.05$, and $\langle\mathrm{SFR}_{\rm GMC}\rangle_{\rm m_{\rm b}}=2.9\cdot10^{-3}~\mathrm{M}_{\odot}~\mathrm{yr}^{-1}$, respectively. The number-density-weighted quantities are lower since the abundance of low-mass GMCs is higher.

As expected ($\mathrm{SFR}_{\rm GMC}\propto m_{\rm b}$), more massive GMCs have slightly higher SFRs. In this random draw, massive GMCs with $m>10^6~\mathrm{M}_{\odot}$ contribute significantly ($\approx40\%$) to the total SFR, consistent with measurements from the Milky Way \citep[e.g.,][]{murray11}. Since these massive GMCs are rare (see mass function), their contribution to the total SFR varies significantly between different draws, but they usually dominate the SFR. It is important to note that Fig.~\ref{fig:GMC_example} shows the GMC population at the time of formation. At later times, after several such draws, the observed quantities will actually differ: the mass function will be shallower because the more massive clouds will live longer (Appendix~\ref{app:birth_vs_observed_MF}). Importantly, this leads to an increase in the share of star formation in high-mass clouds. Specifically, since 
\begin{equation}
\label{eq:massWeightedGMC}
    \int \mathrm{SFR}_{\rm GMC}(m) n_{\rm obs}(m) \mathrm{d}m \propto m^{\alpha_{\rm obs}+ 2.25},
\end{equation} 
low-mass GMCs start to dominate the total SFR only when $\alpha_{\rm obs}<-2.25$, i.e. when the observed mass function is extremely steep (and the birth mass function is even steeper).

\subsubsection{Drawing of GMCs}

How many GMCs need to form within a given time step $\mathrm{d}t$? We assume that the number of newly formed GMCs follows a Poisson process $\mathcal{P}$ and can be written as:

\begin{equation}
    N_{\rm birth}(t) = \mathcal{P}\left(\frac{1}{\langle m \rangle_{\rm n_b}} \frac{M_{\rm gas}(t)}{\tau_{\rm mol}} \mathrm{d}t \right),
    \label{eq:N_birth}
\end{equation}

\noindent
where $\tau_{\rm mol}$ is the timescale for the formation of molecular gas and $\langle m \rangle_{\rm n_b}$ is the $n_{\rm b}$-weighted average GMC birth mass (see Eq.~\ref{eq:GMC_avgM}). We are assuming that molecular gas forms in proportion to the amount of gas available in the reservoir of non-molecular cold gas $M_\mathrm{gas}$, so that on average in each time step a mass of $\mathrm{d}t M_\mathrm{gas}(t) / \tau_\mathrm{mol}$ forms molecular clouds. This mass is then partitioned into molecular cloud masses according to the birth mass function, so that the number of new GMCs is on average the mass of forming GMCs divided by the typical mass of those clouds. Rather than setting $\tau_{\rm mol}$ directly, we can relate it to the overall depletion time $\tau_{\rm dep}$ as we show in the next section. Throughout this paper, we specify $\tau_{\rm dep}$ for each model run and calculate $\tau_{\rm mol}$ self-consistently according to Eq.~\ref{eq:total_sfr_ext}. Typically, $\tau_{\rm mol}$ will be of the order of $10-200~\mathrm{Myr}$, i.e. comparable to the dynamical timescale in the denser parts of galaxies (see also \citealt{semenov18}).

In each time step, we then draw $N_{\rm birth}$ GMCs from the birth mass function, thereby obtaining for each GMC $i$ a birth mass $m_{\rm b,i}$. Following this, we use Eqs.~\ref{eq:GMC_LT}, \ref{eq:GMC_SFE} and \ref{eq:GMC_SFR} to compute the GMCs' SFRs, which they sustain over their lifetimes $\tau_{\rm L}(m_{\rm b,i})$. Only a fraction $\varepsilon(m_{\rm b,i})$ of the initial GMC mass $m_{\rm b,i}$ is converted into stars. We assume that the remaining molecular gas mass, $(1-\varepsilon(m_{\rm b,i}))\cdot m_{\rm b,i}$, is returned to the gas reservoir over the timescale $\tau_{\rm L}(m_{\rm b,i})$.

With these assumptions, we can write for a given time step $\mathrm{d}t$ :

\begin{equation}
    \mathcal{G}(t) = \frac{1}{\mathrm{d}t} \sum_{i=1}^{N_{\rm birth}(t)} m_{\rm b,i} - \sum_{j=1}^{N_{\rm GMC}(t)} (1-\varepsilon(m_{\rm b,j}))\frac{m_{\rm b,j}}{\tau_{\rm L}(m_{\rm b,j})},
    \label{eq:SF_process_full}
\end{equation}

\noindent
where the first term describes the rate at which mass from the gas reservoir is converted into GMCs and, hence, molecular gas (green arrow in Fig.~\ref{fig:cartoon}), while the second term describes the molecular gas mass that is not converted into stars and transferred back to gas reservoir (cyan arrow in Fig.~\ref{fig:cartoon}). The second sum is over all living GMCs $j$ (there is a total of $N_{\rm GMC}(t)$ living GMCs at time $t$) that have $t-t_j < \tau_{\rm L}(m_{\rm b,j})$, where $t_j$ is the time when GMC $j$ was formed. Similarly, the SFR sustained by these GMCs can be written as:

\begin{equation}
    \mathrm{SFR}(t) = \sum_{j=1}^{N_{\rm GMC}(t)} \mathrm{SFR_{\rm GMC}(m_{\rm b,j})} = \sum_{j=1}^{N_{\rm GMC}(t)} \varepsilon(m_{\rm b,j})\frac{m_{\rm b,j}}{\tau_{\rm L}(m_{\rm b,j})}.
    \label{eq:SFR_full}
\end{equation}

In summary, we introduced a self-consistent description for how gas cycles from a reservoir into the molecular phase, forms stars, and gets returned back to the reservoir. We call this the extended regulator model. It allows us to capture galaxy-internal physics related to the star-formation process. 

\subsection{Connecting global and local star formation: basic and extended regulator model}
\label{subsec:connection_basic_full}

In this section, we connect the basic to the extended regulator model, basically connecting global with local star formation. Our considerations are similar to the ones by \citet{semenov17}, who used the idea of gas cycling between star-forming and non-star-forming states on certain characteristic timescales under the influence of dynamical and feedback processes to explain why the depletion timescale is significantly longer than the timescales of processes governing the evolution of interstellar gas \citep[see also][]{semenov18, burkert17}.

In order to connect the basic with the extended regulator model, we impose that the average SFR in the entire galaxy will correspond to the basic gas regulator model's value, which is $\mathrm{SFR}=M_{\rm gas}/\tau_{\rm dep}$. For the total SFR of the regulator model, we can write for an average of $N_{\rm GMC}$ active GMCs:

\begin{equation}
    \langle \mathrm{SFR} \rangle_{\rm t} = \langle N_{\rm GMC} \rangle_{\rm t} \langle \mathrm{SFR}_{\rm GMC} \rangle_{n_\mathrm{obs}}.
    \label{eq:total_sfr_ext}
\end{equation}


\noindent
where $\langle \cdot \rangle_{\rm t}$ is the average over a long time interval and $\langle \cdot \rangle_{\rm n_\mathrm{obs}}$ is the average over the instantaneous GMC population (see Eq.~\ref{eq:massWeightedGMC}). For the remainder of the paper, we use this equation to calibrate $\tau_{\rm mol}$, which sets via Eq.~\ref{eq:N_birth} $N_{\rm birth}$ and therefore $N_{\rm GMC}$. This means that we specify $\tau_{\rm dep}$ for both the basic and extended regulator model, and $\tau_{\rm mol}$ will the be given via Eq.~\ref{eq:total_sfr_ext}. 

On average, in a steady state, the formation and disappearance rate of GMCs should be equal, and the following equation will hold approximately\footnote{The exact expression for the disappearance of GMCs (the right-hand side of Eq.~\ref{eq:relation_Nbirth_Ngmc})} is more complicated.:

\begin{equation}
    \frac{\langle N_{\rm birth} \rangle_{\rm t}}{\mathrm{d}t} \approx \left\langle \frac{N_{\rm GMC}}{ \tau_{\rm L} } \right\rangle_{\rm t}
    \label{eq:relation_Nbirth_Ngmc}
\end{equation}

\noindent
which leads to:

\begin{equation}
    \tau_{\rm mol} \approx \tau_{\rm dep} \cdot \frac{\langle \mathrm{SFR}_{\rm GMC} \rangle \langle \tau_{\rm L} \rangle}{\langle m \rangle_{\rm n_b}} \approx \tau_{\rm dep} \cdot \langle \varepsilon \rangle.
    \label{eq:relation_tau_mol_dep}
\end{equation}

\noindent
As expected, the depletion time on average is directly related to the timescale molecular gas forms and the efficiency of converting molecular gas into stars. 

The depletion timescale $\tau_{\rm dep}$ is defined as the ratio of the mass in the gas reservoir and the SFR (see Eq.~\ref{eq:depletion_time_basic}). This definition leads to slightly larger values for $\tau_{\rm dep}$ than quoted in the observational literature, since the latter case usually only considers the molecular gas mass, which we call molecular depletion times $\tau_{\rm dep, mol}$. In a similar fashion, we can estimate $\tau_{\rm dep, mol}$:

\begin{equation}
    \tau_{\rm dep,mol} \approx \frac{\langle M \rangle_{\rm n_b}}{\langle \mathrm{SFR}_{\rm GMC} \rangle} \approx \frac{\langle \tau_{\rm L} \rangle}{\langle \varepsilon \rangle}.
    \label{eq:depletion_time_mol}
\end{equation}

\noindent
For a star-formation efficiency of a few $\%$ and $\tau_{\rm L}$ of a few to a few tens Myr, this leads to a molecular depletion time of $0.3-1$ Gyr, which is consistent, but slightly smaller than observational estimates, which show only a weak redshift and stellar mass dependence \citep[e.g.][]{daddi10, genzel15, saintonge17, tacconi18}. In summary, we build our model in a way that the long molecular depletion times in galaxies are due to star formation that is sufficiently inefficient to counteract the relatively short molecular lifetimes. This is consistent with similar investigations by \citet{burkert17} and \citet{semenov17}, as well as recent observations \citep{kruijssen19, chevance20}. Furthermore, Eq.~\ref{eq:depletion_time_mol} implies that $\tau_{\rm dep, mol}\approx \mathrm{const}$ when $\langle \tau_{\rm L} \rangle$ and $\langle \varepsilon \rangle$ are constant, resulting in a linear Kennicutt-Schmidt relation for molecular gas. More generally, the linear slope of molecular Kennicutt-Schmidt relation can be explained by a cancellation of $\langle \tau_{\rm L} \rangle$ and $\langle \varepsilon \rangle$ trends in Eq.~\ref{eq:depletion_time_mol} \citep{semenov19}.

Similarly, we can estimate what the average molecular-to-total gas mass ratio:

\begin{equation}
    R_{\rm mol} \approx \frac{\langle M \rangle_{\rm n_b}}{\langle \mathrm{SFR}_{\rm GMC} \rangle \tau_{\rm dep}} \approx \frac{\langle \tau_{\rm L} \rangle}{\tau_{\rm mol}},
    \label{eq:Rmol}
\end{equation}

\noindent
which is consistent with the expectation that this ratio depends on the timescale on which molecular gas forms and how long these clouds live. For $\tau_{\rm L}$ of a few Myr and $\tau_{\rm mol}$ of a few tens to a hundred Myr, we obtain $R_{\rm mol}\approx0.01-0.3$. Again, this is consistent, but slightly on the low end in comparison with observational estimates \citep{dame01}. 

Important to note is that our molecular gas mass, $M_{\rm mol}$, only includes molecular gas that is in actively star-forming GMCs. However, there are observational indications that molecular gas also exists in a diffuse phase \citep[e.g.][]{wolfire10}, though the quantity depends on the precise definition of molecular clouds \citep[e.g.][]{miville-deschenes17} and in non-star-forming GMCs \citep{kruijssen19}. This can explain why our rough estimates of $\tau_{\rm dep, mol}$ and $R_{\rm mol}$ are slightly smaller than observations.

\subsection{Gas inflow rate $\Phi(t)$}
\label{subsec:phi}

\begin{figure*}
    \centering
    \includegraphics[width=\textwidth]{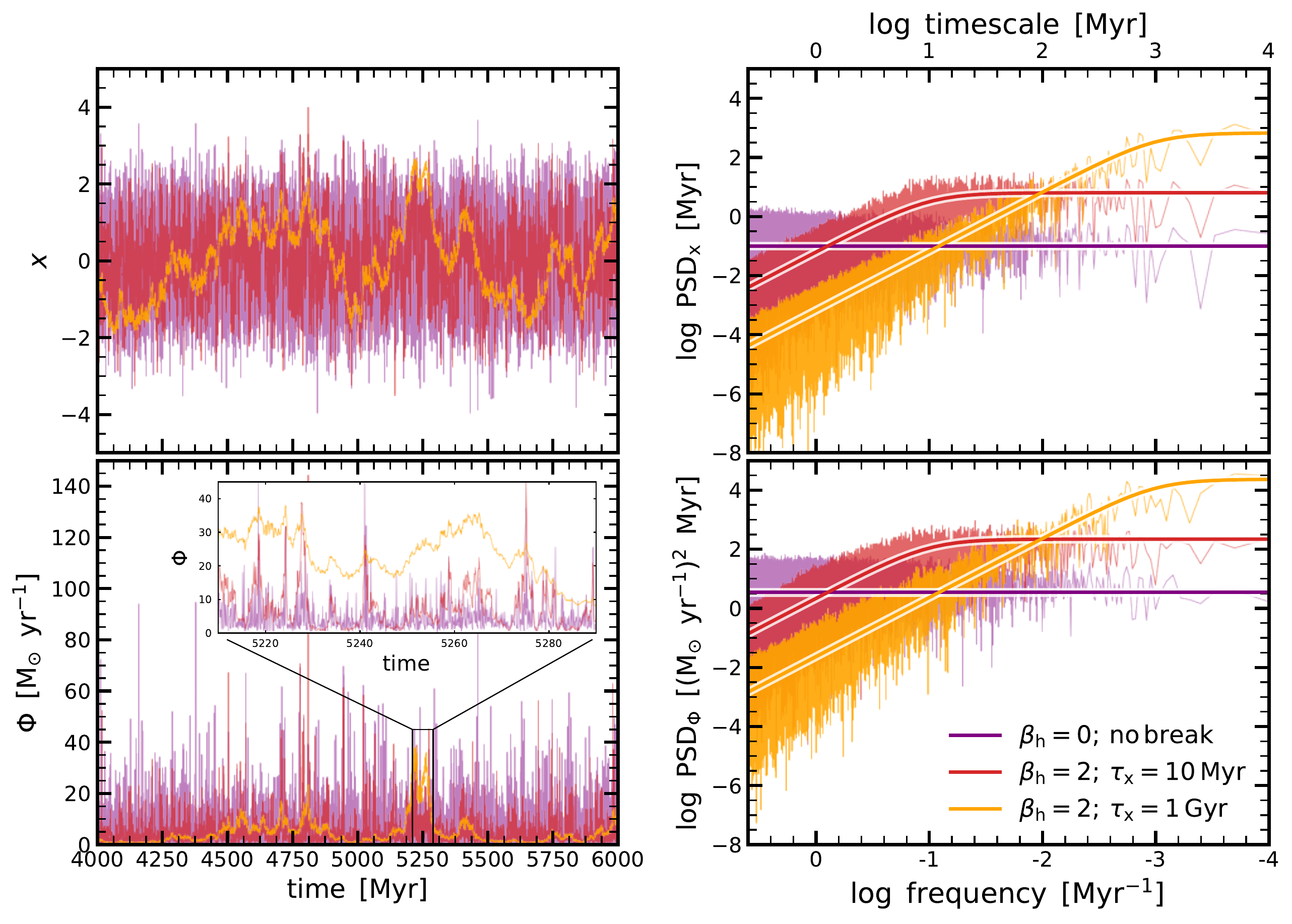}
    \caption{Three examples of stochastic inflow rates. We model the inflow rate $\Phi(t)$ via a stochastic process, which we described as a random variable $x(t)$ with a certain power spectral density PSD$_\mathrm{x}$ (Eq.~\ref{eq:PSD_x}). $\Phi(t)$ can be directly obtained from $x(t)$ (Eq.~\ref{eq:inflow_regulator}) and their PSDs are related as well (Eq.~\ref{eq:inflow_psd}). We assume $\mu=1.0$ and $\sigma=1.0$. The time-evolution of $x(t)$ and $\Phi(t)$ are shown on the left, while their corresponding PSDs are shown on the right. The inset panel zooms in on the inflow rate in the time interval $5210-5290$ Myr. The thick solid lines are given by Eq.~\ref{eq:PSD_x}. The three examples correspond to white noise (purple lines), broken power-law with high-frequency slope $\beta_{\rm h}=2$ and break-timescale $\tau_{\rm x}=10~\mathrm{Myr}$ (red lines), and broken power-law with high-frequency slope $\beta_{\rm h}=2$ and break-timescale $\tau_{\rm x}=1~\mathrm{Gyr}$ (orange lines). The latter two examples correspond to a damped random walk process (low frequency slope of $\beta_{\rm l}=0$). All three examples have the same total variance (i.e. $\int \mathrm{PSD}(f) df$) within the given frequency bounds, but the orange example has the most power on long timescales (low frequencies) relative to short timescales (high frequencies), leading to a more continuous evolution with time (i.e. less bursty).}
    \label{fig:inflow_psd}
\end{figure*}

The inflow rate $\Phi(t)$ and its variability with time, as well as its dependence on galaxy properties, is uncertain observationally as well as theoretically. As mentioned above, in this work, the gas inflow rate $\Phi(t)$ includes not only pristine gas inflow into the gas reservoir $M_{\rm gas}$ from the \textit{outside} of the halo, but also complex physical processes such as gas cooling within the halo, cold streams feeding the galaxy, galaxy-galaxy mergers and recycling of outflows. This means that we do not necessarily want to assume that the gas inflow rate solely arises from the growth rate of the dark matter halo \citep[e.g.][]{lilly13_bathtub, dekel14_bathtub, forbes14b}, so we take a somewhat more general approach. Many physical variables related to star formation (including the scatter about the star-forming main sequence) follow a log-normal distribution. Numerical simulations show that this might also be the case for the inflow rate \citep[e.g.][]{goerdt15, mitchell20}. We therefore generalize the assumption made by \citet{forbes14b}, that the inflow rate $\Phi(t)$ is a log-normal distribution with fixed median and scatter:

\begin{equation}
    \Phi(t) = \exp(\mu+\sigma x(t)).
    \label{eq:inflow_regulator}
\end{equation}

\noindent
As in \citet{forbes14b} $x(t)$ is a random variable with values distributed as a standard normal (zero mean, unit variance). While   \citet{forbes14b} drew a new value of $x$ at fixed time intervals so that during these time intervals $x$ was constant, i.e. perfectly correlated, we generalise this by assuming a spectrum of timescales over which $x$ varies. This comes at the cost of some simplicity and analytical tractability, but has the virtue of being both more general and more realistic. We concentrate on a particular family of PSD$_{x}$ described a broken power-law:

\begin{equation}
    \begin{split}
    \mathrm{PSD_{\rm x}}(f) & = \frac{C}{(\tau_{\rm x} f)^{\beta_{\rm l}}+(\tau_{\rm x} f)^{\beta_{\rm h}}},
    \end{split}
    \label{eq:PSD_x}
\end{equation}

\noindent
where $f$ denotes frequency, $C$ is a normalization constant with units $1/f$, $\tau_{\rm x}$ is the break time-scale, and $\beta_{\rm l}$ and $\beta_{\rm h}$ describes the power-law slopes at low and high frequencies relative to $1/\tau_{\rm x}$, respectively. The normalization constant $C$ is chosen so that $\mathrm{Var}(x)=1$ over the defined frequency regime. We choose this parametrization because it allows us to describe a large variety of stochastic processes (see discussion in \citealt{caplar19} and examples below). Furthermore, Iyer et al. (in prep.) show that the PSD of dark matter accretion histories and of star-formation histories can indeed be well described by (broken) power-laws in cosmological simulations.

If the stochasticity of $x(t)$ can be described with PSD$_{x}$, then the PSD of $\Phi(t)$ is given by the following renormalization of PSD$_{x}$:

\begin{equation}
    \mathrm{PSD}_{\Phi} = \frac{\mathrm{Var}(\Phi)}{\mathrm{Var}(x)} \mathrm{PSD}_{\rm x} = \exp(2\mu+\sigma^2)[\exp(\sigma^2)-1.0] \mathrm{PSD}_{\rm x}.
    \label{eq:inflow_psd}
\end{equation}

We generate the time-series $x(t)$ using the GPU-accelerated framework presented in \citet{sartori19}, which is based on the approach presented in \citet{emmanoulopoulos13}. The high efficiency of the algorithm enables us to generate timeseries $x(t)$ with a given power spectral density, while covering the whole age of the Universe with dense (0.1 Myr) sampling.

We show three examples of $x(t)$ and their corresponding $\Phi(t)$ in Fig.~\ref{fig:inflow_psd}. Specifically, the left panels show the time-evolution of these quantities, while the right panels display their PSDs. We convert $x(t)$ to $\Phi(t)$ using Eq.~\ref{eq:inflow_regulator}, arbitrarily adopting $\mu=1.0$ and $\sigma=1.0$ for the moment. As discussed above, the PSD of $\Phi(t)$ is related to the PSD of $x(t)$ via a re-normalization (Eq.~\ref{eq:inflow_psd}).

The simplest parameter choice of the model is where $\beta_{\rm l}=\beta_{\rm h}=0$, i.e. PSD$_{\rm x}$ is a constant. This timeseries is generated by the white noise and it is indicated as purple lines in Fig.~\ref{fig:inflow_psd}. The red and yellow lines correspond to broken power-law PSDs with $\beta_{\rm l}=0$ and $\beta_{\rm h}=2$. In these two cases, also called damped random walk, the high-frequency slope $\beta_{\rm h}$ of the PSD determines how quickly $x(t)$ (and hence $\Phi(t)$) changes on short timescales and the break timescale $\tau_{\rm x}$ sets timescale on which $x(t)$ and $\Phi(t)$ lose ``memory'' of previous accretion history (\citealt{caplar19}; this is similar to the coherence timescale in the \citealt{forbes14b} approach). Specifically, the red line corresponds to an example with $\beta_{\rm h}=2$ and $\tau_{\rm x}=50~\mathrm{Myr}$, while the yellow line corresponds to $\beta_{\rm h}=2$ and $\tau_{\rm x}=1000~\mathrm{Myr}$. Since the timeseries' variance, i.e. $\int {\rm PSD}(f) df$, is the same in all three examples, the orange PSD has the most power on long timescales and the least power on short timescales. This makes this example less bursty than the others. We refer the reader to \citet{caplar19} for more examples of PSDs.

\section{Variability of the star-formation rate on different timescales}
\label{sec:PSD}

In the previous section, we have introduced the regulator model (a basic and an extended version) to study how gas cycles through galaxies and drives star formation. We now make a step forward and link the physical parameters and processes (Table~\ref{tab:parameters}) to the variability of the SFR. We quantify the variability (or ``burstiness'') with the PSD, which is a measure of the amount of power contained in SFR fluctuations on a given timescale. We show how different features, breaks in particular, in the PSD correspond to different timescales governing the regulation of star formation. We start with some simple considerations (Sections~\ref{subsec:PSD_basic_regulator} and \ref{subsec:only_gmc}) to build up to the general case in Section~\ref{subsec:combined}. In Section~\ref{subsec:case_studies}, we show and analyse PSDs of galaxies in different regimes (from dwarf to high-redshift galaxies).

\subsection{Basic regulator model}
\label{subsec:PSD_basic_regulator}

\begin{figure*}
    \centering
    \includegraphics[width=\textwidth]{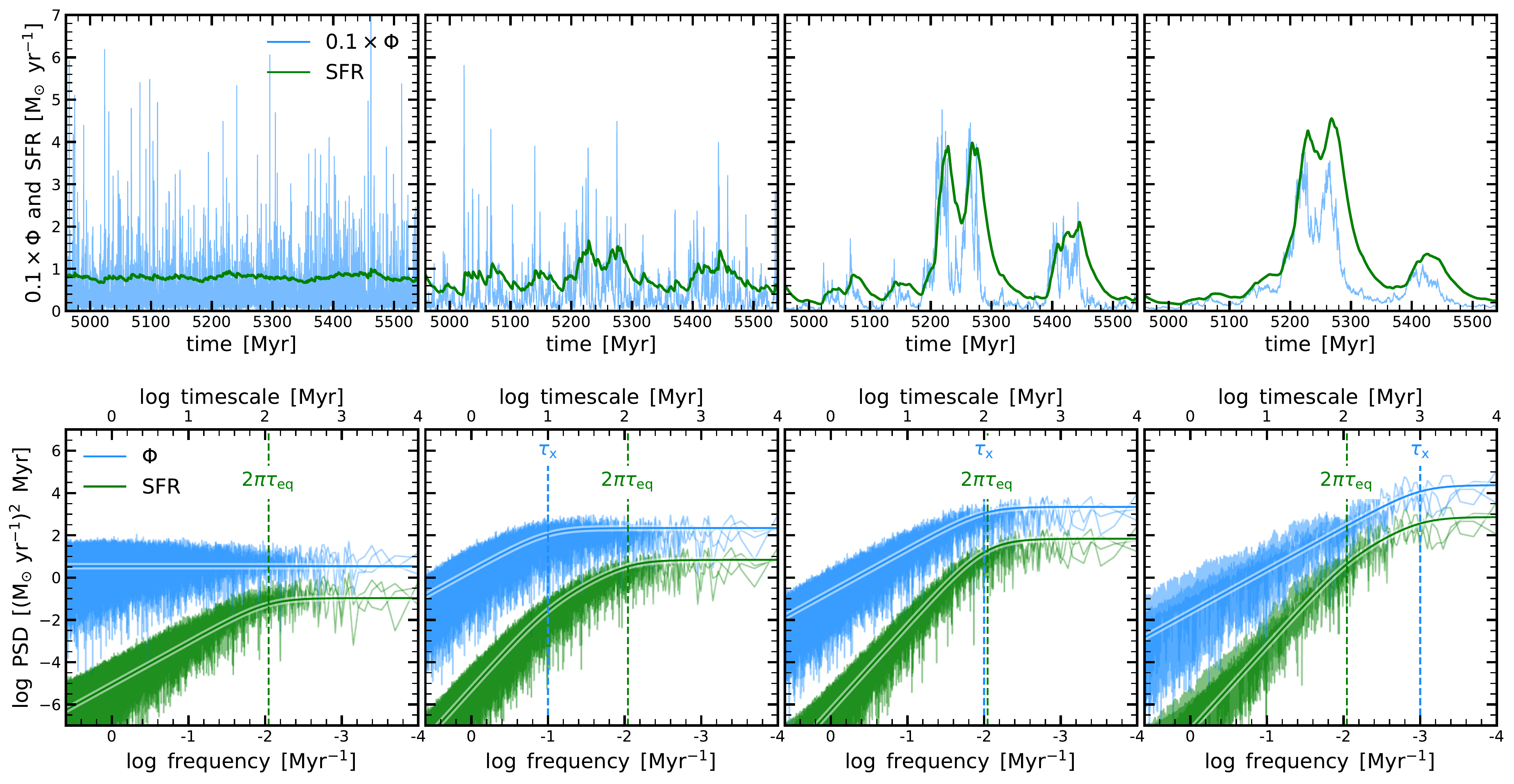}
    \caption{How does the basic regulator model shape the PSD? The upper panels show the time evolution of the inflow rate $\Phi$ (re-normalised by 0.1) and SFR, while the lower panels show the corresponding PSDs. We hold the parameters of the basic regulator fixed, while changing only the PSD of $\Phi$ from right to left. In the left panels, $\mathrm{PSD}_{\Phi}$ is constant (white noise process), while in the panels to the right, we show broken power-law PSDs with high frequency slopes $\beta_{\rm h}=2$, low frequency slopes $\beta_{\rm l}=0$, and increasing break timescales ($\tau_{\rm x}=10$, 100, and 1000 Myr). The parameters of the basic regulator model were fixed in all four runs: the depletion time is $\tau_{\rm dep}=100~\mathrm{Myr}$, the star-formation efficiency $\varepsilon_0=0.1$ with no mass dependence ($\alpha_{\rm e}=0$), and the mass-loading $\lambda=5$. These numbers lead to an equilibrium timescale of $\tau_{\rm eq}=18~\mathrm{Myr}$. In addition to the break at $\tau_x$, the PSD of the basic regulator model has a break at $2\pi\tau_{\rm eq}\approx110~\mathrm{Myr}$. Above this timescale (towards lower frequencies), the PSD of the regulator model follows the shape of $\mathrm{PSD}_{\Phi}$, while below this timescale, the slope is steeper by 2, as in  Eq.~\ref{eq:PSD_drw_triple} (solid green lines). The overall strength of the PSD is given by $\sigma_{\rm reg}^2$, which is set by the stochasticity of the inflow rate $\Phi(t)$.}
    \label{fig:example_basic_regulator}
\end{figure*}

We start by considering the basic regulator model. Specifically, we solve Eq.~\ref{eq:sfr_regulator} for the SFR($t$), assuming that the driving term $\Phi(t)$ is a stocastic process as described in Section~\ref{subsec:phi}. The inflow $\Phi(t)$ sets the amount of ``randomness'' entering the system, while the equilibrium timescale $\tau_{\rm eq}$ describes the timescale by which the process converges towards a stable-state, equilibrium solution. In the simplest case, where $\Phi(t)$ is a continuous white-noise process (PSD with power-law slope of 0, i.e. $\beta_{\rm l}=\beta_{\rm h}=0$, and with no break), the solution to this equation is a damped random walk\footnote{This process is also known as Orstein-Uhlenbeck process in physics, or the Vasicek model in financial literature.} with the following PSD \citep{kelly14}:

\begin{equation}
    \mathrm{PSD}(f) = \frac{\sigma^{2}_{\rm int}}{ 1/\tau^{2}_{\rm eq}+(2 \pi  f)^{2} }  = \frac{\sigma^2_{\rm reg}}{1+(2\pi \tau_{\rm eq} f)^{2}},
    \label{eq:PSD_drw}
\end{equation}

\noindent 
where $\sigma_{\rm int}^{2}=C\cdot\mathrm{Var}(\tilde{\Phi})$ and, hence, $\sigma^2_{\rm reg} = \tau_{\rm eq}^2 \sigma_{\rm int}^2 = C\frac{\tau_{\rm eq}^2}{\tau_{\rm dep}^2} \mathrm{Var}(\Phi)$ employing the definition of $\tilde{\Phi}$. From Eq.~\ref{eq:inflow_psd}, the variance of $\Phi(t)$ is given by $\mathrm{Var}(\Phi)=\exp(2\mu+\sigma^2)[\exp(\sigma^2)-1.0]$ as specified in Eq.~\ref{eq:inflow_psd}. As pointed out in \citealt{wang20}, we see that the response of the regulator is given by PSD of the inflow and the frequency dependent response (for more detailed treatment of a non-stochastic case see also \citealt{wang19_manga}). The PSD of a damped random walk (Eq.~\ref{eq:PSD_drw}) is a broken power law. The timescale of the break is at $2\pi \tau_{\rm eq}$. The ``long-term'' variability is given by $\sigma^2_{\rm reg}$, which is set by the stochasticity of the inflow rate $\Phi(t)$ . We can transform this PSD to the auto-correlation function (ACF), i.e. translate from frequency to time-domain, via the Wiener-Khinchin theorem \citep{wiener30, khinchin34, emmanoulopoulos10}:

\begin{equation}
    \mathrm{ACF}(t) = \exp(-t/\tau_{\rm eq}).
    \label{eq:ACF_drw}
\end{equation}

The meaning of this is the following: the evolution of the SFR of the basic regulator is highly correlated on short timescales / high frequencies with $\mathrm{ACF}\sim1$ and described with a PSD with a slope that equals 2. Then, the process becomes rapidly decorrelated at a timescale $\tau_{\rm eq}$, and at long time-scales (low frequencies), the ACF drops quickly ($\mathrm{ACF}\sim0$) and the PSD is well described with a flat slope. 

We show examples of the PSD resulting from the basic regulator model in Fig.~\ref{fig:example_basic_regulator}. Specifically, the left panels show the result for the case where $\mathrm{PSD}_{\Phi}$ is a white-noise process (constant PSD). The upper panels show one example of the inflow rate and SFR (the inflow rate has been re-normalized by 0.1 in order to increase visibility), while the lower panels show the associated PSDs for three examples. In all panels, we assume for the inflow rate $\Phi$ a normalisation $\mu=1$ and dispersion of $\sigma=1.0$, which set the overall strength of the PSD. For the basic regulator model, we assume a depletion timescale of $\tau_{\rm dep}=100~\mathrm{Myr}$, a star-formation efficiency of $\varepsilon_0=0.1$ with no mass dependence ($\alpha_{\rm e}=0$), and a mass-loading of $\lambda=5$. These numbers give an equilibrium timescale of $\tau_{\rm eq}=18~\mathrm{Myr}$. We choose this rather short $\tau_{\rm eq}$ just for pedagogical reasons, i.e. to be able to clearly differentiate the cases where $\tau_{\rm x}$ is smaller than, similar to, and larger than $\tau_{\rm eq}$. We see that the PSD of the SFR indeed shows a break at $2\pi\tau_{\rm eq}\approx110~\mathrm{Myr}$. Below this break, the PSD has a slope of 2, as given by Eq.~\ref{eq:PSD_drw}.

We can now move to a more complex case, where the inflow $\Phi$ is a correlated process itself, i.e. we assume that $\mathrm{PSD}_{\Phi}$ is a broken power-law, as described in Eq.~\ref{eq:inflow_psd} and shown in Fig.~\ref{fig:inflow_psd}. Consistent with the above picture, the basic regulator model correlates the star-formation history on timescales shorter than $\tau_{\rm eq}$, giving rise to a break in the PSD in addition to the one from $\Phi$ at $\tau_{\rm x}$. This gives rise to a triple power-law PSD\footnote{This solution for the damped random walk process is motivated by examining the stochastic differential equation (Equation \ref{eq:sfr_regulator}) in Fourier space, namely  $i\omega \mathcal{F}(\mathrm{SFR}) + (1/\tau_\mathrm{eq}) \mathcal{F}(\mathrm{SFR}) = \mathcal{F}(\tilde{\Phi})$. It follows that $\mathcal{F}(\mathrm{SFR}) = \mathcal{F}(\tilde{\Phi}) (i \omega + 1/\tau_{\rm eq})^{-1}$. In the case of a white noise source term, we re-derive Equation \eqref{eq:PSD_drw}, and in the more general case, any non-white-noise shape of the driver's PSD propagates through without modification.} given by:

\begin{equation}
    \begin{split}
    \mathrm{PSD}_{\rm reg}(f) & = \frac{1}{ 1/\tau_{\rm eq}^2 + (2 \pi f)^{2} } \mathrm{PSD}_{\tilde{\Phi}}\\
    & = \frac{\sigma^{2}_{\rm reg}}{1+(2 \pi \tau_{\rm eq} f)^{2}}\frac{2}{(\tau_{\rm x} f)^{\beta_{\rm l}}+(\tau_{\rm x} f)^{\beta_{\rm h}}} \\
    & = \frac{2\sigma^{2}_{\rm reg}}{(\tau_{\rm x} f)^{\beta_{\rm l}}+(\tau_{\rm x} f)^{\beta_{\rm h}} + (2 \pi \tau_{\rm eq})^2(\tau_{\rm x}^{\beta_{\rm l}} f^{\beta_{\rm l} + 2} + \tau_{\rm x}^{\beta_{\rm h}} f^{\beta_{\rm h} + 2})}.
    \end{split}
    \label{eq:PSD_drw_triple}
\end{equation}

\noindent
This equation describes a PSD with two breaks. One break lies at a timescale $\tau_{\rm x}$, the other at $2\pi\tau_{\rm eq}$. The order of the slopes depend on the relative locations of $\tau_{\rm x}$ and $2\pi\tau_{\rm eq}$. In the case where $\tau_{\rm x}>2\pi\tau_{\rm eq}$, the slope on low frequencies (long timescales, i.e. $f<1/\tau_{\rm x}$) is given by $\beta_{\rm l}$. At intermediate ($1/\tau_{\rm x}<f<1/(2\pi\tau_{\rm eq})$) and high ($f>1/(2\pi\tau_{\rm eq})$) frequencies, the slopes of the PSD are $\beta_{\rm h}$ and $\beta_{\rm h}+2$, respectively. In the case where $\tau_{\rm x}<2\pi\tau_{\rm eq}$, the slopes are $\beta_{\rm l}$, $\beta_{\rm l}+2$ and $\beta_{\rm h}+2$ at low, intermediate and high frequencies, respectively. Essentially, the basic regulator model correlates the star-formation history on timescales shorter than $2\pi\tau_{\rm eq}$, leading to a factor of 2 steeper slope at high frequencies.

We show examples of this Fig.~\ref{fig:example_basic_regulator}: the three column to the right show an inflow $\mathrm{PSD}_{\Phi}$ with increasing break timescale of $\tau_{\rm x}=10$, 100 and 1000 Myr. In all panels, the basic regulator model is run with the same input parameters as described above, giving rise to the same $\tau_{\rm eq}$ of 18 Myr. In the second panel from the left, $\tau_{\rm x}$ is shorter than $2\pi\tau_{\rm eq}$: the PSD of the SFR shows two breaks, where for intermediate timescales, the slope of the PSD is 2, while on shorter timescales than $\tau_{\rm x}$, the slope changes to 4. The panel on the right shows the opposite case, where $\tau_{\rm x} > 2\pi\tau_{\rm eq}$.

In summary, we have shown that the basic regulator model introduces an additional feature in the PSD of the SFR. In particular, the basic regulator model correlates the SFR on timescales shorter than $\tau_{\rm eq}$, giving rise to a break in the PSD at $2\pi\tau_{\rm eq}$. Below this break, the PSD has a slope that is steeper by 2 than what is present in $\mathrm{PSD}_{\Phi}$. This implies that measuring this break will give us a handle on the equilibrium timescale of the systems, and hence on the depletion time and the mass-loading factor $\lambda$. Importantly, the normalization, which is proportional to $\left(\frac{\tau_{\rm eq}}{\tau_{\rm dep}}\right)^2\approx (1-f_{\rm R}+\lambda)^{-2}$, will give us an additional joint constraint on the mass-loading factor and the variance of the forcing $\Phi$.

\subsection{Imprint of the lifetime of molecular clouds on the PSD}
\label{subsec:only_gmc}

\begin{figure*}
    \centering
    \includegraphics[width=\textwidth]{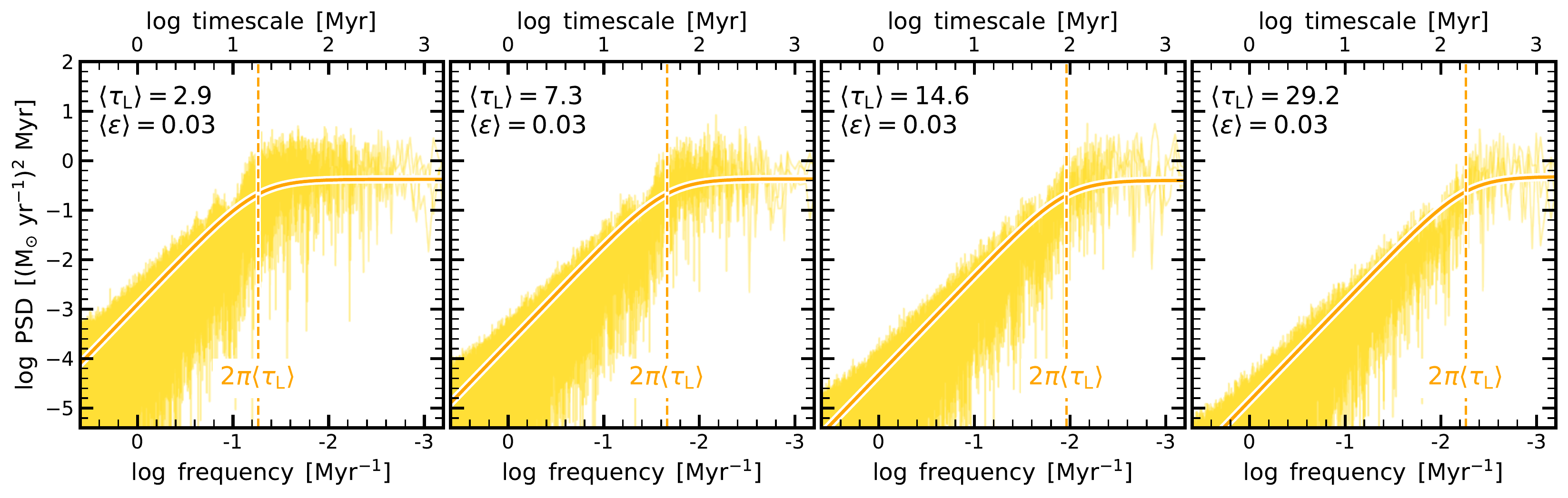}
    \caption{PSDs of example star-formation histories obtained by sampling GMCs. In this simplified model, we assume a constant mass of the reservoir of $M_{\rm gas}=10^{9}~M_{\odot}$ and sample from this reservoir GMC according to the equations given in Section~\ref{subsec:GMC}. Specifically, we assume a depletion timescale of $\tau_{\rm dep}=500~\mathrm{Myr}$, a GMC birth mass function with a slope of $\alpha_{\rm b}=-2$ and $m_{\rm min},m_{\rm max}=[10^4, 10^7]$, and a star-formation efficiency of $\varepsilon_0=0.02$ and $\alpha_{\rm e}=0.25$. The panels from left to right show the PSDs of the resulting star-formation histories with increasing GMC lifetimes with $\tau_0=2$, 5, 10 and 20 Myr with $\alpha_{\rm l}=0.25$. In all panels, we plot the PSDs of three example star-formation histories. The PSDs are consistent with a damped random walk: the star-formation histories that originate from sampling the molecular cloud mass function are correlated below a timescale of $\langle\tau_{\rm L}\rangle$ (with a powerlaw slope of 2), resulting in a PSD that has a break at $2\pi\langle\tau_{\rm L}\rangle$ (vertical dashed line in all panels). On longer timescales, the star-formation histories are uncorrelated. The orange solid lines show the expected PSD from Eq.~\ref{eq:PSD_gmc}, which is in good agreement with the numerically obtained PSDs.}
    \label{fig:example_gmc_only}
\end{figure*}

In a second step, before moving to our extended regulator model, we want to constrain the PSD imprint of GMCs as described in Section~\ref{subsec:GMC}. To first order, since the the lifetime of GMCs, $\tau_{\rm L}$, correlates the SFR on this timescale, we expect an imprint of $\tau_{\rm L}$ on the PSD. To show this, we run a simplified model. Specifically, we assume a constant gas mass in the reservoir. At each timestep, we then draw a population of GMCs as specified by the equations in Section~\ref{subsec:GMC}. Importantly, we do not solve the equations of the regulator model, i.e. we assume that there is no mass exchange between the GMC population and the gas reservoir: $M_{\rm gas}$ is constant throughout, so the only source of variability is the random draws from the distribution of GMC masses.

We analytically derive the solution for a simplified case, with a single GMC mass, lifetime, and SFR$_\mathrm{GMC}$, in Appendix~\ref{app:full_derivation}. This idealized case is a good approximation to our full GMC model when the SFR is dominated by the high-mass end of the GMC mass function, but in general the PSD will be smoother than what we derive analytically as the result of averaging over the full population of GMCs. We therefore use a broken powerlaw that has the same asymptotic behaviour as the analytically derived PSD, namely a damped random walk process:

\begin{equation}
    \begin{split}
    \mathrm{PSD}_{\rm GMC}(f) & = \frac{\sigma^{2}_{\rm GMC}}{ 1 + (2\pi\langle\tau_{\rm L}\rangle f)^{2} },
    \end{split}
    \label{eq:PSD_gmc}
\end{equation}
with 

\begin{equation}
    \begin{split}
    \sigma_{\rm GMC}^2 & = \pi\langle\tau_{\rm L}\rangle \mathrm{Var}(\mathrm{SFR}) 
    = \pi\langle\tau_{\rm L}\rangle \mathrm{Var}(\mathrm{N}_{\rm GMC}\cdot\mathrm{SFR}_{\rm GMC}) \\
    & = \pi\langle\tau_{\rm L}\rangle [\mathrm{Cov}(\mathrm{N}_{\rm GMC}^2, \mathrm{SFR}_{\rm GMC}^2) + \langle\mathrm{SFR}_{\rm GMC}^2\rangle \cdot \langle\mathrm{N}_{\rm GMC}^2\rangle \\
    &~~~ - (\mathrm{Cov}(\mathrm{N}_{\rm GMC}, \mathrm{SFR}_{\rm GMC}) +\langle\mathrm{SFR}_{\rm GMC}\rangle \cdot \langle\mathrm{N}_{\rm GMC}\rangle)^2] \\
    & \approx \pi\langle\tau_{\rm L}\rangle [\mathrm{Var}(\mathrm{N}_{\rm GMC}) \langle\mathrm{SFR}_{\rm GMC}\rangle^2 + \mathrm{Var}(\mathrm{SFR}_{\rm GMC})\cdot\langle \mathrm{N}_{\rm GMC} \rangle_{\rm t}^2 \\
    &~~~ + \mathrm{Var}(\mathrm{N}_{\rm GMC}) \cdot \mathrm{Var}(\mathrm{SFR}_{\rm GMC})],
    \end{split}
\end{equation}

\noindent
where $\langle \cdot \rangle$ is the average over all living GMC. The last step in the aforementioned equation usually holds, since the covariance between $\mathrm{N}_{\rm GMC}$ and $\mathrm{SFR}_{\rm GMC}$ is negligible in the cases we consider. This means that the break of the PSD is sensitive to the average lifetime of GMCs: the SFR is correlated on timescales below $2\pi\langle\tau_{\rm L}\rangle$, while it is uncorrelated on longer timescales (as expected from a Poisson process). The overall normalisation depends on the number of GMCs drawn and the SFR of GMCs. These numbers depend themselves on the GMC mass function, the star-formation efficiency $\varepsilon(M)$, as well as the molecular gas formation timescale $\tau_{\rm mol}$. 

Fig.~\ref{fig:example_gmc_only} shows PSDs of example star-formation histories obtained by sampling the GMC population. In all panels, we assume a constant gas mass reservoir of $M_{\rm gas}=10^9~\mathrm{M}_{\odot}$, a depletion timescale of $\tau_{\rm dep}=500~\mathrm{Myr}$ and a GMC birth mass function with a slope of $\alpha_{\rm b}=-2$ with cutoffs at $m_{\rm min},m_{\rm max}=[10^4, 10^7]$. Furthermore, we set $\varepsilon_0=0.02$ and $\alpha_{\rm e}=0.25$. In panels from left to right, we increase the GMC lifetimes: $\tau_0=2$, 5, 10 and 20 Myr with $\alpha_{\rm l}=0.25$. As expected from Eq.~\ref{eq:PSD_gmc}, we find in all panels a PSD consistent with a damped random walk. The break is related to the average lifetime $\langle\tau_{\rm L}\rangle$: increasing the lifetime of GMCs leads to a star-formation history that is correlated on longer timescales.

\subsection{Extended regulator model}
\label{subsec:combined}

\begin{figure*}
    \centering
    \includegraphics[width=\textwidth]{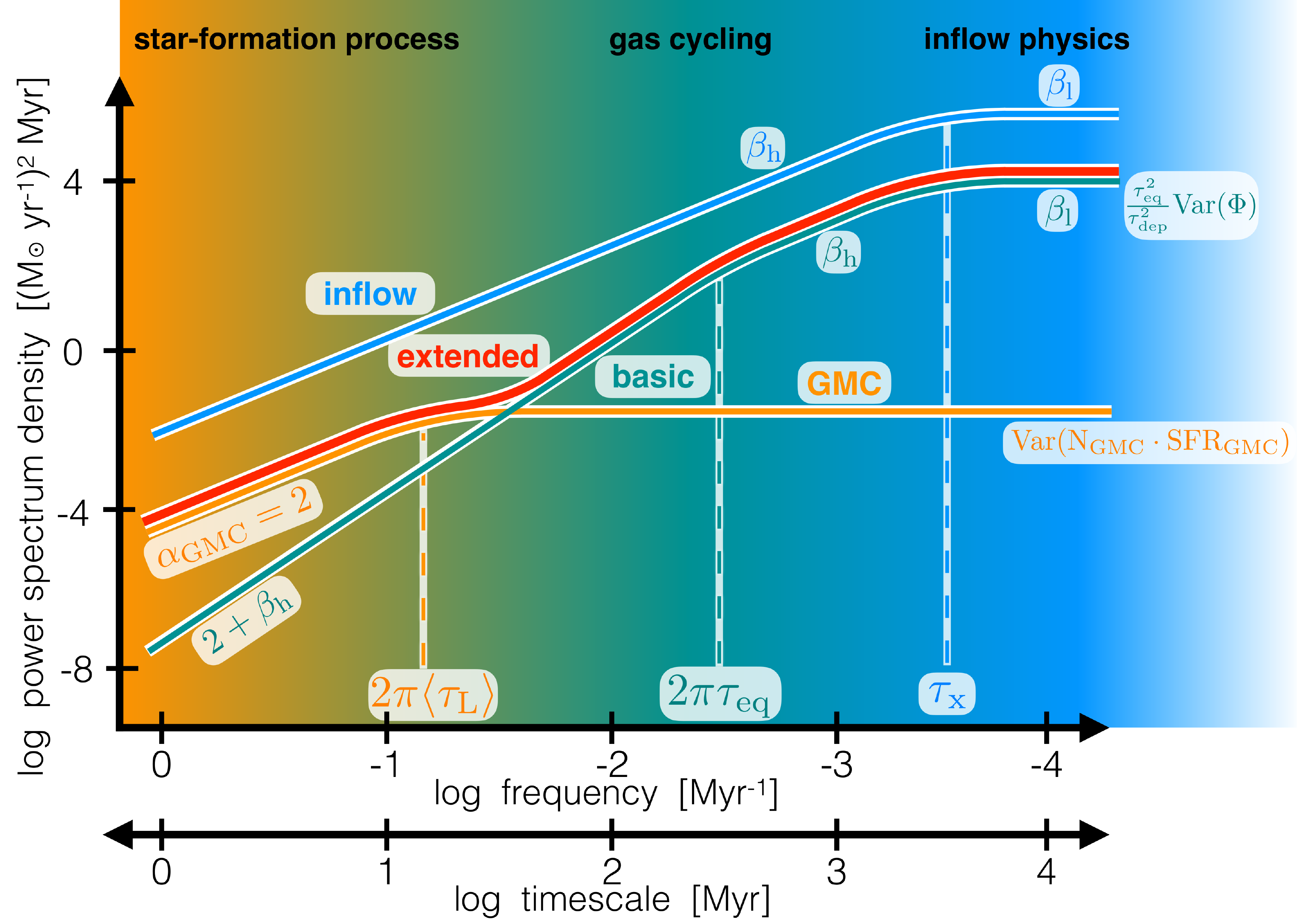}
    \caption{What can we learn from the power spectrum density (PSD) of the SFR? This schema summarizes the key conclusion of the paper. In the extended regulator model, the PSD of the SFR (red line) is shaped by the combined effect of the inflow process (blue line), processes  related to the regulation of gas flow (green line), and the star-formation process related to the formation of GMCs (orange line). Specifically, the PSD can be described by Eqs.~\ref{eq:PSD_drw_triple}, \ref{eq:PSD_gmc} and \ref{eq:PSD_extended}. The inflow rate to the gas reservoir describes the variability of the SFR on the longest timescales on the order of $\tau_{\rm x}$. A combination of the depletion timescale, the mass-loading factor, and the average star-formation efficiency are shaping the intermediate timescales around the equilibrium timescale $\tau_{\rm eq}$. The shortest timescales are dominated by the star formation process: particularly, the average lifetime of GMCs $\langle \tau_{\rm L} \rangle$ correlate the SFR on the shortest timescales and imprint an additional break in the PSD.}
    \label{fig:PSD_schema}
\end{figure*}

After discussing the basic regulator model (Section~\ref{subsec:basic_regulator}) and the star-formation prescription (i.e. GMC model, Section~\ref{subsec:only_gmc}), we are now ready to study the PSD of the extended regulator model. The PSD of the extended regulator model can be described by the following combination of the PSD of the basic regulator model, the PSD of the GMC model, and their cross PSD\footnote{While the PSD is defined as the Fourier Transform of the auto-correlation function, the cross PSD is defined as the Fourier Transform of the cross-correlation function between two signals.}: 

\begin{equation}
    \mathrm{PSD}_{\rm ext}(f) = \mathrm{PSD}_{\rm GMC}(f) + \mathrm{PSD}_{\rm reg}(f) - 2\mathrm{Re}[\mathrm{PSD}_{\rm GMC\times reg}],
    \label{eq:PSD_extended}
\end{equation}

\noindent
where $\mathrm{PSD}_{\rm reg}(f)$ and $\mathrm{PSD}_{\rm GMC}(f)$ are given by Eqs.~\ref{eq:PSD_drw_triple} and \ref{eq:PSD_gmc}, respectively, and $\mathrm{PSD}_{\rm GMC\times reg}$ is the cross PSD. This follows from the fact that, roughly speaking, the SFR in the extended regulator model is the SFR from the basic regulator model modulated by the stochasticity of randomly drawing GMCs, i.e. $\mathrm{SFR} \approx \mathrm{SFR}_\mathrm{reg}\sum \mathrm{SFR}_\mathrm{GMC} / ( \langle N_\mathrm{GMC} \rangle \langle {\rm SFR}_\mathrm{GMC} \rangle) $. 

Fig.~\ref{fig:PSD_schema} shows a schematic diagram of the PSD of the extended regulator model (red line). The PSD of the extended regulator model can be decomposed into the PSD of the GMC model (orange line) and the PSD of the basic regulator model (green line), which itself depends on the PSD of the inflow rate (blue line). In general, we expect the PSD to have three breaks, corresponding to the timescale of the average lifetime of GMCs, the equilibrium timescale of the regulator model, and the correlation timescales of the inflow rate. Typically, we expect the short timescales (high frequencies) to be determined by the GMC model, while the intermediate and long timescales are dominated by the physics related to outflows and inflows. Hence, the slope of the PSD at short timescales is $\beta=\beta_{\rm GMC}=2$, while at longer timescales, the PSD slope depends on the shape of the inflow PSD. We discuss four practical case studies (galaxy regimes) in Section~\ref{subsec:case_studies}.

The cross PSD, $\mathrm{PSD}_{\rm GMC\times reg}$, is likely to be important on timescales where the average number of GMCs is determined by the SFR of the basic regulator model. On shorter timescales, the individual GMC draws will entirely determine the power spectrum, while on longer timescales the SFR is determined by the PSD of the gas supply. Therefore this cross-term only contributes in the relatively small range of timescales where the PSD is transitioning from one regime to another.

\subsection{Case studies: different galaxy regimes}
\label{subsec:case_studies}

\begin{figure*}
    \centering
    \begin{tabular}{c}
    \includegraphics[width=0.7\textwidth]{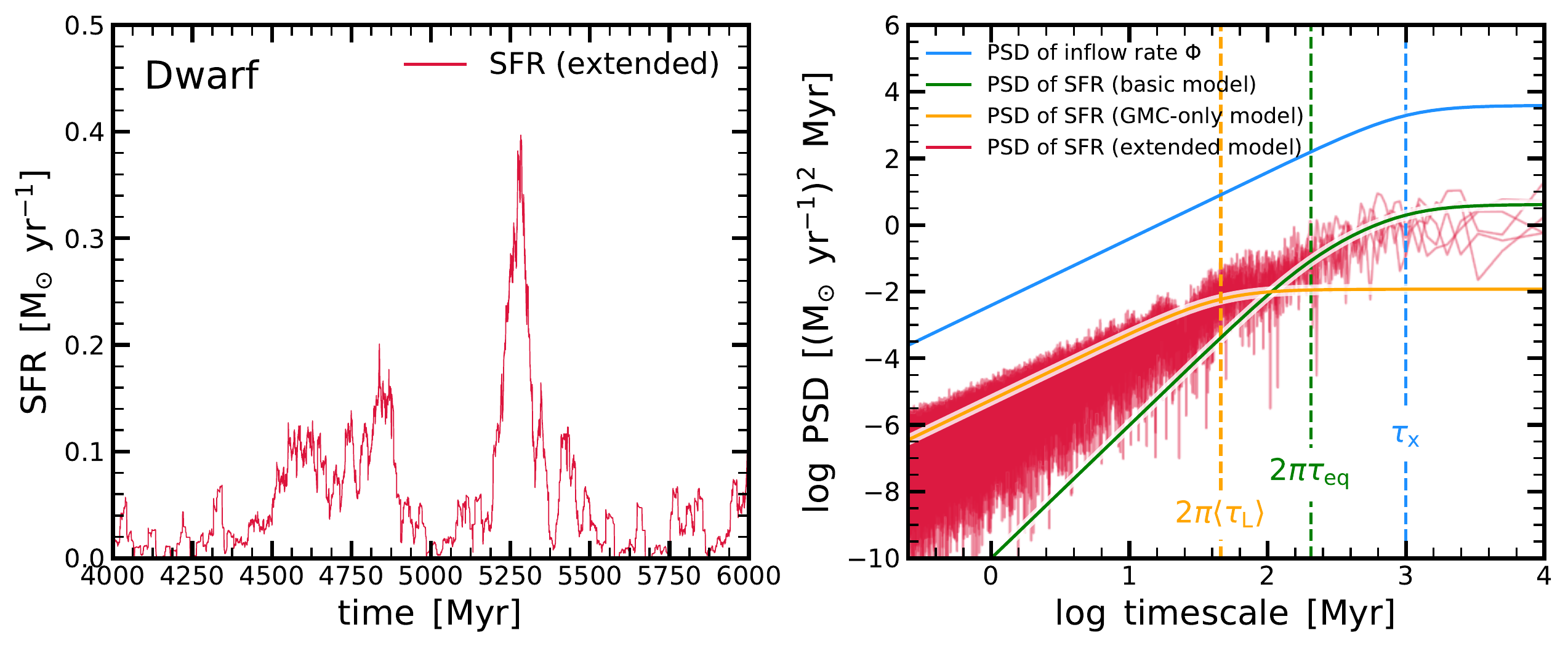} \\ 
    \includegraphics[width=0.7\textwidth]{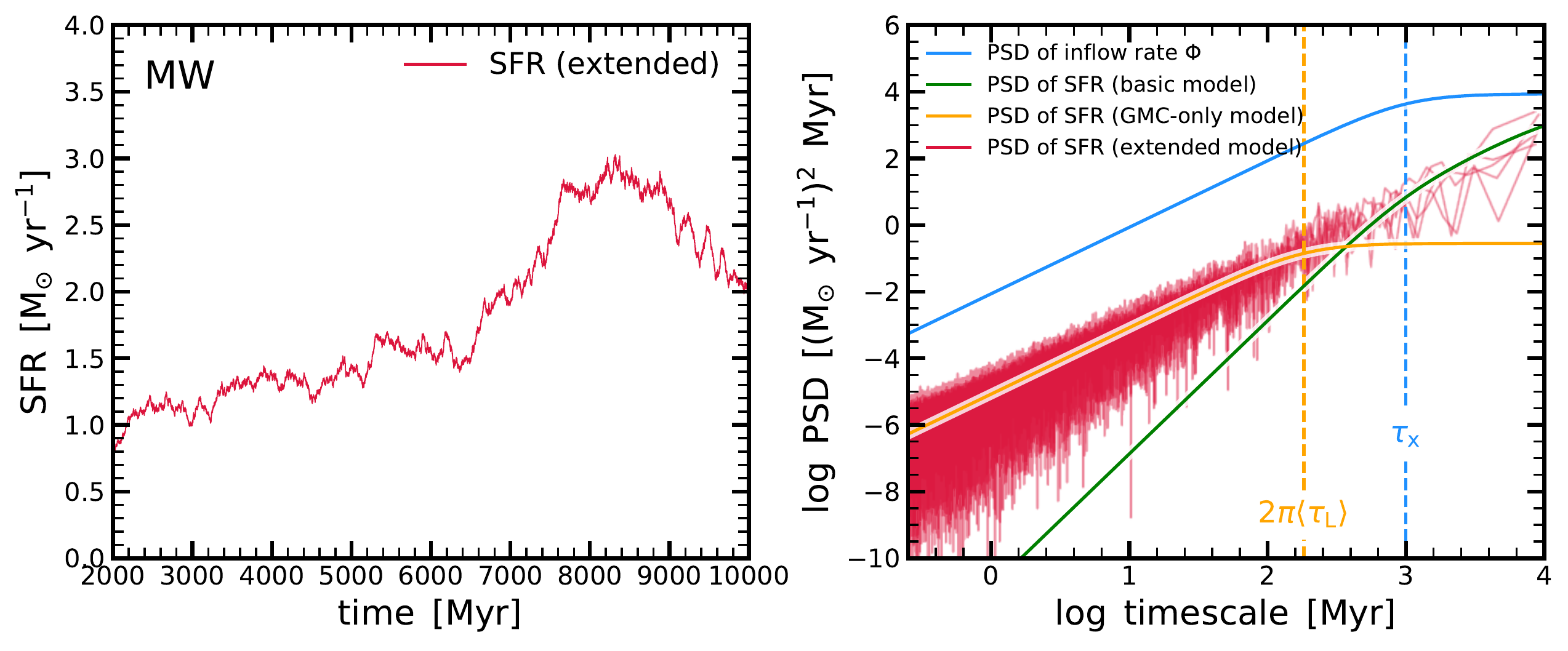} \\
    \includegraphics[width=0.7\textwidth]{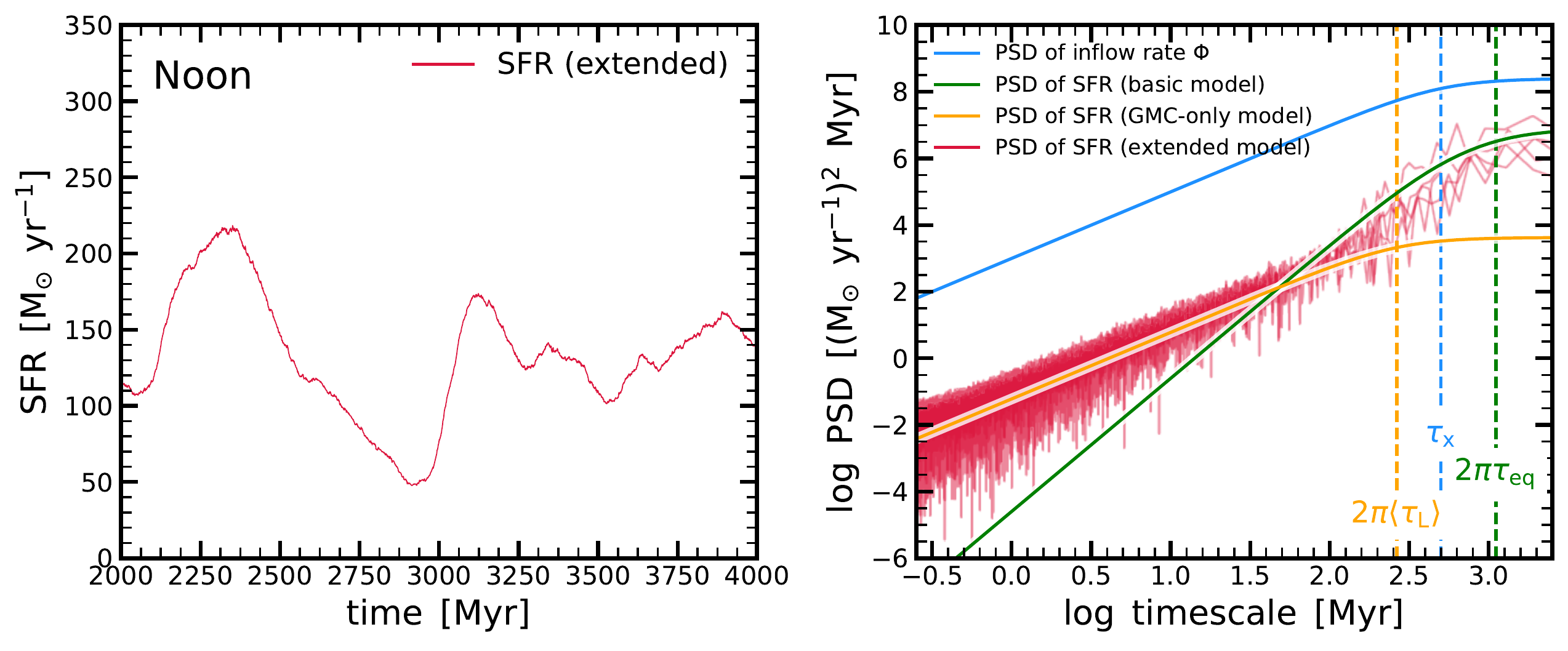} \\ 
    \includegraphics[width=0.7\textwidth]{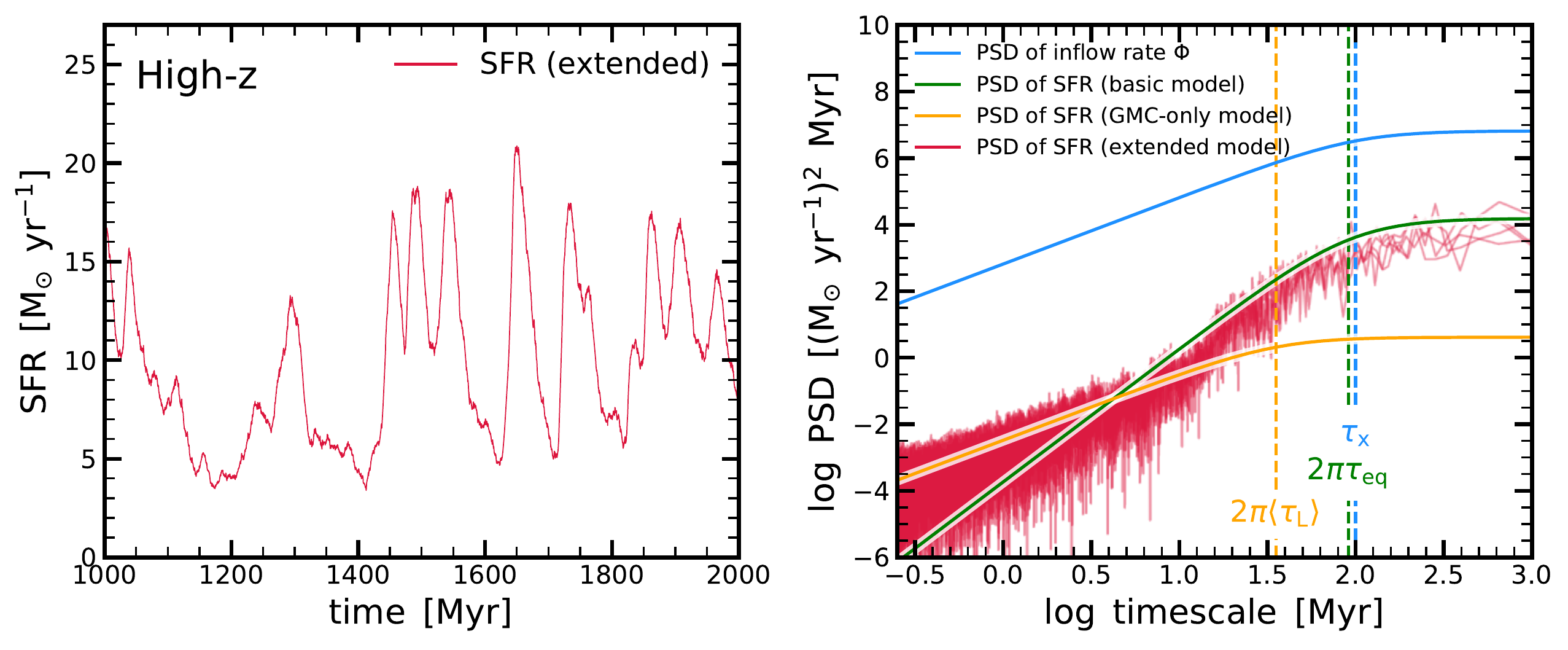} 
    \end{tabular}
    \caption{Star-formation histories and PSDs for four different galaxy regimes. From top to bottom: dwarf galaxy (``Dwarf''), Milky Way-like galaxy (``MW''), massive galaxy at $z\sim2$ (``Noon''), and galaxy at $z>6$ (``High-z'') regime. The parameters for these four regimes are given in Table~\ref{tab:parameters_cases}. Note that the ranges of the axes are different in the different panels. For each example, we plot the star-formation history of the extended model in the left panel. In the right panel, the blue line shows the PSD of the inflow rate $\Phi$, while the red line shows the PSD of the SFR from five realisations of the extended model (only one of those star-formation histories is shown in the left panel). The green and orange line show the predictions for the basic regulator and the GMC-only models, respectively. The vertical dashed line indicate the relevant timescales of the breaks in the PSD, related to the average GMC lifetime, the equilibrium timescale, and the decorrelation timescale of the inflow. We find that the PSD encodes interesting physical processes and that the different galaxy regimes span a wide range of PSD shapes. }
    \label{fig:combined_PSD}
\end{figure*}

\begin{table*}
	\centering
	\caption{Choice of parameters and derived properties for the case studies presented in Section~\ref{subsec:case_studies} and Fig.~\ref{fig:combined_PSD}. The four cases include a low-mass ($M_{\star}\sim10^{7}-10^{8}$), dwarf-like galaxy at $z\sim0$ (Dwarf), a Milky Way-like galaxy at $z\sim0$ (MW), a massive ($M_{\star}\sim10^{10}-10^{11}$), star-forming galaxy at $z\sim2$ (Noon), and a high-$z$ ($z>6$) galaxy (High-$z$). The numbers in parentheses show the standard deviations of these quantities.}
	\label{tab:parameters_cases}
	\begin{tabular}{llcccc}
 \hline \hline
 Parameter & Definition & Dwarf & Milky Way (MW) & Noon & High-$z$ \\ 
 \hline \hline
 \textbf{Input:} & & & & & \vspace{0.1cm} \\
 $\tau_{\rm x}$ [Myr] & Eq.~\ref{eq:PSD_x} & 1000 & 1000 & 500 & 100 \\
 $\beta_{\rm l}$ & Eq.~\ref{eq:PSD_x} & 0 & 0 & 0 & 0 \\
 $\beta_{\rm h}$ & Eq.~\ref{eq:PSD_x} & 2 & 2 & 2 & 2 \\
 $\mu$ & Eq.~\ref{eq:inflow_regulator} & 0.1 & 0.5 & 6.0 & 5.0 \\
 $\sigma$ & Eq.~\ref{eq:inflow_regulator} & 1.0 & 1.0 & 1.0 & 1.0 \\
 $\lambda$ & Eq.~\ref{eq:outflow} & 30 & 1.0 & 5.0 & 20.0 \\
 $\tau_{\rm dep}$ [Myr] & Eq.~\ref{eq:depletion_time_basic} & 1000 & 4000 & 1000 & 300 \\
 $\alpha_{\rm b}$ & Eq.~\ref{eq:GMC_MF_birth} & $-2.0$ & $-2.0$ & $-2.0$ & $-2.0$ \\
 $m_{\rm min}, m_{\rm max}$ [M$_{\odot}$] & Eq.~\ref{eq:GMC_MF_birth} & 10$^4$, 10$^7$ & 10$^4$, 10$^7$ & 10$^5$, 10$^9$ & 10$^4$, 10$^8$ \\
 $\varepsilon_0$ & Eq.~\ref{eq:GMC_SFE} & 0.02 & 0.02 & 0.02 & 0.02 \\
 $\alpha_{\rm e}$ & Eq.~\ref{eq:GMC_SFE} & 0.25 & 0.25 & 0.2 & 0.0 \\
 $\tau_0$ [Myr] & Eq.~\ref{eq:GMC_LT} & 5.0 & 20.0 & 20.0 & 5.0 \\ 
 $\alpha_{\rm l}$ & Eq.~\ref{eq:GMC_LT} & 0.25 & 0.25 & 0.2 & 0.1 \\
	\hline
 \textbf{Derived:} & & & & & \vspace{0.1cm} \\
 $\langle \Phi \rangle_{_{\rm t}}~[\mathrm{M_{\odot}}~\mathrm{yr}^{-1}]$ & Eq.~\ref{eq:inflow_regulator} & 1.0 (0.6) & 1.5 (0.8) & 231.6 (152.2) & 170.1 (109.4)  \\
 $\langle M_{\rm gas} \rangle_{_{\rm t}}~[\mathrm{M_{\odot}}]$ & Eq.~\ref{eq:inflow_regulator} & 3.2 (1.6) $\times10^7$ & 6.6 (2.7) $\times10^9$ & 6.6 (3.3) $\times10^{10}$ & 3.8 (2.3) $\times10^{9}$  \\
 $\langle \mathrm{SFR} \rangle_{_{\rm t}}~[\mathrm{M_{\odot}}~\mathrm{yr}^{-1}]$ & Eq.~\ref{eq:SFR_full} & 0.03 (0.02) & 1.7 (0.1) & 80.1 (36.9) & 13.0 (7.8) \\
 $\tau_{\rm mol}$ [Myr] & Eq.~\ref{eq:depletion_time_mol} & 52 & 211 & 121 & 47 \\
 $\tau_{\rm eq}$ [Myr] & Eq.~\ref{eq:eq_timescale} & 33 & 2448 & 177 & 15 \\
 $\tau_{\rm dep, mol}$ [Myr] & Eq.~\ref{eq:depletion_time_mol} & 124 & 498 & 493 & 203 \\
 $\langle \tau_{\rm L} \rangle$ [Myr] & Eq.~\ref{eq:GMC_LT} & 7.3 & 29.1 & 42.1 & 5.6 \\
 $\langle \varepsilon \rangle$ & Eq.~\ref{eq:GMC_SFE} &  0.03 & 0.03 & 0.04 & 0.03 \\
 $\langle \mathrm{N_{\rm GMC}} \rangle_{_{\rm t}}$ & --- & 60 (31) & 11862 (459) & 36672 (18551) & 39466 (23694) \\
 $\langle R_{\rm mol} \rangle_{_{\rm t}}$ & Eq.~\ref{eq:Rmol} & 0.16 (0.10) & 0.12 (0.01) & 0.62 (0.09) & 0.70 (0.09) \\
 $\langle f_{R} \rangle_{_{\rm t}}$ & --- & 0.38 & 0.37 & 0.34 & 0.30 \\
	\hline \hline
	\end{tabular}
\end{table*}

After the rather theoretical discussion of PSDs in Section~\ref{subsec:combined}, we now turn to some practical case studies. We want to demonstrate how the star-formation histories and PSDs look in different galaxy regimes. Specifically, we focus on the following regimes:

\begin{itemize}
    \item low-mass ($M_{\star}\sim10^{7}-10^{8}$), dwarf-like galaxy at $z\sim0$ (``Dwarf'');
    \item Milky Way-like galaxy at $z\sim0$ (``MW'');
    \item massive ($M_{\star}\sim10^{10}-10^{11}$), star-forming galaxy at $z\sim2$ (``Noon''); and
    \item high-$z$ ($z>6$) galaxy (``High-$z$'').
\end{itemize}

The choices of the input parameters are listed in the first half of Table~\ref{tab:parameters_cases}. The second half of Table~\ref{tab:parameters_cases} lists the derived properties from our extended regulator model, such as the average number of active GMCs. The numbers in the brackets show the standard deviations of these quantities.

These parameter choices are motivated by reproducing the observed properties of typical galaxies in these regimes. Clearly, some of the observational constraints are quite uncertain (see Section~\ref{subsec:regulator}) and galaxies in these regimes also have diverse properties. Therefore, these choices are necessarily somewhat arbitrary. In particular, the PSD of the inflow rate is uncertain. Since the inflow rate is likely related to the circumgalatic gas in the halo, we assume that the inflow rate is correlated up to the halo dynamical time:

\begin{equation}
    \tau_{\rm DM,dyn} = \left( \frac{3\pi}{32G\rho_{\rm 200,crit}} \right)^{1/2} \sim 0.1\tau_{\rm H}
\end{equation}

\noindent
where $\tau_{\rm H}$ is the Hubble time (at the epoch of observation). This means we set $\tau_{\rm x}\approx0.1\tau_{\rm H}$. We assume that on longer timescales than $\tau_{\rm x}$, the inflow rate is uncorrelated, i.e. $\beta_{\rm l}=0$, while on shorter timescales, it is correlated with $\beta_{\rm h}=2$. 

The depletion time $\tau_{\rm dep}$ and mass-loading factor are motivated by observations of galaxies across cosmic time \citep[][and reference therein]{tacconi20}. Finally, the GMC parametrisation is motivated by observations of the Milky Way and local galaxies (see Section~\ref{subsec:GMC}, including \citealt{murray11} and \citealt{kruijssen19}). We assume $m_{\rm max}=10^7~\mathrm{M_{\odot}}$ in local galaxies \citep{rice16} and $m_{\rm max}=10^{8-9}~\mathrm{M_{\odot}}$ at higher redshifts \citep[e.g.][]{escala08, reina-campos17}.

We then run our model with these input parameters, sampling the star-formation history in steps of 0.1 Myr, which is enough to resolve the ages of the youngest GMCs with at least 30 temporal resolution elements. Furthermore, the length of the produced star-formation history corresponds to roughly the Hubble time of the different galaxy regimes, i.e. the length of the star-formation history of the Dwarf, MW, Noon and High-$z$ regimes is 10 Gyr, 10 Gyr, 4 Gyr, and 2 Gyr, respectively. 

The resulting average quantities are shown in the second half of Table~\ref{tab:parameters_cases}. The most important quantity predicted by our model is the SFR. The time-averaged SFR is in excellent agreement with observational estimates for all of the four galaxy regimes. The depletion time of the molecular gas ($\tau_{\rm dep,mol}$) is a well-constrained quantity in observations. However, as we mention in Section~\ref{subsec:connection_basic_full}, the molecular gas mass -- and therefore $\tau_{\rm dep,mol}$ and $R_{\rm mol}$ -- may be underestimated in our model, because our model only accounts for the molecular gas in GMCs, and misses diffuse molecular gas and molecular gas in non-star-forming clouds (or equivalently early phases of GMC formation with little star formation present).

Fig.~\ref{fig:combined_PSD} shows the star-formation histories (for one realisation of our model) and PSDs (five realisations) of the four galaxy regimes. From top to bottom, we show an example of the Dwarf regime, the MW regime, the Noon regime, and the High-$z$ regime. We find that the resulting star-formation histories are qualitatively different (left panels), both in absolute normalization and variability. In the Dwarf regime, the star-formation history looks bursty and, consistently, the PSD shows substantial power on short timescales relative to longer timescales. In this regime, PSD does not exhibit two breaks, i.e. the PSD transitions smoothly from the GMC regime to the white noise regime. This is because the equilibrium timescale is short and comparable to the average lifetime of GMCs. 

In the MW regime (second panel from the top in Fig.~\ref{fig:combined_PSD}), the star-formation history and PSD indicate that longer timescales play a more important role in the star-formation variability. Although the PSD of the GMC component is similar to the Dwarf regime, fluctuations on longer timescales are more pronounced because the equilibrium timescale is significantly longer than the average GMC lifetime. This leads to a clear separation of the GMC and the basic regulator component, allowing in principle a constraint on the average lifetime of GMCs from the galaxy-integrated star-formation history or PSD alone.

In the Noon regime, the strength of the basic regulator component relative to the GMC model further increases. The main reason for this is the shorter depletion time and higher mass loading with respect to the MW regime, leading to a shorter equilibrium timescale. In this regime, the break of the GMC model is invisible, and the break of the regulator model is clearly present. 

Finally, in the High-$z$ regime, the equilibrium timescale is even shorter. The inflow rate contributes significant power on short timescales since $\tau_{\rm x}$ is now comparable to the average GMC lifetime and the equilibrium timescale. The PSD of the GMC component is clearly sub-dominant over most of the timescale considered, only dominating around timescales of a few Myr. 

In summary, we show that the different galaxy regimes give rise to different PSDs. As expected, dwarf galaxies today and high-$z$ galaxies are bursty, i.e. showing significant power on short timescales. However, the reason for this burstiness is different: in dwarf galaxies, GMCs are responsible for the power on short timescales, while in high-$z$ galaxies, burstiness is related to short equilibrium timescales and large variability of the inflow rate. Consistently, \citet{faucher-giguere18} studies the origin of bursty star formation in galaxies using a simple analytic model. They identify two regimes in which galaxy-scale star formation should be bursty: at high redshift for galaxies of all masses and at low masses (depending on gas fraction) for galaxies at any redshift. At high redshift, burstiness is enhanced because of elevated gas fractions in the early Universe and because the galactic dynamical time-scales become too short for supernova feedback to effectively respond to gravitational collapse in galactic discs. In dwarf galaxies star formation occurs in too few bright star-forming regions to effectively average out, leading to bursty star formation.

\section{Discussion}
\label{sec:discussion}

We showed in the previous section how the SFR variability, i.e. the shape and normalisation of the PSD, depends on different physical processes, such as gas inflow, gas regulation within galaxies, and the star formation process (GMC-related physics). We now use this framework to highlight that the lifetime of GMCs can in principle be measured in certain regimes of galaxy-integrated constraints on the variability. Furthermore, we discuss the main sequence scatter and gradients of galaxy properties across the main sequence. We end by discussing caveats and the next steps ahead.

\subsection{Constraining the time evolution of the GMC life-cycle}
\label{subsec:lifetime_gmc}

\begin{figure*}
    \centering
    \includegraphics[width=\textwidth]{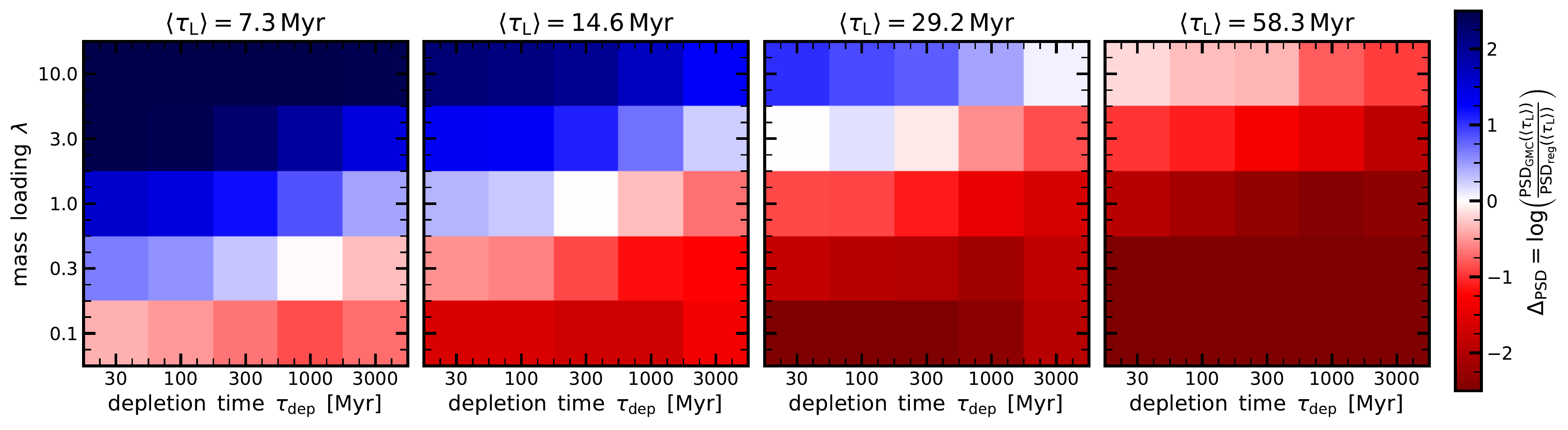} \\ 
    \caption{Measurability of the GMC lifetime: is the break of the GMC PSD visible? We quantify the measurability of the GMC lifetime by the excess of the GMC PSD relative to the PSD of the basic regulator at $\langle\tau_{\rm L}\rangle$ (break of the GMC PSD): $\Delta_{\rm PSD}=\log(\mathrm{PSD}_{\rm GMC}(\langle\tau_{\rm L}\rangle)/\mathrm{PSD}_{\rm reg}(\langle\tau_{\rm L}\rangle))$. We plot $\Delta_{\rm PSD}$ as a function of depletion time $\tau_{\rm dep}$ and mass-loading factor $\lambda$, for four different lifetime distributions (panels from left to right show $\tau_0=5$, 10, 20, and 40 Myr with while $\alpha_{\rm l}=0.25$, leading to $\langle\tau_{\rm L}\rangle=7.3$, 14.6,29.2 and 58.3 Myr). The colour coding is shown with the colour bar on the right: red indicates that the GMC PSD has a smaller amplitude than the PSD of the basic regulator model, i.e. the break of the GMC break is hidden below PSD of the basic regulator model, making it impossible to measure the average lifetime of GMCs. On the other hand, blue indicates that the GMC of the PSD lies above the PSD of the basic regulator model, implying that the break and hence the average lifetime can be inferred. We see that the average GMC lifetime can be better measured if the lifetimes and depletion times are short, and the mass-loading is high.}
    \label{fig:measure_GMC_lt}
\end{figure*}

We highlighted in the introduction that it remains a major challenge to derive a theory of star formation and feedback, because there are only a few robust empirical constraints on the GMC lifecycle. Currently, the best observational constraints are obtained from $z\sim0$, spatially resolved measurements. For example, \citet{engargiola03} infer an upper limit of $10-20$ Myr on GMC lifetimes based on positions of GMC along HI filaments in M33: most GMCs still are associated with their HI filaments, which suggests that they do not live long enough to drift across the filament. Another approach is based on classifying GMCs according to their evolutionary states (GMCs, HII regions, and young stellar objects): such studies find that GMC lifetimes are $20-40$ Myr \citep[e.g.][]{kawamura09, miura12}. Furthermore, \citet[][see also \citealt{kruijssen18}]{kruijssen14} developed a statistical method to measure GMC lifetime based on the ``uncertainty principle for star formation,'' which assumes that peaks in star formation tracers and molecular gas tracers correspond to an evolutionary sequence (GMCs to young star clusters) leading to an anti-correlation between the two at small spatial scales. Applying this method to spatially-resolved ($\sim100$ pc) CO and H$\alpha$ maps, \citet{kruijssen19} and \citet{chevance20} were able to probe the physical variation of the underlying GMC lifecycle in a large sample of galaxies (10 and growing). Thanks to a single homogeneous methodology and unprecedented precision ($<30\%$ uncertainties), they infer GMC lifetimes of $10-30$ Myr, where the range reflects environmental variation. 

Similar measurements are challenging at earlier cosmic time, because spatial resolution and depths are lower than at $z=0$. Since the PSD can be measured from integrated galaxy properties \citep{caplar19}, we now investigate in which regimes constraining the PSD will help us to measure the lifetime of spatially unresolved GMCs. As shown in Fig.~\ref{fig:example_gmc_only} and Eq.~\ref{eq:PSD_gmc}, the PSD of the GMC component depends on the average lifetime of the GMC population $\langle\tau_{\rm L}\rangle$. Specifically, the break of the PSD is at $2\pi\langle\tau_{\rm L}\rangle$. Hence, if one is able to measure the break of the PSD at short timescales (i.e. if the GMC PSD dominates), one is able to infer the average lifetime of GMCs. We find in the previous section that the GMC PSD is more prominent in certain regimes (i.e. Dwarf and MW regime). We now investigate in which regimes a constraint on the average GMC lifetime can be obtained, i.e. in which regimes the break of the GMC PSD is visible. Note that in our work, the GMC lifetime is defined as the time interval during which the GMC is forming stars. Other methods might use a different definition.

We quantify the visibility of the break of the GMC PSD by taking the ratio of the amplitudes of the GMC PSD and the basic regulator model at the timescale of the break, i.e. $2\pi\langle\tau_{\rm L}\rangle$: $\Delta_{\rm PSD}=\log(\mathrm{PSD}_{\rm GMC}(f_{\rm L})/\mathrm{PSD}_{\rm reg}(f_{\rm L}))$, where $f_{\rm L}=1/(2\pi\langle\tau_{\rm L}\rangle)$. Positive $\Delta_{\rm PSD}$ indicates that GMC PSD is the dominating PSD and therefore the break timescale (and the average lifetime) is in principle measurable. On the other hand, a negative $\Delta_{\rm PSD}$ implies that the PSD of the basic regulator model is dominant and the break of the GMC PSD is hidden. 

The value of $\Delta_{\rm PSD}$ depends on several of our input parameters. For the analysis here, we focus on varying the GMC lifetime, the depletion time, and the mass-loading factor. We assume an inflow rate similar to the dwarf/MW regime: $\tau_{\rm x}=1000$ Myr, $\beta_{\rm l}=0$, $\beta_{\rm h}=2$, $\mu=1.0$, and $\sigma=1.0$. Furthermore, we fix the following parameters of the GMC prescription: $\alpha_{\rm b}=-2.0$, $m_{\rm min}=10^4~\mathrm{M_{\odot}}$, $m_{\rm max}=10^7~\mathrm{M_{\odot}}$, $\varepsilon_{0}=0.02$, and $\varepsilon_{\rm e}=0.25$. We vary the depletion time from $\tau_{\rm dep}=30$ Myr to 3000 Myr, and the mass-loading factor from $\lambda=0.1$ to 10.0. For the GMC lifetimes, we investigate $\tau_0=5$, 10, 20 and 40 Myr, all with $\alpha_{\rm l}=0.25$, which gives rise to average lifetimes of $\langle\tau_{\rm L}\rangle=7.3$, 14.6, 29.2, and 58.3 Myr, respectively. 

Fig.~\ref{fig:measure_GMC_lt} shows $\Delta_{\rm PSD}$ in the $\tau_{\rm dep}-\lambda$ plane. The panels from left to right show increasing GMC lifetimes. The colour coding is shown in the colour bar on the right: blue corresponds to positive $\Delta_{\rm PSD}$, the break of the GMC PSD is visible and the lifetime is measurable, while red corresponds to negative $\Delta_{\rm PSD}$, implying that the lifetimes are not measurable. In this latter case, a lower limit on the GMC lifetime can in principle be obtained.

Fig.~\ref{fig:measure_GMC_lt} highlights that shorter GMC lifetimes are easier to measure. For a given setup of $\tau_{\rm dep}$ and $\lambda$, shorter GMC lifetimes move the GMC PSD break to shorter timescales, making it more apparent. Furthermore, GMC lifetimes are easier to measure if the mass loading is high. A higher mass loading leads to a lower equiliribum timescale $\tau_{\rm eq}$ (Eq.~\ref{eq:eq_timescale}), and therefore a lower normalisation ($\sigma_{\rm reg}$) and a shorter break timescale of the PSD of the basic regulator. Finally, the depletion time mainly sets the break timescale of the PSD of the regulator model, thereby amplifying aforementioned trends. 

In summary, we show that the PSD on timescales of the order to the lifetime of GMCs comprises information about the GMC lifetime. In certain galaxy regimes, in particular for galaxies with high mass-loading factor and short depletion times, the PSD of the star-formation history shows a break, which corresponds to $2\pi\langle\tau_{\rm L}\rangle$. Important to note is that this analysis has been done by solely considering theoretical limitations. Clearly, observational limitations are at least as important to consider. A detailed investigation of the constraining power of observations concerning GMC lifetimes is beyond the scope of this work. Nevertheless, we discuss current observational constraints on the PSD and ideas on how to make progress in Section~\ref{subsec:future}.

\subsection{Evolution about the main sequence ridge-line}
\label{subsec:MS}

\begin{figure*}
    \includegraphics[width=\textwidth]{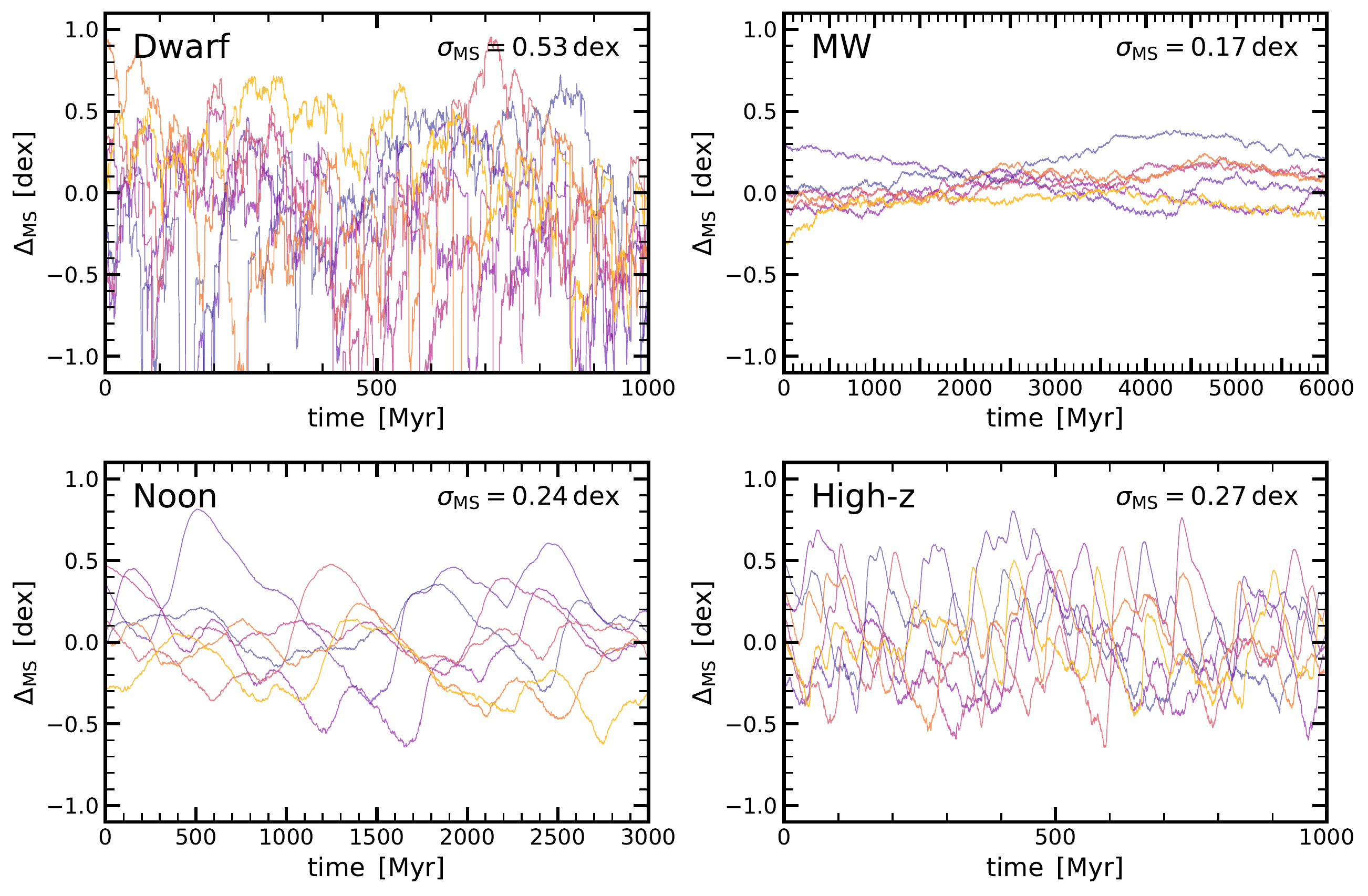} \\ 
    \caption{Movement of galaxies about the main sequence ridgeline. Each panel shows 7 example galaxies from the four different galaxy regimes (Table~\ref{tab:parameters_cases}). The time axis (x-axis) in each of the panels is different. The y-axes show the log distance from the star-forming main sequence ($\Delta_{\rm MS}=\log(\mathrm{SFR}/\mathrm{SFR}_{\rm MS})$, where the SFR of the main sequence, $\mathrm{SFR}_{\rm MS}$, is the overall average SFR. We find that galaxies in the four different regimes move about the MS ridgeline ($\Delta_{\rm MS}=0$) very differently. The width of the star-forming main sequence ($\sigma_{\rm MS}$, measured by averaging the SFR over 1 Myr) is also distinct in the different cases, as indicated in the upper right corners of the panels.}
    \label{fig:MS_movement}
\end{figure*}

\begin{figure*}
    \includegraphics[width=\textwidth]{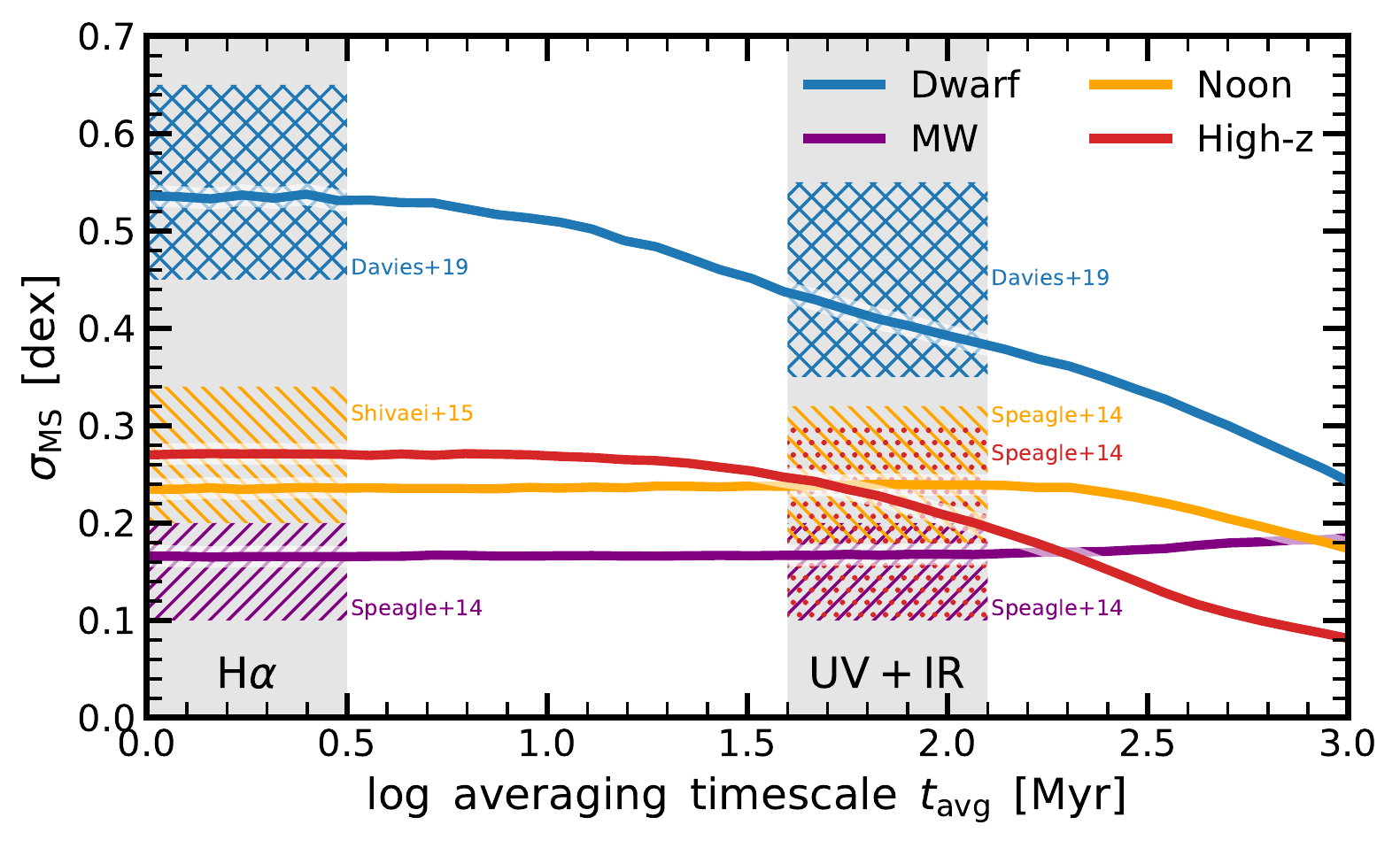} \\ 
    \caption{Scatter of the main sequence for galaxies in the four different regimes (Table~\ref{tab:parameters_cases}). The scatter of the main sequence, $\sigma_{\rm MS}$, is the standard deviation of the logarithm of the measured SFRs. We plot $\sigma_{\rm MS}$ as a function of the timescale $t_{\rm avg}$ over which the SFR is averaged. SFRs measured over longer timescales typically lead to a smaller $\sigma_{\rm MS}$, with the steepness of the decline depending on the timescale over which the SFR is correlated. Bursty star-formation, such as seen in dwarf and high-$z$ galaxies, lead to a larger $\sigma_{\rm MS}$ for SFRs measured on short timescales, but then $\sigma_{\rm MS}$ declines rapidly to longer timescales since one averages over several ``bursts'' of star formation. The vertical grey regions indicate typical averaging timescales for H$\alpha$ and UV+IR SFR indicators; those are approximate since these timescales depend on the burstiness itself. The coloured regions (hatched according to the different galaxy regions) show $\sigma_{\rm MS}$ measurements from observations \citep{speagle14, shivaei15, davies19}. We find overall good agreement for the regimes.}
    \label{fig:scatter_MS}
\end{figure*}

\begin{figure*}
    \includegraphics[width=\textwidth]{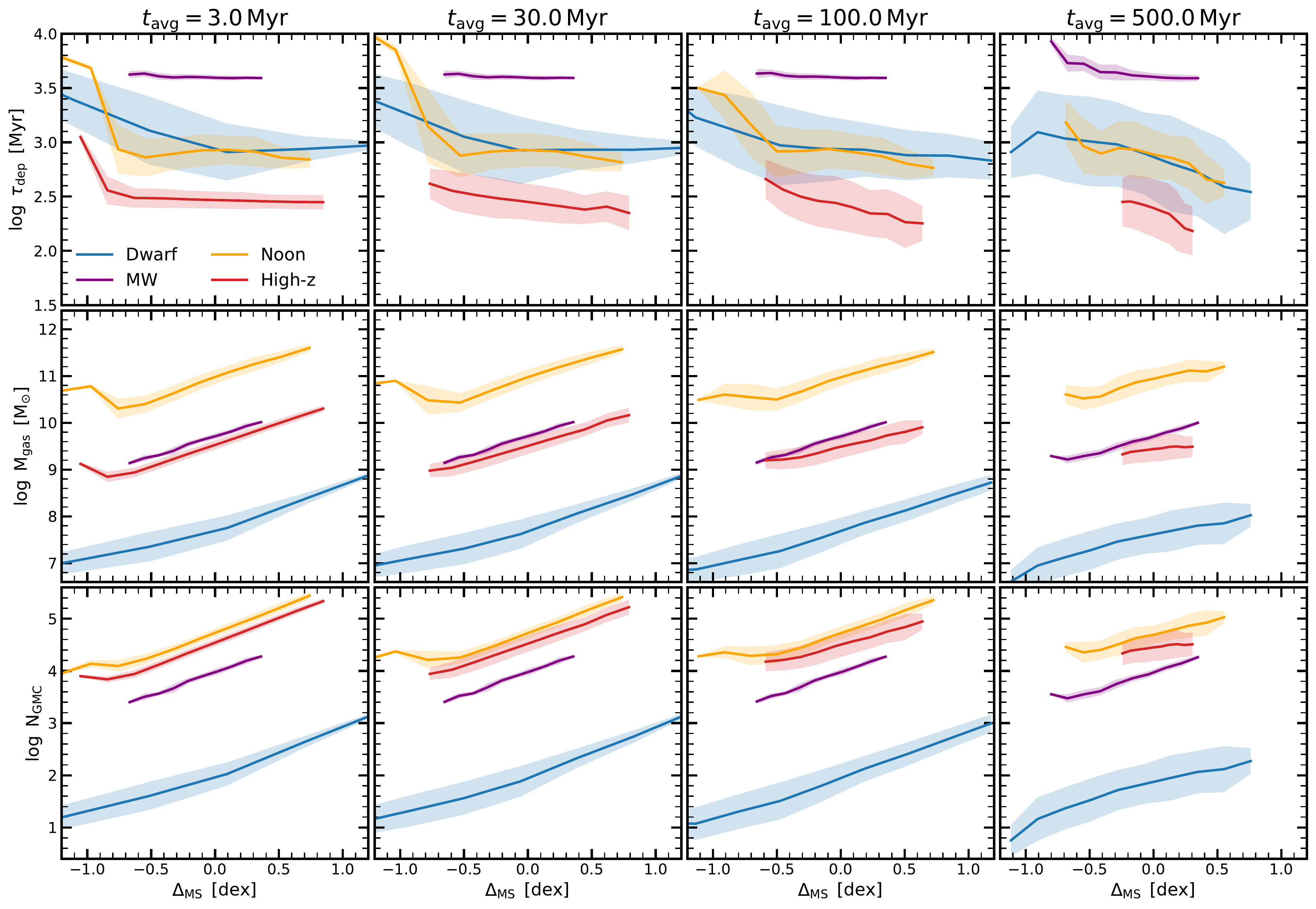} \\ 
    \caption{Gradient across the main sequence. In all panels, the solid lines indicate the median of a certain galaxy property as a function of distance from the main sequence ridgeline ($\Delta_{\rm MS}=\log(\mathrm{SFR}/\mathrm{SFR}_{\rm MS})$), while the shaded region shows the 16th to 84th percentiles. The panels from left to right show the gradients where the SFR is averaged over longer timescales: $t_{\rm avg}=3$, 30, 100 and 500 Myr. The top, middle and bottom panels show the depletion time $\tau_{\rm dep}$, the gas mass $M_{\rm gas}$, and the number of active GMCs $N_{\rm GMC}$. In all panels, the solid lines indicate the median, while the shaded region shows the 16th to 84th percentiles. The blue, purple, orange and red colours indicate the results for the four different galaxy regimes (Table~\ref{tab:parameters_cases}). The gradients depend on both the galaxy regime and the averaging timescale. For example, for High-$z$ galaxies, main sequence gradients in the depletion time are only significantly present if the SFR is average over long timescales ($t_{\rm avg}>10~\mathrm{Myr}$), while the opposite is true for $M_{\rm gas}$ and $N_{\rm GMC}$. This shows that in the High-$z$ regime, short-term fluctuations about the main sequence are mainly set by GMC-related physics, while the longer-time oscillations by depletion time variations.}
    \label{fig:gradient_MS}
\end{figure*}

We mentioned in the Introduction that the main sequence ridgeline (normalisation, slope, scatter and evolution with redshift) is naturally reproduced by a wide range of different galaxy evolution models. However, the key question -- which timescale is encoded in the main sequence scatter ($\sigma_{\rm MS}$)? -- remains both observationally and theoretically unanswered. 

In this work, we model galaxies as self-regulating entities, giving rise to PSDs with breaks related to GMC and gas-regulation physics. These breaks introduce preferred scales related to the underlying physics of star formation. In this picture, the scatter of the main sequence is due to physics related to self-regulation, including feedback processes \citep[e.g.][]{forbes14a, forbes14b, tacchella16_MS}. Interestingly, models that assume a scale-free PSD can equally well describe the evolution of the main sequence evolution and its scatter. Specifically, \citet{kelson20} show that dark matter accretion rates behave as a scale-free, $1/f$ stochastic process -- the same kind of process that was identified at the galaxy level by studying the main sequence of star formation and the stellar mass function \citep{kelson16}. We would argue that this works well because of (i) on short timescales, the coincidence that the scatter of the star-forming main sequence is comparable to the scatter of dark matter accretion rates \citep[e.g.,][]{rodriguez-puebla16}; and (ii) on long timescales, star-formation histories are well described by dark matter accretion histories \citep[e.g.,][]{behroozi19}.

Clearly, progress can be made by studying the PSD of SFR fluctuations of main sequence galaxies, measuring the slope and identifying breaks \citep{caplar19}. Another probe that could potentially shed light onto main sequence oscillations are gradients across the main sequence. For example, observations indicate the the distance from the main sequence is correlated with the depletion time \citep[e.g.][]{genzel15, tacconi18, freundlich19}.

\subsubsection{Scatter of the main sequence}

We first study the scatter of the main sequence. Fig.~\ref{fig:MS_movement} shows how galaxies evolve relative to the main sequence ridgeline. Each panel shows 7 example galaxies from the four different galaxy regimes (Table~\ref{tab:parameters_cases}). The y-axis shows the log distance from the star-forming main sequence ($\Delta_{\rm MS}=\log(\mathrm{SFR}/\mathrm{SFR}_{\rm MS})$, where the SFR of the main sequence, $\mathrm{SFR}_{\rm MS}$, is the overall average SFR\footnote{We ensure that that there is no imprint of the initial conditions.}. We find that galaxies in the four different regimes move about the MS ridgeline ($\Delta_{\rm MS}=0$) very differently. Firstly, the amplitudes of the oscillations about the ridgeline vary between the different galaxy regimes. This is directly reflected in the main sequence scatter, $\sigma_{\rm MS}$, which is the standard deviation of $\sigma_{\rm MS}$ after the SFR is averaged over 1 Myr (for more on this see below). We find a scatter of $\sigma_{\rm MS}=0.17$ dex, 0.24 dex, 0.27 dex, and 0.53 dex for the MW, Noon, High-$z$ and Dwarf regimes, respectively. Secondly, the asymmetry between the excursions above and below the main sequence is different in the different regimes. For bursty star formation, where the scatter is large, galaxies are able to dive significantly further below the ridgeline than above it. No galaxies in the Dwarf regime have $\Delta_{\rm MS}>1$ dex during the considered time intervals, but they have several episodes with $\Delta_{\rm MS}<-1$ dex. On the other hand, galaxies in the Noon and MW regimes are moving about the ridgeline more symmetrically and can also spend long periods (several ten of dynamical times) above or below the main sequence. 

The drivers of the scatter in the main sequence are discussed further in the following subsection by analysing how the gas mass, depletion time, and number of GMCs vary across the main sequence. On short timescales, a galaxy's position relative to the main sequence, and hence ultimately the overall scatter of the main sequence in that regime, is driven primarily by the number of GMCs. Because so few GMCs are active at any one time in the dwarf galaxy regime, their SFRs may deviate asymmetrically far below the main sequence, and their scatter may substantially exceed the variance in the accretion rate (which we have set to be the same in all four galaxy regimes). In the basic regulator model, this cannot happen \citep[as in][where the scatter in the main sequence was always $\le$ the scatter in the accretion rate]{forbes14b}. Remarkably this means that the physics of individual GMCs has a direct impact on $\sigma_\mathrm{MS}$, a quantity typically viewed as the domain of larger-scale galaxy evolution processes.

Clearly, the resulting scatter of the main sequence is very different in the different regimes. \citet{caplar19} discuss extensively how the scatter (and also normalisation) of the main sequence depends on the averaging timescale \textit{as well as} the PSD itself. Therefore, the scatter of the main sequence measured with different SFR indicators (i.e. averaging timescales\footnote{As discussed in \citet{caplar19}, SFR indicators are not strictly speaking simple averages, but actually convolutions, of the past star-formation history.}) allows constraints on the PSD. As an example, we plot in Fig.~\ref{fig:scatter_MS} the scatter of the main sequence $\sigma_{\rm MS}$ as a function of averaging timescale over which the SFR has been measured. We show the four galaxy regimes (Dwarf, MW, Noon, and High-$z$) that we introduced in Section~\ref{subsec:case_studies}. The vertical grey region indicate typical averaging timescales for the H$\alpha$ and UV+IR SFR indicators; those are approximate since these timescale depend on the burstiness itself. Since the dwarf and high-$z$ galaxies have the largest variability on short timescales, $\sigma_{\rm MS}$ is the largest for those cases when measured on short ($<10~\mathrm{Myr}$) timescales. However, because their star-formation history quickly decorrelates, $\sigma_{\rm MS}$ declines with increasing averaging timescale: one basically averages over several bursts, i.e. the measured SFR is not able to trace these short-term bursts anymore. On the other hand, the star-formation histories in the MW and Noon cases are correlated on longer timescales, which gives rise to a weaker decline of $\sigma_{\rm MS}$ with averaging timescale. This decline can be derived from first principles (see Eq. 16 in \citealt{caplar19}), i.e. one can relate the measured $\sigma_{\rm MS}$ to the intrinsic scatter of the main sequence via the auto-correlation function of the underlying stochastic star-formation processes. This decline of $\sigma_{\rm MS}$ with longer averaging timescale is also consistent with results of numerical zoom-in simulations, see Fig. 12 of \citet{hopkins14}.

The coloured regions in Fig.~\ref{fig:scatter_MS} indicate observational constraints from the literature. Measuring the scatter of the main sequence from observational data is difficult, since the intrinsic scatter of the main sequence ($\sigma_{\rm MS}$) is convolved with errors in observing/deriving relevant physical quantities (i.e., photo-$z$, stellar mass, and SFR) and, when the data is binned in $z$-intervals, with the evolution of the main sequence ridgeline itself within the given time interval. These corrections have been considered in detail in \citet{speagle14}. We therefore rely on those $\sigma_{\rm MS}$ measurements wherever available\footnote{The work by \citet{speagle14} is based on a large literature compilation, including \citet{elbaz07, elbaz11, karim11, lee12, reddy12a, rodighiero11, salim07, santini09, sobral14, steinhardt14_splash, whitaker12}.}. We supplement these observational constraints by measurements of \citet{davies19} for the Dwarf regime and of \citet{shivaei15} for the Noon regime for the H$\alpha$ SFR tracer, applying corrections as outlined in  \citet{speagle14}. 

Although the observational constraints are quite broad, we can find clear indications that the scatter of the main sequence depends on the galaxy regime as well as the tracer used. Our model regimes lie in the ballpark of the observations, reproducing the observations both qualitatively and quantitatively. For the High-$z$ regime, we are still missing direct estimates of H$\alpha$-based SFRs. Our model predicts a scatter of $\sigma_{\rm MS}(\mathrm{H}\alpha)\approx0.3$ dex, which can be tested with the upcoming James Webb Space Telescope.

\subsubsection{Gradients about the main sequence}

Secondly, we now turn to main sequence gradients. In general, variation in the SFR for a given set of galaxies with the same underlying PSD may be driven by variation in the gas mass, or variation in the depletion time. In the basic regulator model, depletion time gradients do not exist unless they are imposed by hand since the depletion time is assumed to be constant and the SFR is given by $\mathrm{SFR}=M_{\rm gas}/\tau_{\rm dep}$ at all times. Hence all variability in the SFR comes from variability in the gas reservoir. However, in the extended regular model, the latter equation only holds on long timescales. Therefore we expect some depletion time variations across the main sequence. 

Fig.~\ref{fig:gradient_MS} shows how the median depletion time (top panels), gas mass (middle panels), and number of active GMCs (bottom panels) vary across the main sequence. The distance from the main sequence is defined as $\Delta_{\rm MS}=\log(\mathrm{SFR}/\mathrm{SFR}_{\rm MS}$), where $\mathrm{SFR}_{\rm MS}$ is the SFR of the main sequence. In the context of our model, each galaxy case study is run for a long time; the resulting star-formation history is then filtered with a top-hat window of the given averaging timescale. $\mathrm{SFR}_{\rm MS}$ is then the average SFR of this filtered time series, and each point in the filtered time series may then be assigned a value of $\Delta_\mathrm{MS}$. In each panel, we show the gradients for the four different galaxy regimes. From left to right, the columns show how the gradients depend on increasing values of the averaging timescale of the SFR ($t_{\rm avg}$). 

Fig.~\ref{fig:gradient_MS} may be interpreted to show which galaxy properties are responsible for main sequence oscillations on certain timescales for the different galaxy regimes. When averaging over short timescales ($t_{\rm avg}<10~\mathrm{Myr}$), we only find weak $\tau_{\rm dep}$ gradients across the main sequence in all galaxy regimes (top left panel of Fig.~\ref{fig:gradient_MS}). Averaging over longer timescales increases the strength of the $\tau_{\rm dep}$ gradients: averaging the SFR over $>100$ Myr gives rise to a depletion time gradient of $\tau_{\rm dep} \propto \Delta_{\rm MS}^{-\gamma}$ with $\gamma\approx0.1-0.5$, depending on the galaxy regime. Recent observations of massive galaxies at $z=0-3$ reveal gradients consistent with these estimates \citep{genzel15, tacconi18, freundlich19}, the exact number depending on how upper limits and incomplete data are treated (Feldmann in prep.).

For the $M_{\rm gas}$ gradients across the MS (middle panels of Fig.~\ref{fig:gradient_MS}), we find the opposite behaviour. Specifically, shorter averaging timescales lead to stronger gradients in all galaxy regimes. For $t_{\rm avg}=3$ Myr, we find $M_{\rm gas} \propto \Delta_{\rm MS}^{\sim 1}$ in the MW, Noon, and High-$z$ regimes and a slightly shallower dependence in the Dwarf regime. Averaging over longer timescales weakens these gradients in all galaxy regimes, except the MW and Noon regimes, where the gradient roughly remains. Looking at the number of active GMCs ($N_{\rm GMC}$; bottom panels of Fig.~\ref{fig:gradient_MS}), we find a similar behaviour as for the $M_{\rm gas}$. This is not too surprising since $M_{\rm gas}$ and $N_{\rm GMC}$ are closely related to each other. This effect may be amplified by considering the dynamical influence of galaxy morphology on GMC formation. For instance \citet{martig09} and \citet{gensior20} show that simulated galaxies have less star formation (and presumably fewer GMCs) if they have more massive spheroids, which tend to stabilize the disk against local gravitational collapse.

In summary, the gradients across the main sequence depend on the averaging timescale of the SFR (i.e. the SFR tracer) and, hence, can be used to pin down which physical process is responsible for main sequence oscillations on certain timescales. The exact dependence of galaxy properties on the distance from the main sequence depend on the galaxy regimes. Typically, we find that the short-term oscillations ($<10$ Myr) are mainly set by the number of active GMCs and the gas mass in the reservoir, while the longer-term oscillations ($>10$ Myr) contain a contribution from depletion time variations, which come from variability in the inflow $\Phi$.

\subsection{Clumpy galaxies at $z\approx1.5-3.0$}

\label{subsec:clumpy_galaxies}
\begin{figure}
    \includegraphics[width=\linewidth]{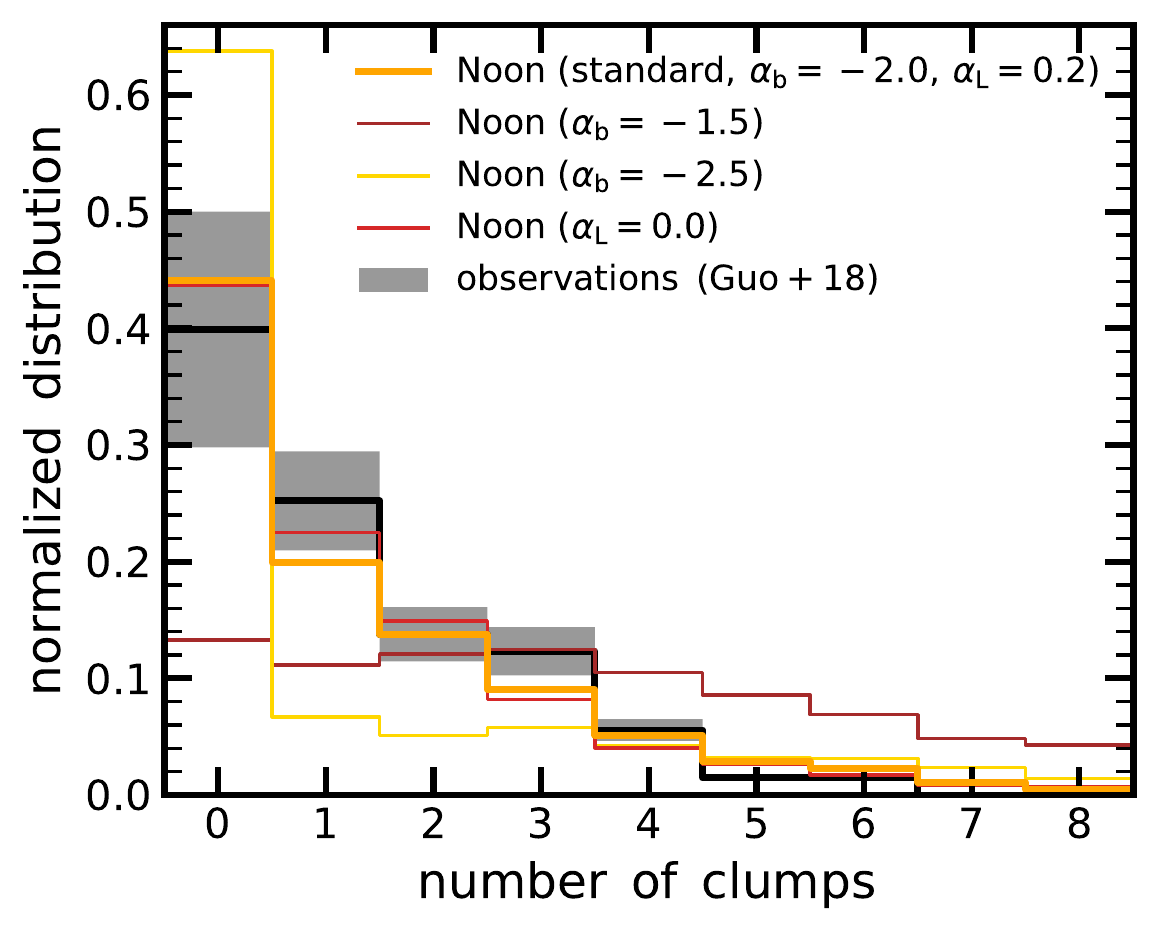}
    \caption{Distribution function of number of giant star-forming clumps in galaxies at $z\approx1.5-3.0$. The orange line shows our model prediction for massive, star-forming galaxies at $z\sim2$ (Noon regime as specified in Table~\ref{tab:parameters_cases}). We indicate variants of the Noon regime with brown (shallower mass function with $\alpha_{\rm b}=-1.5$), yellow (steeper mass function with $\alpha_{\rm b}=-2.5$), and red lines (lifetimes have no mass dependence, i.e. $\alpha_{\rm l}=0.0$). We define giant star-forming clumps as entities with a molecular gas mass of $>5\times10^8~M_{\odot}$. We find that $\sim65\%$ of the galaxies have at least one clump and that galaxies have on average 1.5 clumps. We compare our model predictions with observations by \citet{guo18}, focusing on star-forming galaxies at $z\approx1.5-3.0$ with $\log(M_{\star}/\mathrm{M}_{\odot})=10-11$. The standard Noon model agrees well with the observational data, indicating that these observed star-forming clumps might indeed be individual star-forming regions. }
    \label{fig:clump_count}
\end{figure}

Observations of star-forming galaxies at $z\approx1.5-3.0$ show evidence of giant kpc-scale clumps of star formation \citep[e.g.][]{conselice04, elmegreen06, genzel08, forster-schreiber11a, guo12, wuyts12}, which are extremely rare in massive, low-redshift galaxies. These star-forming clumps are of great interest because their occurrence and lifetimes are highly sensitive to the stellar feedback prescription in galaxy formation models \citep[e.g.,][]{mandelker16, oklopcic17}. However, their exact nature and properties are still debated. For example, it is unclear whether clumps are single entities or rather clusters of small star-forming regions, blurred into kpc-size clumps due to lack of spatial resolution \citep{behrendt16}.

In this short sub-section, we briefly investigate if such giant, star-forming clumps are present in our model. We study their occurrence, but we postpone a more detailed discussion and comparison to observations to future work. We focus now on massive, star-forming galaxies at $z\approx1.5-3.0$. Specifically, we adopt the model runs from the previously introduced Noon regime (Table~\ref{tab:parameters_cases}), which has a GMC mass function with a high-mass cutoff of $m_{\rm max}=10^{9}~\mathrm{M}_{\odot}$. For simplicity, we define these giant star-forming clumps as entities with molecular gas masses of $>5\times10^{8}~M_{\odot}$. This threshold is motivated by observations that indicate the these clumps have stellar masses of $10^8-10^9~M_{\odot}$ \citep[e.g.][]{forster-schreiber11b, guo18}. We randomly choose a galaxy from the Noon regime and the epoch of observation (timestep in the timeseries), and then count the number of such massive clumps. We reduce the number of clumps by 1 to make it comparable to observations, as a 1-clump galaxy would not be identified as clumpy.

Fig.~\ref{fig:clump_count} shows the normalized distribution of number of giant, star-forming clumps. The distribution peaks at 0 and has a tail up to $\sim8$ clumps. We find that $\sim65\%$ of the galaxies have at least one clump, while the average per galaxy is 1.5 clumps. In our model, these clumps have typically a lifetime of $\mathrm{\tau_{\rm L}}\approx150~\mathrm{Myr}$, an integrated star-formation efficiency of $\mathrm{\varepsilon}\approx0.2$, and a star-formation rate of $\mathrm{SFR}\approx1-10~M_{\odot}/\mathrm{yr}$, which is about $10-20\%$ of the total SFR. Fig.~\ref{fig:clump_count} also shows variants of the Noon model. We find that the adopted birth mass function has the largest impact on the clump distribution. The brown and yellow line indicate the clump distribution assuming a shallower ($\alpha_{\rm b}=-1.5$) and steeper ($\alpha_{\rm b}=-2.5$) mass function. Adopting no mass dependence on the GMC lifetimes ($\alpha_{\rm l}=0.0$, red line) has a negligible effect.

We compare our occurrence of clumps with observations by \citet[][see also \citealt{guo15}]{guo18}. They provided a sample of 3193 clumps detected from 1270 galaxies at $0.5<z<3.0$. We select star-forming galaxies at $z\approx1.5-3.0$ with $\log(M_{\star}/\mathrm{M}_{\odot})=10-11$. We plot the distribution of the number of clumps per galaxy as a black line in Fig.~\ref{fig:clump_count}. We obtained a rough estimate of the error from the uncertainties on the fraction of clumpy galaxies in \citet{guo15}. Our model agrees well with this observational estimate, qualitatively reproducing the trend that $\sim40\%$ of the galaxies have no such massive, off-centred clump. Furthermore, the observed galaxies at this epoch and mass range have on average 1.3 clumps, consistent with our model. Finally, the aforementioned physical properties of the clumps are also in rough agreement with observations \citep[e.g.][]{guo18, zanella19}.

In summary, our model with the parameters from the Noon regime naturally reproduces massive, star-forming clumps. This indicates that the observed giant star-forming clumps may indeed be individual star-forming regions, and not composed of a cluster of small star-forming regions. Specifically, these giant clumps might be just the upper tail of of the GMC mass function and therefore a sensitive probe of it.

\subsection{Caveats}
\label{subsec:caveats}

The strength of our approach lies in the simplicity of the physical model: it allows us to explore a wide range of parameters and understand with analytical arguments the emerging trends. At the same time, this simplicity is also a weakness because we might miss certain physical processes. 

In particular dynamical processes, such as spiral arms or bars inducing star formation, can give rise to correlated star formation on roughly the dynamical timescale, which is longer than the GMC lifetime, but shorter than the equilibrium timescale in most galaxies \citep[e.g.][]{jeffreson18}. For example, bars are able to trigger central star formation \citep[e.g.,][]{athanassoula13, chown19}, while spiral arms can sweep up diffuse gas thereby leading to star formation \citep[e.g.,][]{dobbs14_spiral}. Relatedly, the star-formation activity in the Central Molecular Zone (CMZ, i.e. the central 500 pc) of the Milky Way is consistent with episodic cycles of star formation, where a ring-shaped region at $\sim100~\mathrm{pc}$ undergoes episodic starbursts, with bursts lasting $\sim5-10$ Myr occurring at $\sim20-40$ Myr intervals \citep{kruijssen14_center, krumholz15_center, krumholz17}. Such process can approximately be captured by modifying the inflow rate $\Phi$ and/or the star-formation prescription $\mathcal{G}(t)$ by implementing stochastically varying $\tau_{\rm mol}$. However, the detailed investigation of this is beyond the scope of this paper.

Additionally, although the parameterization of the inflow rate is flexible, it does not directly couple to the outflow rate. Therefore, any connection between the expelled and re-accreted gas (``gas recycling'') might be underestimated. Based on the FIRE simulations, \citet{angles-alcazar17_baryoncycle} show that in galaxies of all masses wind recycling can be the dominant accretion mode, with median recycling times of $\sim100-350~\mathrm{Myr}$. In principle, our model could be extended further to include this mode of accretion explicitly. However, since our inflow rate prescription is flexible, we believe that the stocasticity of the SFR will not be changed significantly beyond what we have explored here.

Related to the outflow prescription are processes that cease the SFR of galaxies (quenching). Throughout this paper, we focus on star-forming galaxies and do not attempt to model quenching. As discussed in Iyer et al. (in prep.), quenching leads to an increase in power on long timescales. In the future, we can expand on the model presented here by including processes such as quenching and by modelling galaxies in a cosmological context, ensuring that the long-timescale PSD is consistent with expectation from $\Lambda$CDM.

Furthermore, our GMC formation and evolution model is rather simplistic. \citet{lee16} argue that the GMC properties, in particular the star-formation efficiency per free-fall time, changes with time for individual GMCs in order to explain the large dispersion in the rate of star formation. On the other hand, we assume the the GMC properties are constant and are set based on the birth mass of the GMC, without any scatter. Again, we can in principle extend our model to make the star-formation efficiency (as well as the mass loss) of GMCs time-dependent. This will change the details of the gas cycling within galaxies, however, not affect the stochasticity of the star formation. 

In this work, we also assume that the GMC population consists of independent star-forming region. As we discuss in Section~\ref{subsec:GMC}, the interstellar medium is turbulent and clouds undergoing collapse generally have several different centres, making it difficult to count \textit{independent} star-forming regions. We could imagine that the birth rate and lifetime distribution of GMC of different masses depend on the previous history. For example, a massive GMC will affect neighbouring, smaller GMCs by for example disrupting them, but also by increasing their formation rate. This correlation in the GMC lifetimes and number density would affect the PSD, but it is difficult to capture in our model. Numerical models could be used to estimate the importance of this effect. Here we basically use the  operational definition that an independent region is one that resides on an evolutionary timeline in a way that is independent of its neighbours on that timescale (see also \citealt{kruijssen14, kruijssen18} for a discussion of the independence of star-forming regions).

Beyond uncertainty in modelling the GMCs, there is stochasticity in the values of SFR indicators produced by random sampling from the IMF \citep{da-silva12, krumholz15}. In particular, most star formation indicators are correlated with the effects of rare, massive stars, so for SFRs below about $10^{-3} M_\odot/\mathrm{yr}$, draws from the IMF may begin to dominate the power spectrum. 

Finally, throughout this work, we represent the variability of star formation with a PSD. The power spectrum is a quadratic descriptor of a random field: it contains information about the amplitudes of the Fourier components, but not about any phase relationships that might have evolved through nonlinear processes \citep[e.g.,][]{jones04}. The PSD characterises fully a Gaussian random field, which has Fourier modes that are independent. However, in principle, a star-formation history could have Fourier components with higher order correlations. The next order descriptors are cubic: the three-point correlation function and its Fourier counterpart, the bi-spectrum. However, since the PSD is already challenging to measure observationally (see next section), we postpone the exploration of the bi-spectrum to later. Nevertheless, it would be interesting to quantify the importance of the bi-spectrum of the inflow rate and SFR in numerical simulations.

\subsection{Way forward: numerical simulations and observations}
\label{subsec:future}

In this work we propose that the variability of star formation can be well described by the PSD, which is shaped by a combination of processes related to the small-scale physics of star formation and the overall, large-scale gas cycle of galaxies. The key predictions of this paper, in particular the shape of the PSD, can be tested in more advanced, numerical simulations, which model the star-formation and feedback processes self-consistently. This could help to address some of the aforementioned caveats.  

Iyer et al. (in prep.) study the PSD of the star-formation histories of galaxies from an extensive set of simulations, ranging from cosmological  hydrodynamical simulations (Illustris, IllustrisTNG, Mufasa, Simba, Eagle), zoom simulations (FIRE-2, g14, and Marvel/Justice League), semi-analytic models (Santa Cruz SAM) and empirical models (UniverseMachine). They find that variability on long timescales accounts for most of the power in galaxy star-formation histories, and is tied to the dark matter accretion histories of their parent haloes and quenching of star formation. However, this alone is insufficient to explain the overall shape of star-formation histories, since the PSDs exhibit a broken power-law behaviour, where the timescale of the breaks as well as the high frequency slope vary significantly between different simulations. They highlight that observational constraints in the PSD space will help constrain the relative strengths of the physical processes responsible for these star-formation fluctuations. 

More generally, zoom-in simulations such as VELA \citep{ceverino14_radfeed, zolotov15} and FIRE \citep{hopkins14, hopkins18_FIRE2} allow for resolving GMC-based star formation in a cosmological context \citep{mandelker16, oklopcic17, benincasa19}. Similarly, idealised simulations allow for studying the formation and disruption of GMCs \citep[e.g.,][]{semenov18, li18}. Such simulations are well suited to test some of our predictions. Specifically, one can indeed study how the GMC lifetime distribution shapes the PSD on short timescales, and how this is connected to the overall gas cycle on longer timescales. Fixed-volume cosmological simulation are making tremendous progress in resolving the sites of star formation \citep[e.g. TNG50;][]{nelson19, pillepich19}, but still miss a factor of a few in resolution to self-consistently model the interstellar medium and hence GMCs. Nevertheless, these simulation can be used to test our predictions on longer timescales, particularly constraining the PSD shape on the timescale of the depletion timescale. Furthermore, it would be interesting to understand the details of how black hole feedback and dark matter accretion shape the PSD on the longest timescales ($>1~\mathrm{Gyr}$).

After showing that the PSD is a powerful measure of star-formation variability that allows us to constrain models of galaxy formation (Iyer et al. in prep), the next step is to move to observations. \citet{caplar19} put forward the idea of how the PSD can be used to quantify the variability of star formation and provide a first observational estimate. They show how the measurements of the normalisation and width of the star-forming main sequence, measured in several pass-bands that probe different time-scales, give a constraint on the parameters of the underlying PSD, assuming that the galaxy sample can be characterized by a single PSD. They first derive these results analytically for a simplified case where they model observations by averaging over the recent star-formation history. However, since SFR indicators are not just simple averages of the past star-formation history, they show that more realistic observational cases need to be treated numerically. As a proof of concept, \citet{caplar19} use observational estimates from the GAMA survey of the main sequence scatter at $z\sim0$ and $M_{\star}\approx10^{10}~\mathrm{M_{\odot}}$ measured in H$\alpha$, UV+IR, and the $u$-band \citep{davies19}. Assuming a high-frequency slope of $\alpha=2$, they find a break timescale of $170^{+169}_{-85}~\mathrm{Myr}$, which implies that star-formation histories of galaxies lose ``memory'' of their previous activity on roughly the dynamical timescale. They highlight several caveats in their analysis, including that the dust attenuation correction for the different SFR indicators is uncertain. 

\citet{wang20} infer the SFR averaged over 5 Myr, 800 Myr, and the ratio of the two by using D$_n$(4000), EW(H$\delta$) and EW(H$\alpha$) from SDSS/MaNGA. They use these quantities to constrain the PSD. They characterise the PSD with a single power law (i.e. without breaks), meaning they assume that the star-formation histories of individual galaxies are effectively correlated over the age of the universe (similar to the approach by \citealt{kelson14}). They also explore the consequences of existence of the variation of the SFR within the population at a given epoch that are unrelated to the temporal variations of individual galaxies which they call the ``intrinsic scatter''. The PSD is strongly degenerate with the assumed value of the intrinsic scatter, but if the intrinsic scatter is subdominant they find that the PSDs of star-forming galaxies with $M_{\star}=10^{8.5}-10^{11.0}~M_{\odot}$ have slopes between 1.0 and 2.0, consistent with the slopes inferred by \citet{caplar19} and with the slopes found in this work. As discussed in \citet{caplar19}, a limited number of observational constraints allows only a limited number of inferences about the properties of star-formation variability.

Moving forward, an important step lies in breaking the degeneracy between the star-formation history of galaxies and other stellar population parameters, such as dust attenuation, metallicity, and the initial mass function. A full Bayesian approach in modelling the spectral energy distribution allow estimates of the co-variance between different parameters and therefore a more robust determination of the star-formation history \citep[e.g.][]{iyer17, iyer19, leja17, leja19_nonparm, carnall19_sfh}. Under the assumption that a galaxy ensemble can be constructed that follows a single PSD, one possibly could constrain the PSD via hierarchical modelling \citep[e.g.][]{curtis-lake20}. Future work will provide guidance of how to best tackle the challenge of measuring the PSD observationally.

\section{Conclusions}
\label{sec:conclusions}

It is challenging to gain further insight on star formation and feedback because these processes act on multiple spatial and temporal scales. In this paper, we address the question of how different processes -- external and internal to galaxies such as formation and distribution of GMCs, spiral arms, galaxy-galaxy mergers, and galaxy- and halo-scale cosmological inflows and outflows -- drive the variability of star formation on a wide range of timescales. We use the power spectral density (PSD) of the star-formation history to quantify the star-formation variability: the PSD shows the strength of SFR variations as a function of frequency. 

In order to probe a wide range of different galaxy regimes and parameter space, we build our analytical model on the previously studied regulator model, which assumes mass conservation of the gas reservoir. We add more physical realism to the model by assuming that the SFR of a galaxy is sustained by a population of GMCs. In this extended regulator model, the variability in the SFR is driven by both the variability of the gas inflow rate and the stochasticity of GMC formation. Our key conclusions are the following:

\begin{itemize}
    \item The PSD of the star-formation history generically has three breaks, corresponding to the correlation time of the inflow rate, the equilibrium timescale of the gas reservoir of the galaxy, and the average lifetime of individual molecular clouds. On long and intermediate timescales (relative to the dynamical timescale of the galaxy), the PSD is typically set by the variability of the inflow rate and the interplay between outflows and gas depletion. On short timescales, the PSD shows an additional component related to the life-cycle of molecular clouds, which can be described by a damped random walk with a power-law slope of $\beta\approx2$ at high frequencies and a break proportional to the average cloud lifetime. 
    \item Since the short-term variability of the SFR is set by GMCs, measuring the PSD at high frequencies (short timescales) allows us to learn about star-formation physics, even without spatially resolved observations. Specifically, the PSD break at short timescales enables us to constrain the average GMC lifetime within a galaxy. Short GMC lifetimes in combination with a high mass loading and short depletion time are easier to constrain.
    \item The PSD provides a framework with which one can rigorously define ``bursty'' star formation (short-term versus long-term variability). We study the PSD in four different galaxy regimes in detail: Milky-Way-like galaxies (``MW''), dwarf galaxies (``Dwarf''), massive galaxies at Cosmic Noon ($z\sim2$; ``Noon''), and high-$z$ galaxies ($z>6$; ``High-$z$''). We find that today's dwarf and high-$z$ galaxies are both bursty, but for different reasons. In dwarf galaxies, GMCs are responsible for the power on short timescales, while in high-$z$ galaxies, burstiness is related to the short equilibrium timescale and large variability of the inflow rate.
    \item Consistently, the scatter of the star-forming main sequence $\sigma_{\rm MS}$ for the Dwarf and High-$z$ galaxy regimes are larger than for the MW and Noon regimes when the SFR is measured over short timescales (i.e. using a SFR tracer that is susceptible to short-term fluctuations, such as H$\alpha$). However, because the star-formation histories of galaxies in the Dwarf and High-$z$ regimes quickly decorrelate, $\sigma_{\rm MS}$ declines with increasing averaging timescale of the SFR: one basically averages over several bursts and the SFR is not able to trace these short-term bursts anymore. On the other hand, the star-formation histories in the MW and Noon cases are correlated on longer timescales,which gives rise to a weaker decline of $\sigma_{\rm MS}$ with averaging timescale. Remarkably we find that the physics of individual GMCs has a direct impact on $\sigma_{\rm MS}$ in the Dwarf regime, a quantity  typically  viewed  as  the  domain  of  larger-scale galaxy evolution processes.
    \item Similarly, gradients of galaxy properties (such as the depletion time or the number of GMC) depend on the SFR averaging timescale (i.e. SFR tracer) used: we find that the short-term oscillations ($<10$ Myr) are mainly set by the number of active GMCs and the gas mass in the reservoir, while the long-term oscillations ($>10$ Myr) are controlled by depletion time variations, which come from variability in the inflow.
    \item An additional outcome of our approach is that we model the GMC population of galaxies. We demonstrate that the giant star-forming clumps in massive $z\sim1.5-3.0$ galaxies are naturally produced in our model in the Noon regime. This indicates that these observed star-forming clumps might indeed be individual star-forming regions. Furthermore, we show that the GMC birth mass function can be significantly different from the observed mass function, depending mainly on the GMC lifetime distribution. We show that a single power-law for the birth mass function gives natural rise to an observed mass function that is double power-law, similar to a Schechter function. 
\end{itemize}

Clearly, several of the ideas and results outlined here need to be analysed in more detail with more complex numerical models. Specifically, learning about the GMC-scale star-formation physics from integrated galaxy measurement is of great potential and interest, since it would allow us to shed new light onto star formation within the first galaxies with the upcoming James Webb Space Telescope.

\section*{Acknowledgements}

We thank the referee for carefully reading the manuscript and providing constructive feedback that helped to enhance clarity. We are very thankful to Andreas Burkert for the inspiration of this work. We thank Mike Grudi\'{c}, Charlie Conroy, Margaret Geller, Daniel Eisenstein, Lars Hernquist, Kartheik Iyer, Scott Kenyon, Diederik Kruijssen, Hui Li, Vadim Semenov, and Enci Wang for insightful discussions and useful feedback. S.T. and N.C. thank J.C.F. and the Center for Computational Astrophysics (CCA) of the Simons Foundation for the hospitality during their visit. This research made use of NASA's Astrophysics Data System (ADS), the arXiv.org preprint server, the Python plotting library \texttt{matplotlib} \citep{hunter07}, and \texttt{astropy}, a community-developed core Python package for Astronomy \citep{astropycollaboration13}. The background image of Fig.~\ref{fig:cartoon} is M101 (credit: NASA, ESA, CXC, SSC, and STScI). S.T. is supported by the Smithsonian Astrophysical Observatory through the CfA Fellowship. 

\section*{Data Availability Statement}

Data available on request.




\bibliographystyle{mnras}



\appendix

\section{Mass returned to the gas reservoir}
\label{app:mass_return}

\begin{figure*}
    \centering
    \includegraphics[width=\textwidth]{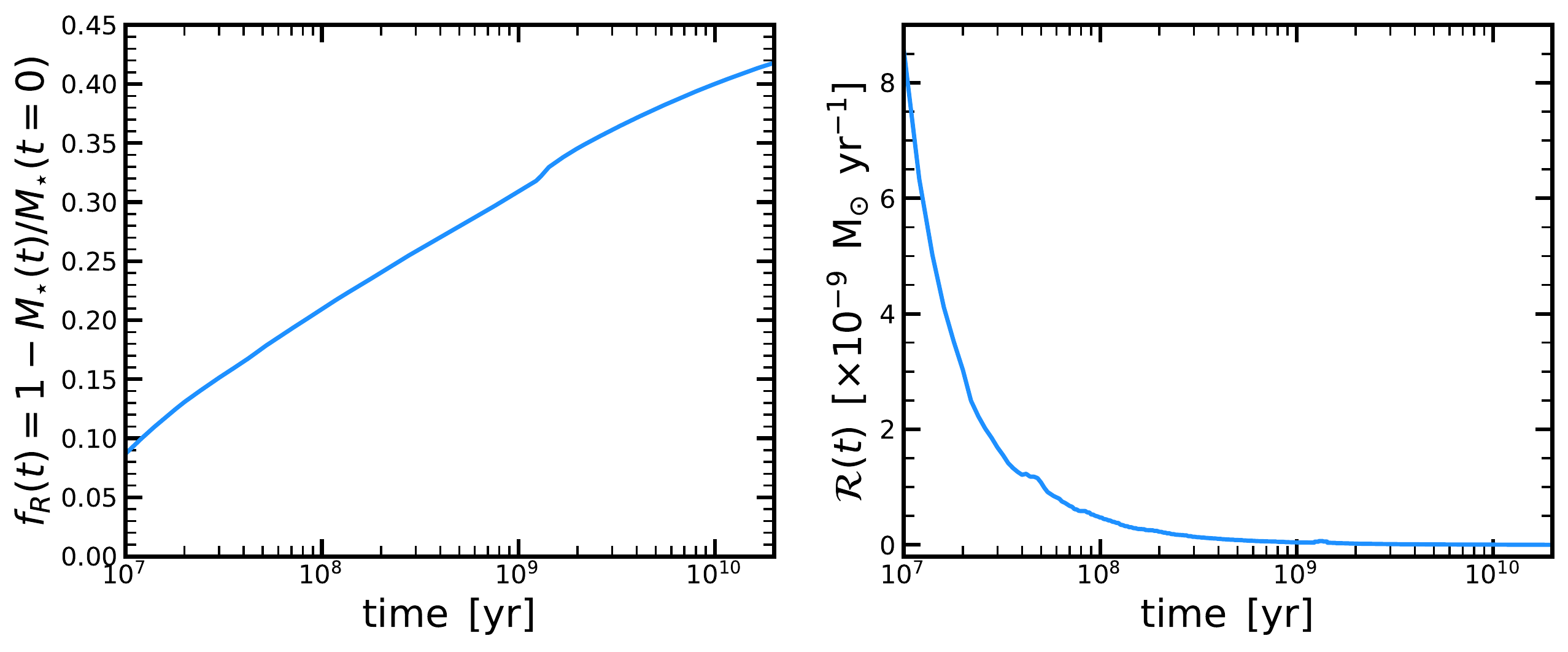}
    \caption{Stellar mass loss from a Simple Stellar Population (SSP). Left panel: fraction of mass returned back to the gas phase ($f_R$) due to winds and supernovae as a function of the SSP's age. Right panel: return rate $\mathcal{R}(t)$ at which mass is transferred from the stellar to the gas reservoir, normalized to an initial SSP mass of $1~\mathrm{M_{\odot}}$.}
    \label{fig:mass_loss}
\end{figure*}

We define the stellar mass $M_{\star}$ to be the mass in stars and remnants, i.e., after subtracting stellar mass loss due to winds and supernovae. This mass returned to the gas reservoir can again be available for star formation. In our model (see Section~\ref{subsec:regulator}), we track this mass return self-consistently and abandon the assumption of instantaneous recycling. Specifically, at each time $t^{\prime}$, we compute from the newly formed stellar mass, $\mathrm{SFR}(t^{\prime})\times\mathrm{d}t$, the mass that will be returned to the gas reservoir at $t>t^{\prime}$. This self-consistent treatment is important in certain regimes, such as at high redshifts, when galaxies are dominated by young stellar population \citep{tacchella18}, or in low-redshift late-type galaxies where the gas from stellar mass loss can provide most or all of the fuel required to sustain the today's level of star formation in \citep{leitner11}.

We estimate the stellar mass loss due to winds and supernovae with the stellar population synthesis code \texttt{Flexible Stellar Population Synthesis} \citep[\texttt{FSPS};][]{conroy09a, conroy10}. In particular, we follow the mass loss of stars due to winds as they evolve along the MIST isochrones \citep{choi16, choi17}. We follow the stars until the end of their lifetime, where we then assign remnant masses according to \citet{renzini93}. We assume solar metallicity and a \citet{chabrier03} intial mass function. Stellar mass loss depends only weakly on metallicity, and therefore changing the metallicity does not affect our results significantly.

Fig.~\ref{fig:mass_loss} shows the fraction of mass returned to the gas reservoir for a Simple Stellar Population (SSP) as a function of time (left panel) and the return rate $\mathcal{R}(t)$ at which this mass transfer takes place (right panel). Most of the mass return takes place within less than a Gyr. After $\sim1$ Gyr, the mass locked in long-lived stars ($=1-f_{\rm R}$) converges to roughly $60\%$. 

\section{Relation between birth and observed GMC mass function}
\label{app:birth_vs_observed_MF}

\begin{figure}
    \includegraphics[width=\linewidth]{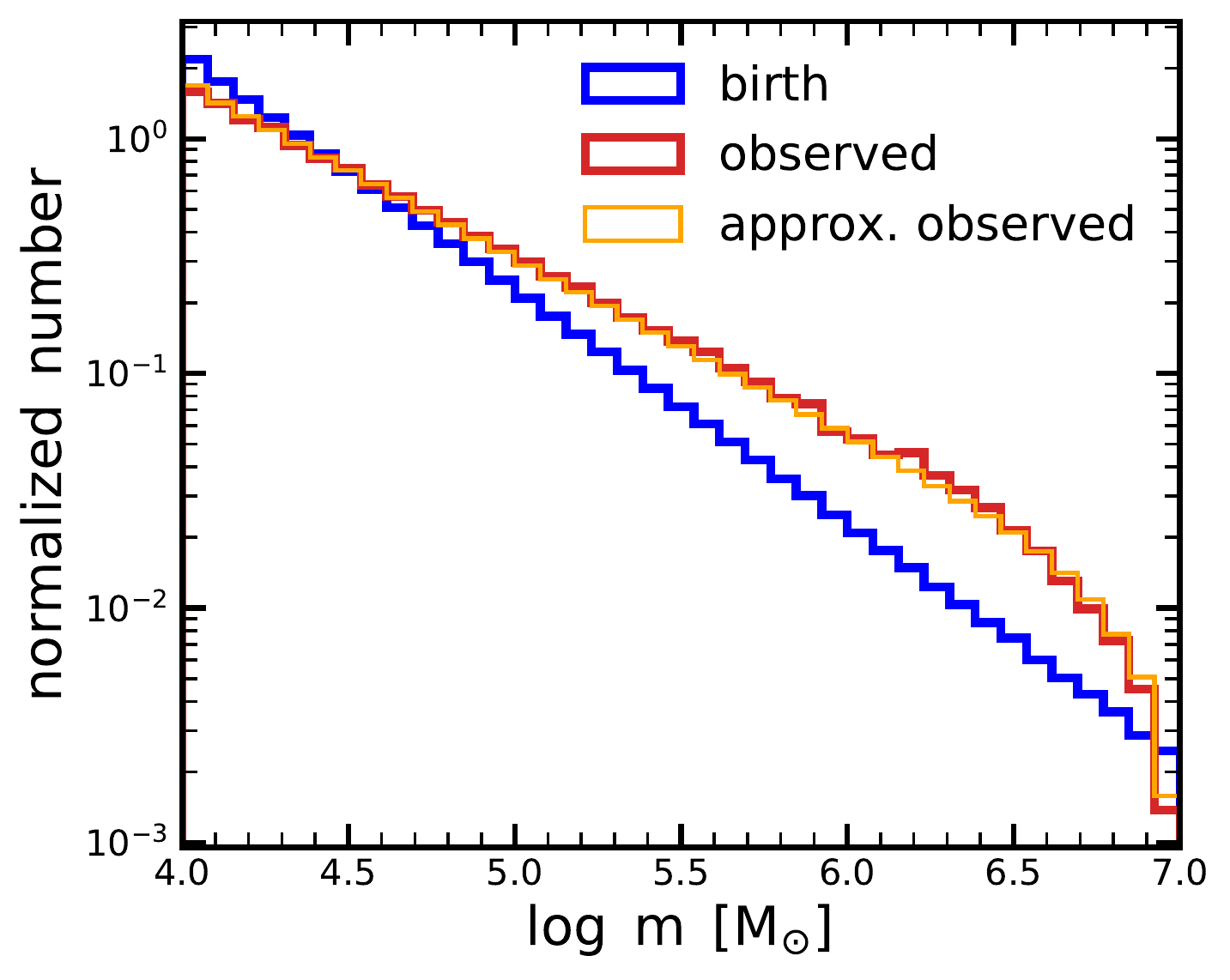}
    \caption{Comparison of the birth and observed GMC mass functions. The blue histogram shows the normalized mass distribution for a population of GMC drawn from the birth mass function (Eq.~\ref{eq:GMC_MF_birth}). The red line shows the observed mass function, computed by sampling the birth mass function over 1000 Myr (1000 time steps) and evolving the GMC population forward in time, assuming the same GMC parameters as in Fig.~\ref{fig:GMC_example}. Since more massive GMC live longer, the observed mass function is shallower than the birth mass function. The orange histogram shows the observed mass function approximated by Eq.~\ref{eq:MF_conversion}, which is in excellent agreement with the numerically calculated observed mass function. The slopes of the birth and observed mass functions are related via the power-law index of the lifetime function, i.e. $\alpha_{\rm obs}=\alpha_{\rm b}+\alpha_{\rm l}$.}
    \label{fig:MF_evolution}
\end{figure}

As discussed in the main text, the birth mass function and the observed mass function differ in the general case, because the GMC lifetime $\tau_{\rm L}$ depends on the birth mass $m_{\rm b}$. As shown in Eq.~\ref{eq:GMC_MF_birth}, we assume that the birth mass function is a simple powerlaw with slope $\alpha_{\rm b}$ and remains constant with time. Therefore, the birth rate function is $\dot{n}_{\rm b}(m_{\rm b})=n_{\rm b}^{\prime}m_{\rm b}^{\alpha_{\rm b}}$. Furthermore, each formed GMC loses mass according to Eq.~\ref{eq:mass_GMC} due to the formation of stars and gas return to the gas reservoir. We can then derive the observed mass function $n_{\rm}$ by integrating over the whole lifetime of the GMCs at different masses:

\begin{equation}
    \begin{split}
        n_{\rm obs}(m) &= \int_{0}^{\infty} \dot{n}_{\rm b}(m_{\rm b}) \mathrm{d}t \\
        &= \int_{0}^{\infty} \dot{n}_{\rm b}(m_{\rm b}) \frac{\mathrm{d}m}{\mathrm{d}m_{\rm b}}  \mathrm{d}t \\
        &= n_{\rm b}^{\prime} m^{\alpha_{\rm b}} \int_{0}^{T_{\rm max}(m)} \left(1-\frac{t}{\tau_{\rm L}(m_{\rm b})} \right)^{1-\alpha_{\rm b}} \mathrm{d}t \\
    \end{split}
\end{equation}

\noindent
where the birth mass $m_{\rm b}$ and the observed mass $m$ at some later time are related via Eq.~\ref{eq:mass_GMC} and $T_{\rm max}(m)$ is the maximal lifetime of a GMC of mass $m$. This maximal lifetime is determined by the upper cutoff of the birth masses, $m_{\rm max}$, i.e., by
\begin{equation}
    T_{\rm max}(m)= \tau_{L}(m_{\rm max}) \left( 1 - \frac{m}{m_{\rm max}}\right).
\end{equation}

The analytical solution of the full integral is unwieldy given the time dependence within the $\tau_{\rm L}$ factor. However, the integral can be easily evaluated in the case when $\alpha_{\rm l}=0$, i.e., when the lifetimes of GMCs are not mass dependent. In this special case:

\begin{equation}
    n_{\rm obs}(m) = n_{\rm b}^{\prime} m^{\alpha_{\rm b}} \frac{\tau_{0}}{-\alpha_{\rm b}} \left[1- \left( \frac{m}{m_{\rm max}} \right)^{-\alpha_{\rm b}} \right].   
\end{equation}

Numerical experiments shows that the final expression, in the cases when $\alpha_{\rm l}$ is small ($\alpha_{\rm l}\lesssim1.0$) can be well approximated by simply replacing $\tau_{0}$ with the full $\tau_{\rm L}$, i.e., 

\begin{equation}
    n_{\rm obs}(m) = n_{\rm b}^{\prime} m^{\alpha_{\rm b}} \frac{\tau_{\rm L}(m)}{-\alpha_{\rm b}} \left[ 1- \left( \frac{m}{m_{\rm max}} \right)^{-\alpha_{\rm b}} \right].
    \label{eq:MF_conversion}
\end{equation}

\noindent
This function has a low-mass power-law slope of $\alpha_{\rm obs}=\alpha_{\rm b}+\alpha{\rm L}$ and a cutoff at high masses, similar to a \citet{schechter76} function. 

Fig.~\ref{fig:MF_evolution} compares the birth mass function with the observed mass function. The blue histogram shows the normalised mass distribution for a population of GMC drawn from the birth mass function (Eq.~\ref{eq:GMC_MF_birth}). The red line shows the observed mass function, computed by sampling the birth mass function over 1000 Myr (1000 time steps) and evolving the GMC population forward in time, assuming the same GMC parameters as in Fig.~\ref{fig:GMC_example}, i.e., $\alpha_{\rm b}=-2.0$, $\tau_0=10.0~\mathrm{Myr}$, and $\alpha_{\rm l}=0.25$. The orange line shows the observed mass function approximated by Eq.~\ref{eq:MF_conversion}, which is in excellent agreement with the numerically calculated observed mass function. Since more massive GMC live longer, the observed mass function is shallower than the birth mass function.

\section{Analytic derivation of the GMC power spectrum}
\label{app:full_derivation}

The contribution to the PSD introduced by having stars form from a fixed reservoir of gas through GMCs is well-described by a broken power-law, with short timescales proportional to $1/f^2$, and long timescales constant. In this appendix we derive this result analytically for the case of a single GMC mass and lifetime, and as in the rest of the paper, assuming that GMCs form stars at a constant rate throughout their lifetimes.

In this regime, SFR $\propto N_\mathrm{GMC}$, so we focus on deriving the PSD of $N_\mathrm{GMC}$. Because the gas reservoir is assumed to be constant, the expected value of $N_\mathrm{GMC}$ is also constant. To simplify the notation, we hereafter drop the GMC subscript on $N$. To compute the PSD, we first compute the auto-covariance, namely

\begin{equation}
    R(\tau) = E[ (N(t)-\langle N \rangle)(N(t+\tau)-\langle N \rangle].
\end{equation}

\noindent
When $\tau > \tau_{\rm L}$, the number of clouds will be completely independent, because any clouds that were active at time $t$ will have vanished by $t+\tau$. The ACF is therefore zero in this case.

When $\tau < \tau_{\rm L}$, some clouds may exist in common between the two epochs. To describe this possibility, we can split $N(t)$ as follows: $N(t) = N_{\mathrm{ind},1} + N_\mathrm{common}$, where $N_{\mathrm{ind},1}$ is the number of clouds that formed before $t+\tau-\tau_{\rm L}$, and are hence guaranteed to be independent of the number of clouds at time $t+\tau$. Similarly, we set $N(t) = N_{\mathrm{ind},2} + N_\mathrm{common}$ where $N_{\mathrm{ind},2}$ is the number of clouds formed after $t+\tau_{\rm L}$. The auto-covariance can then be written

\begin{equation}
\label{eq:acf_gmc}
\begin{split}
    R(\tau) = \langle N_{\mathrm{ind},1} \rangle\langle N_{\mathrm{ind},2} \rangle + \langle N_{\mathrm{common}}\rangle (\langle N_{\mathrm{ind},1}\rangle+\langle N_{\mathrm{ind},2}\rangle) \\ 
    ~ + \langle N_\mathrm{common}^2 \rangle - \langle N \rangle^2 \\ \mathrm{for}\ 0\le \tau<\tau_{\rm L}
    \end{split}
\end{equation}

\noindent
where we have used the fact that $N_{\mathrm{ind},1}$ and $N_{\mathrm{ind},2}$ are independent of each other by construction. These two variables also have the same expectation value, $\langle N \rangle  \tau/\tau_{\rm L}$, i.e. the average fraction of clouds which are not common between the two epochs. The number of common clouds between the two epochs is Poisson distributed with expectation value $\langle N \rangle (1- \tau/\tau_{\rm L})$, so that $\langle N_\mathrm{common}^2 \rangle = \langle N \rangle^2 (1 - \tau^2/\tau_{\rm L}^2) +  \langle N \rangle(1 - \tau/\tau_{\rm L})$.

Inserting each of these averages into Eq.~\eqref{eq:acf_gmc}, every term cancels except $\langle N \rangle(1 - \tau / \tau_{\rm L}$). Finally, to compute the autocorrelation function, we divide out the variance of $N$, which for a Poisson process is just $\langle N \rangle$. Thus the ACF decreases linearly from 1 at $\tau=0$ to 0 at $\tau=\tau_{\rm L}$, and remains 0 thereafter. This ACF is similar to the exponentially-declining case of the damped random walk, and so has similar limiting powerlaw slopes (see below), but the behavior at intermediate frequency values is quite different, with, unsurprisingly, the exponential ACF (damped random walk) case yielding a smoother transition between the two regimes.

Via the Wiener-Khinchin theorem, we can compute the PSD as the Fourier transform of this function,

\begin{equation}
    \mathrm{PSD}_N = \int_{-\infty}^\infty R(\tau)d\tau =  \int_0^{\tau_{\rm L}}\langle N \rangle (1 - \tau/\tau_{\rm L}) e^{-2\pi i f \tau}d\tau.
\end{equation}

\noindent
Integrating by parts and taking the real component yields

\begin{equation}
    \mathrm{PSD}_N = \langle N \rangle \frac{1 - \cos(2\pi f \tau_{\rm L})}{\tau_{\rm L}(2\pi f)^2}
\end{equation}

\noindent
In the limit of high frequencies, i.e. $f \rightarrow \infty$, the PSD behaves as $1/f^2$, while for low frequencies, the leading-order expansion of the Taylor series of cosine yields an $f^2$ term that cancels the denominator's, yielding a constant white noise spectrum. In the general case with a whole spectrum of GMC properties, the higher-order aliases from the cosine term are largely washed out, though they are still visible in the parts of the PSD dominated by GMCs.


\bsp	
\label{lastpage}
\end{document}